\documentclass[epj]{svjour}

\usepackage{amsmath}
\usepackage{amssymb}
\usepackage{mathrsfs}
\usepackage{xspace}
\usepackage{braket}

\usepackage{ifpdf}
\ifpdf
\fi

\newcommand{\eg}{{e.g.,\/}\xspace}
\newcommand{\ie}{{i.e.,\/}\xspace}

\newcommand{\eq}[1]{(\ref{#1})}
\newcommand{\Eq}[1]{Eq.~(\ref{#1})}
\newcommand{\Eqs}[1]{Eqs.~(\ref{#1})} 
 
\newcommand{\Ref}[1]{Ref.~\cite{#1}}
\newcommand{\Refs}[1]{Refs.~\cite{#1}}
\newcommand{\Sec}[1]{Sect.~\ref{#1}}
\newcommand{\Secs}[1]{Sects.~\ref{#1}}
\newcommand{\App}[1]{Appendix~\ref{#1}}

\newcommand{\mc}[1]{\mathcal{#1}}
\newcommand{\mcc}[1]{\mathfrak{#1}}
\newcommand{\msf}[1]{\mathsf{#1}}
\newcommand{\mcu}[1]{\mathscr{#1}}

\newcommand{\favr}[1]{\langle #1 \rangle}
\newcommand{\mvec}[1]{{\boldsymbol{\rm #1}}} 
\newcommand{\oper}[1]{\hat{\mvec{#1}}}
\newcommand{\extd}[1]{\underline{#1}}
\newcommand{\moyal}[2]{\{\!\{#1, #2\}\!\}}
\newcommand{\ph}{\underline{\hspace{2mm}}}
\newcommand{\rdoteq}{\text{ \d = }}

\begin{document}

\title{Geometric view on noneikonal waves}
\author{I.~Y. Dodin}
\institute{Princeton Plasma Physics Laboratory, Princeton, New Jersey 08543, USA}
\date{\today} 

\abstract{
An axiomatic theory of classical nondissipative waves is proposed that is constructed based on the definition of a wave as a multidimensional oscillator. Waves are represented as abstract vectors $\ket{\psi}$ in the appropriately defined space $\Psi$ with a Hermitian metric. The metric is usually positive-definite but can be more general in the presence of negative-energy waves (which are typically unstable and must not be confused with negative-frequency waves). The very \textit{form} of wave equations is derived from properties of $\Psi$. The generic wave equation is shown to be a quantumlike Schr\"odinger equation; hence one-to-one correspondence with the mathematical framework of quantum mechanics is established, and the quantum-mechanical machinery becomes applicable to classical waves ``as is''. The classical wave action is defined as the density operator, $\ket{\psi}\bra{\psi}$. The coordinate and momentum spaces, not necessarily Euclidean, need not be postulated but rather \textit{emerge} when applicable. Various kinetic equations flow as projections of the von Neumann equation for $\ket{\psi}\bra{\psi}$. The previously known action conservation theorems for noneikonal waves and the conventional Wigner-Weyl-Moyal formalism are generalized and subsumed under a unifying invariant theory. Whitham's equations are recovered as the corresponding fluid limit in the geometrical-optics approximation. The Liouville equation is also yielded as a special case, yet in a somewhat different limit; thus ray tracing, and especially nonlinear ray tracing, is found to be more subtle than commonly assumed. Applications of this axiomatization are also discussed, briefly, for some characteristic equations.
\PACS{
 {52.35.-g}{Waves, oscillations, and instabilities in plasmas and intense beams} \and
 {03.50.Kk}{Other special classical field theories} \and
 {45.20.Jj}{Lagrangian and Hamiltonian mechanics} \and
 {02.40.Yy}{Geometric mechanics}
 }
}

\maketitle

\section{Introduction}
\label{sec:bg}

\subsection{Motivation} 

Basic theorems of classical wave physics, such as typical field equations and conservation of the wave action (a linear measure of the ``number of photons''), are well known to describe a tremendously wide variety of physical systems. In literature, these theorems commonly emerge in the context of specific applications, leaving such universality somewhat miraculous. One may wonder then whether a more consistent, \textit{axiomatic} formulation of wave physics is possible that would not appeal to empirical arguments and be abstracted from the wave nature and equations describing particular environments (\eg Maxwell's equations). Apart from the promise to be intellectually stimulating and aesthetically appealing, such a theory could also have a tangible practical value. In particular, it would standardize methods of searching for conservation theorems, otherwise commonly done \textit{ad~hoc}, and find most natural representations of wave dynamics in specific media.

It is the purpose of this paper to propose such an axiomatic formulation by expanding on a similarly-spirited study \cite{my:amc} of the geometrical-optics (GO), or ``eikonal'', limit toward general, or ``noneikonal'', waves. (We will specifically focus on nondissipative linear waves here, but dissipation and nonlinearity could be included too, much like in \Ref{my:amc,my:itervar}, and will also be discussed below, albeit briefly.) The idea is to replace their conventional understanding, which comes through studying properties of characteristic but nevertheless specific models, with a theory that formalizes the wave concept and thereby, for the first time, offers the advantage of true generality. Within this approach, a basic theory, as it turns out, can be constructed deductively and using nothing more than geometric arguments. Extraneous mathematical tools like the Fourier and Wigner-Weyl transforms, which are commonly considered as pillars of the wave kinetic theory (WKT) \cite{ref:mcdonald85,ref:mcdonald91,ref:tracy93,ref:mcdonald88,ref:littlejohn86}, then happen to be redundant for deriving any of the fundamental theorems. Once a wave is actually defined, these transforms rather \textit{emerge}, in a generalized form, and so, in fact, does the very space in which the wave propagates. The axiomatic formalism also happens to reproduce the mathematical framework of quantum mechanics (QM) as a special case. Hence classical and QM waves can be treated on exactly the same footing, which facilitates cross-fertilization of the two fields.

Below we elaborate on the utility and historical context of this approach and describe our specific findings.

\subsection{Historical background} 

Notwithstanding the long history of research in the field of general wave physics, the field still remains in the developing phase and continues to be studied actively on the level of basic formalism  \cite{tex:hirota10}. This situation is in striking contrast to similarly-spirited disciplines like QM that has enjoyed numerous formalizations. The difference might stem from the fact that QM deals with relatively simple Hermitian operators that give rise to conservation laws manifestly, whereas classical waves are much more general. (Another difference, of course, is that the classical-physics community has its traditions rooted in days when an abstract language was not an immediate necessity, unlike in QM, and thus did not seem advantageous in the long run either.) Hence the question regarding the existence and the specific form of their conserved quantities is not at all trivial; \eg see a related discussion in \Refs{ref:stepanov96,my:iorec}.

It is more or less a consensus today in fundamental theory, slowly but steadily penetrating also into applied calculations, that complex dynamics is often understood most efficiently through geometric arguments rather than brute-force algebra. (The methodological advantages of such arguments are well known, \eg from QM\footnote{See, \eg \Ref{tex:ashtekar99}. One may also find parallels between our results and the mentioned paper (see papers cited therein too), except that the latter focuses on finding classical mechanics in QM, whereas we will do the opposite.} and need not be restated here.) Same applies to the understanding of classical waves. A number of theories were proposed over the years that offer \textit{ad~hoc} geometric structure for specific wave equations, such as the linearized Vlasov-Maxwell system, that lead to explicit derivation of the action conservation theorem (ACT) and properly define the energy for waves beyond the GO approximation; see, \eg \Refs{tex:hirota10,ref:larsson91,ref:brizard93,ref:hirota10} and references therein. These theories are yet not entirely complete, specifically, for two reasons. First, they assume waves propagating on manifolds and thus cannot treat localized modes on the same footing, even though one could expect their physics to be not too different. In this sense, the commonly adopted formalism for differentiable fields could be expected to flow as a limit of a general theory rather than be a part of its foundation. Second, one may find the existing approaches to be not sufficiently motivated from the physical point of view; rather they just ``happen to work''. (One example is the approach proposed in \Refs{ref:larsson91,ref:brizard93} that involves field complexification and introduction of a pseudo-Hilbert space with inner product that is zero on real fields. Another example is the already mentioned reliance of the modern WKT on the Fourier and Wigner-Weyl transforms that, as formal mathematical operations applied \textit{ad~hoc}, do not have a physical meaning of their own.) It stands to reason that there exists basic physics behind such intriguing but formal mathematical tricks. Identifying this physics could explain why the tricks actually work and, through that, render the wave theory something more fundamental, and further-reaching, than it appears as is. But, in order to do that, the concept of the wave needs to be \textit{formalized} first.

\subsection{Geometric quantumlike approach} 

A suitable way to define a general (nondissipative) wave is to think of it as interference of eigenmode oscillations. This definition may not cover strongly nonlinear perturbations like turbulent eddies but, in return, helps one analyze the basic physics of linear and weakly nonlinear waves efficiently. Whether or not eigenmodes are easy to find explicitly, the very knowledge that they are what comprises the wave permits considering any wave simply as a multidimensional (harmonic or weakly nonlinear) classical oscillator. But such oscillators can be easily studied in general, so all fundamental theorems of wave physics then flow \textit{ordine geometrico demonstrata}. This means, in particular, that the very form of the wave Lagrangian can be inferred from basic geometric considerations (as opposed to empirical arguments common in literature \cite{book:zakharov-b}) and in terms of an invariant vector language that is not restricted to concrete settings, choice of coordinates, and number of dimensions. Specifically this is done as follows.

We will mostly deal with linear waves below. Such waves map to strictly harmonic oscillators, and the dynamics of each harmonic oscillator, in turn, is well known \cite{my:nlinphi,arX:kauffmann10,ref:schenk01} to map exactly to a linear Schr\"odinger equation (LSE) for the appropriately normalized complex amplitudes of the eigenmodes. It then remains to express the set of these amplitudes as a vector $\ket{\psi}$, by introducing a certain Hilbert space, $\Psi$, with an appropriate metric. In contrast to \Ref{ref:larsson91}, the metric that we introduce has a transparent physical meaning familiar from QM; absent negative-energy waves (which are typically unstable and must not be confused with negative-frequency waves), it is, in fact, simply Euclidean in the energy basis. Hence the LSE is obtained in an invariant quantumlike form, so a one-to-one correspondence is established between the classical-wave and quantum dynamics. This extends the conventional, semi-qualitative understanding of this correspondence \cite{ref:mendonca08b} and brings well-established methods of QM to one's disposal for studying classical waves too.

It may be worth repeating here that we are considering \textit{classical} waves, and formal similarity with QM is obtained only through our choice of notation. This similarity, however, is by itself sufficiently powerful and leads to a new interpretation of the classical-wave action. Within the new theory, the action is naturally defined as the density matrix, or, more generally, density operator $\ket{\psi}\bra{\psi}$, so the ACT flows in the form of the von Neumann equation, except it does not contain the Planck constant, $\hbar$. The wave coordinate is also naturally defined as an operator, $\oper{x}$. In case when the space $X$ formed by the eigenvalues of $\oper{x}$ is a manifold, a general wave Hamiltonian $\hat{H}$ cannot be expressed through $\oper{x}$ alone (unlike when $X$ is a discrete set). This invites, \textit{in that particular case}, introduction of an auxiliary operator $\oper{k}$ as, loosely speaking, the gradient on $X$. (Alternative, ``noncanonical'' auxiliary operators can be chosen too, but at the expense of complicating calculations.) It can be naturally termed the momentum operator, and the well-known commutation relation [\Eq{eq:maincommut}] between $\oper{x}$ and $\oper{k}$ is then satisfied by construction, so it does not need to be postulated. At least for scalar waves, $\oper{x}$ and $\oper{k}$ then form a complete basis of operators on $\Psi$, so the wave Hamiltonian that governs the dynamics of $\ket{\psi}$ is \textit{proved} to have a general form $\hat{H} = H(t, \oper{x}, \oper{k})$. 

As the next step, one can construct WKT from scratch, without specifying $H$ and without appealing to the Fourier and Wigner-Weyl transforms. All it takes is to \textit{project} the von Neumann equation for $\ket{\psi}\bra{\psi}$ on the space of interest. This is done by applying the standard projectors $\ket{\mvec{x}}\bra{\mvec{x}}$ and $\ket{\mvec{k}}\bra{\mvec{k}}$ in appropriate combinations. Projecting on the coordinate space, for instance, yields the spatial representation of the ACT for general, or noneikonal, waves, constructively generalizing the result reported in \Ref{ref:brizard93}. Projecting on the momentum space similarly yields the ACT in the momentum representation, and so on; in fact, infinitely many equivalent equations can be produced. A particular class of such equations, called kinetic equations, also describes the system dynamics in projection on the ``phase space'', that is, the space formed by eigenvalues of $\oper{x}$ and $\oper{k}$. Phase space coordinates are kept general curvilinear canonical coordinates, and various combinations of the mentioned projectors lead to various scalar ``distribution functions'', including the Wigner function \cite{ref:wigner32,ref:case08} as a special case. (However, contrary to a common practice, identifying such distributions as ``quasiprobabilities'' is somewhat misleading, as will be discussed below.) This new, invariant representation readily leads also to the general quantitative correspondence between the classical WKT and the phase-space formulation of QM \cite{book:zachos}, so far explored mainly \textit{ad~hoc} \cite{ref:mendonca11b,ref:santos05,ref:santos07}.

The GO limit of this theory yields standard Whitham's equations \cite{my:amc,book:whitham}, including the Hamilton-Jacobi equation for the wave phase and the eikonal limit of the ACT. The Liouville equation is also yielded as a special case, but, contrary to a common misconception, in a somewhat different limit. The associated ray equations, still in generalized canonical coordinates, are then shown to satisfy Hamilton's equations of discrete, and generally nonlinear, classical mechanics. In this sense, the general classical mechanics can be considered as a byproduct of WKT, obtained as a certain projection of the dynamics of a linear multidimensional harmonic oscillator. The possibility of treating broad-spectrum classical waves as gases of quasiparticles (photons, plasmons, driftons) \cite{ref:trines10,ref:trines06,ref:reitsma06,ref:tsintadze04,ref:mendonca03,ref:tsintadze98,ref:bingham97} is subsumed under this theory too. 

Finally, the extension to vector waves and weakly nonlinear waves also can be done straightforwardly within the same geometrical language, as will be explained later.

\subsection{Outline} 

Below we will show how the steps that were described above are realized in detail, with the focus on straightening out definitions and explaining the ubiquity of certain \textit{types} of equations. The plan is to do so without cutting too many corners, albeit without complicating the text beyond necessary.\footnote{The paper is intended as only physically-rigorous. Because of this, we also avoid referencing too-mathematically oriented works on geometrization of classical physics.} We thereby work in generalized coordinates and address a number of subtleties that are rarely covered in literature but are essential for keeping the presentation coherent. A discussion of specific applications does not reasonably fit into this logic and is left to future publications. To anticipate possible misunderstanding, we emphasize that \textit{the present paper is not about electromagnetic or any other specific waves; it is rather about axiomatization of the general wave theory.} Nevertheless, some examples will be considered, mainly to accentuate advantages of representing the dynamics of (nondissipative linear) classical waves in the LSE form rather than in the form of arbitrary partial differential equations (PDEs).

The obvious advantages of this approach (apart from elegance, which could be called subjective) are as follows:
\begin{enumerate}
\item[(i)]
Casting the wave dynamics in the LSE form permits studying all waves on the same footing and ignore insignificant details specific to particular media. The ACT and kinetic equations then need not be rederived for each given wave separately, in contrast to how it is commonly done in literature. This allows us, for instance, to generalize the ACT for noneikonal waves derived in \Ref{ref:brizard93}. 
\item[(ii)]
Same applies to derivation of kinetic equations. We show, as an example, how our formalism unambiguously yields kinetic equations for linear drift waves in inhomogeneous magnetized plasmas, which topic used to stir a controversy in other approaches \cite{ref:smolyakov99,ref:krommes00}. (Having said that, it is not our goal to review WKT here, also because the related literature has become prohibitively extensive to be surveyed even briefly.)
\item[(iii)]
The LSE comes with a straightforward variational principle, \ie has a simple Lagrangian. This makes the associated conservation laws (such as the ACT) manifest and also ensures that they are preserved even when simplifying approximations are made, if the approximations are made in the Lagrangian. In contrast, making approximations directly in PDEs does not enjoy this property and must be done with more precautions; \eg dropping a term only because it is small compared to others in a PDE may be justified locally, but the error can accumulate globally and give rise to nonphysical results. See, \eg related discussions in \Refs{my:mquanta,my:dense,my:iorec} on waves in plasmas undergoing compression, ionization, and recombination.
\end{enumerate}

The paper is organized as follows. In \Sec{sec:basic}, we introduce the underlying system as a multidimensional harmonic oscillator and define the appropriate variables that permit us casting the dynamic equations in a convenient, and \textit{physically-motivated}, complex form. We also describe exact and approximate conservation laws that are readily seen in this representation. In \Sec{sec:fs}, we define the appropriate geometric structure to express the complex-coordinate set in the form of an invariant vector. In \Sec{sec:invrep}, we derive the general wave equation in the vector form and introduce the classical-wave action as an operator. In \Sec{sec:ps}, we define the generalized coordinate operator and the generalized coordinate space, $X$, and focus on the special case when $X$ is a differentiable manifold. We \textit{motivate} the introduction of the momentum operator and derive the generic form of the (scalar) wave Hamiltonian and the associated LSE. In \Sec{sec:dmeq}, we introduce scalar equations for the density-operator projection on various spaces, including $X$ and the phase space. We also discuss the ``quasiprobability'' concept in the context of our geometric approach. In \Sec{sec:waves}, we discuss standard properties of scalar waves, including the following: (i)~formal definition of a homogeneous stationary wave; (ii)~quasioptical approximation in the operator form; (iii)~the Liouville limit (LL) and the wave kinetic equation (WKE), including kinetic ray tracing; (iv)~``hydrodynamic'' equations for wave dynamics in the GO limit, including hydrodynamic ray tracing and the point-particle limit; (v)~statistical kinetic equation (SKE), and, as an example, its applications to linear Hasegawa-Mima equations in the context of the drift-turbulence theory for inhomogeneous magnetized plasmas. In \Sec{sec:glw}, we generalize our formalism to waves in the extended space (where the time is treated as yet another coordinate) and discuss its application to the Klein-Gordon equation (KGE). Then we also discuss generalization to vector waves. In \Sec{sec:nlin}, we explain how our theory is extended to nonlinear waves and contemplate subtleties of the nonlinear ray tracing that often go unnoticed in literature. In \Sec{sec:rqm}, we expand on the relation between our theory and QM. In \Sec{sec:conc}, we summarize our main results. Some auxiliary calculations are also presented in appendixes.

\subsection{Notation}
\label{sec:notation}

We use the symbol $\doteq$ to denote definitions; namely, ``$a \doteq b$'' will mean ``$a$ is defined as $b$'', and ``$a \rdoteq b$''  will mean ``$b$ is defined as $a$''. We also adopt the standard summation notation for repeating indexes; namely, $a_n b^n$ will mean $\sum_n a_n b^n$, whereas $a_n b_n$ and $a^n b^n$ imply no summation. The symbol $^*$ will denote complex conjugation, and $^\dag$ will denote duality relation. Finally, the abbreviations we use are summarized as follows:\\[10pt]
\begin{tabular}{@{\quad} r@{\quad -- \quad} l @{\quad}}
ACT & action conservation theorem,\\
FKE & full kinetic equation,\\
GO & geometrical optics,\\
KGE & Klein-Gordon equation,\\
LAP & least action principle.\\
LL & Liouville limit,\\
LSE & linear Schr\"odinger equation,\\
NLSE & nonlinear Schr\"odinger equation,\\
QM & quantum mechanics,\\
QNW & quasimonochromatic nonlinear waves,\\
PDE & partial differential equation,\\
PSI & phase space image,\\
SKE & statistical kinetic equation,\\
WKE & wave kinetic equation,\\
WKT & wave kinetic theory.\\
\end{tabular}\\

\section{Basic equations}
\label{sec:basic}

\subsection{General Lagrangian}
\label{sec:physsys}

Suppose a nondissipative dynamical system described by some nondegenerate Lagrangian of the form $L = L(\xi, \dot\xi, t)$, where $\xi \doteq (\xi^1, \ldots \xi^N)$ are some real coordinates, $\dot\xi$ are the corresponding velocities, and $t$ is time. The system trajectory is derived as the trajectory satisfying the least action principle (LAP), $\delta_\xi \bar{\mc{S}} = 0$, where the action integral is given by $\bar{\mc{S}} \doteq \int^{t_2}_{t_1} \bar{L}\,dt$, $\bar{L}$ is called a Lagrangian, and $\xi$ is assumed fixed at the ends of the time interval; \ie
\begin{gather}\label{eq:dxi0}
\delta \xi(t_1) = \delta \xi (t_2) = 0.
\end{gather}
This leads to Euler-Lagrange equations \cite[Sec.~2]{book:landau1}
\begin{gather}\label{eq:ELxipi}
\dot{\pi}_n = \partial \bar{L}/\partial \xi^n, \quad \pi_n \doteq \partial \bar{L}/\partial {\dot\xi^n},
\end{gather}
where $(\pi_1, \ldots \pi_N) \rdoteq \pi$ are called canonical momenta.

Equivalent equations are obtained from the LAP if one expresses the Lagrangian as $\bar{L} = \pi_n \dot\xi^n - H(t, \xi, \pi)$, where \textit{both} $\xi$ and $\pi$ are treated as independent variables \cite[Sec.~43]{book:landau1}. The function $H$ is called a Hamiltonian, and the corresponding Euler-Lagrange equations are given by
\begin{gather}\label{eq:hameq0}
\dot{\xi}^n = \partial H/\partial \pi_n, \quad \dot{\pi}_n = - \partial H/\partial \xi^n.
\end{gather}
Equations \eq{eq:hameq0} are known as Hamilton's equations. Their only difference from \Eq{eq:ELxipi} is that what was the definition of $\pi_n$ now serves as an independent equation.

Notice that \Eqs{eq:hameq0} can be obtained just as well if, in addition to \Eqs{eq:dxi0}, one imposes constrains also on $\pi$: 
\begin{gather}\label{eq:dpi0}
\delta \pi(t_1) = \delta \pi (t_2) = 0.
\end{gather}
This helps as follows. Let us write
\begin{gather}
\bar{L} = L + \frac{d}{dt}\left(\frac{\pi_n \xi^n}{2}\right), \\ 
\bar{\mc{S}} = \mc{S} + \left(\frac{\pi_n \xi^n}{2}\right)\bigg|^{t_2}_{t_1}, \label{eq:bS}
\end{gather}
where we introduced $L \doteq (\pi_n \dot{\xi}^n - \dot{\pi}_n \xi^n)/2 - H$ and
\begin{gather}
\mc{S} \doteq \int^{t_2}_{t_1} L\,dt.
\end{gather}
The second term in \Eq{eq:bS} is constant due to \Eqs{eq:dxi0} and \eq{eq:dpi0} and thus can be omitted. Hence the LAP turns into
\begin{gather}
\delta_z \mc{S} = 0,\label{eq:dSdz}\\
\delta z(t_1) = \delta z(t_2) = 0,\label{eq:dSdzf}
\end{gather}
where $z \doteq (\xi^1, \ldots \xi^N, \pi_1, \ldots \pi_N)$ is a $2N$-dimensional vector. We will use the notation $Z$ to denote the space of all such vectors and $z^\alpha$ for their individual components. (Accordingly, Greek indexes henceforth span from 1 to $2N$, as opposed to Latin indexes, which span from to 1 to $N$.) Then $L$ has a form that is naturally symmetric with respect to transformations $(\xi, \pi) \leftrightarrow (\pi, - \xi)$; namely,
\begin{gather}\label{eq:L}
L = \varpi_{\alpha\beta} z^\alpha \dot{z}^\beta/2 - H(t, z^\alpha).
\end{gather}
Here $\varpi_{\alpha\beta}$ is a $2N \times 2N$ constant antisymmetric matrix,
\begin{gather}\label{eq:omega}
\varpi_{\alpha\beta} \doteq 
\left(
\begin{array}{c @{\quad} c}
0 & - \delta^\beta_\alpha\\[3pt]
 \delta^\alpha_\beta & 0
\end{array}
\right),
\end{gather}
with the indexes on the right-hand sides taken as modulo $N$. Hence \Eq{eq:dSdz} readily leads to another representation of Hamilton's equations, equivalent to \Eqs{eq:hameq0},
\begin{gather}\label{eq:hameq}
\varpi_{\alpha\beta} \dot z^\beta = \partial H/\partial z^\alpha,
\end{gather}
and $\varpi_{\alpha\beta}$ is recognized as the canonical symplectic form.

\subsection{Reference modes}
\label{eq:ls}

We will now adopt that the system is linear and that there are no external forces (which could be added straightforwardly if needed). The energy $H$, to be denoted $h$ for such a system, must hence be bilinear in $z^\alpha$; \ie
\begin{gather}\label{eq:Hlin}
h = h_{\alpha\beta} z^\alpha z^\beta/2,
\end{gather}
where we assume, without loss of generality, that $h_{\alpha\beta} = h_{\beta\alpha}$. Equation \eq{eq:hameq} will then take the form
\begin{gather}\label{eq:aux8}
\varpi_{\alpha\beta} \dot z^\beta = h_{\alpha\beta} z^\beta,
\end{gather}
and we assume that $h_{\alpha\beta}$ is such that \Eq{eq:aux8} describes a multidimensional nondissipative linear \textit{oscillator}. This is formalized as follows. At any given time, the system described by \Eq{eq:aux8} can be assigned a set of instantaneous eigenmodes, which we will call reference eigenmodes, whose ``polarization vectors'' $\bar{z}_\nu$ and eigenfrequencies $\Omega_\nu$ are found as solutions of\footnote{Equation \eq{eq:aux1} can be understood as a standard eigenvector problem, ${L^\alpha}_\beta \bar{z}_\nu{}^\beta = -i \Omega_\nu \bar{z}_\nu{}^\alpha$, where ${L^\alpha}_\beta \doteq J^{\alpha\lambda}h_{\lambda\beta}$, and $J^{\alpha\beta}$ is the skew-symmetric matrix inverse to $\omega_{\alpha\beta}$.}
\begin{gather}\label{eq:aux1}
-i \Omega_\nu\varpi_{\alpha\beta} \bar{z}_\nu{}^\beta = h_{\alpha\beta} \bar{z}_\nu{}^\beta.
\end{gather}
We will require that there are $N$ such modes with $\Omega_\nu > 0$ and thus some associated~$\bar{z}_\nu$ too. Yet for each mode $(\Omega_\nu, \bar{z}_\nu)$ there exists another, conjugate mode $(-\Omega_\nu, \bar{z}_\nu^*)$, so there are exactly $2N$ eigenmodes with $2N$ nonzero frequencies overall. For clarity, we adopt that $\Omega_\nu > 0$ for $\nu = 1, \ldots N$, and $\Omega_\nu < 0$ for $\nu = (N + 1), \ldots N$.

We will now utilize these eigenmodes to represent the system dynamics in an equivalent but more easily tractable representation, which is done as follows. First of all, let us introduce ${S^\beta}_\nu \doteq \bar{z}_\nu{}^\beta$ and express \Eq{eq:aux1} as a matrix equation
\begin{gather}\label{eq:aux2}
-i \varpi_{\alpha\beta} {S^\beta}_\lambda \Omega^\lambda_\nu = h_{\alpha\beta} {S^\beta}_\nu,
\quad
\Omega^\lambda_\nu \doteq \Omega_\nu\delta^\lambda_\nu.
\end{gather}
Multiplying this by ${S^{\alpha}}^*_\mu$, one further gets
\begin{gather}\label{eq:aux4}
G_{\mu\alpha} \Omega^\alpha_\nu = h_{\alpha\beta} {S^{\alpha}}^*_\mu {S^\beta}_\nu, \quad
G_{\mu\nu} \doteq - i \varpi_{\alpha\beta}{S^{\alpha}}^*_\mu {S^\beta}_\nu.
\end{gather}
Since the matrices $G_{\alpha\beta}$, $\Omega^\alpha_\beta$, and $h_{\alpha\beta}$ are Hermitian, taking the conjugate transpose of \Eq{eq:aux4} yields that
\begin{gather}
(\Omega_\mu - \Omega_\nu)G_{\mu\nu} = 0.
\end{gather}
For any pair of modes with different $\Omega_\mu$ and $\Omega_\nu$ the associated $G_{\mu\nu}$ is therefore zero. Moreover, if eigenfrequencies coincide within some set of eigenmodes, the corresponding blocks on the diagonal of $G_{\mu\nu}$ can be further diagonalized, since each of them is a Hermitian matrix by itself. Hence we can adopt that $G_{\mu\nu}$ is diagonal.

Notice now that \Eq{eq:aux4} yields $G_{\mu\mu} = \bar{h}_\mu/\Omega_\mu$, where $\bar{h}_\mu \doteq h_{\alpha\beta}\bar{z}^*_\mu{}^{\alpha} \bar{z}_\mu{}^\beta$ are real on the score of $h_{\alpha\beta}$ being symmetric. Since $\bar{h}_\mu$ equals the energy $\bar{h}_m$ of the real eigenmode $\bar{Z}_m(t) \doteq \bar{z}_m \exp(- i \Omega_m t) + \bar{z}^*_m \exp(i \Omega_m t)$ with index $m \doteq \mu\, (\mbox{mod}\, N)$, then $G_{\mu\mu}$ can be expressed as $G_{\mu\mu} = \sigma_\mu |\bar{I}_m|$. Here $\sigma_\mu = \mbox{sgn}\,(\bar{h}_m \Omega_\mu)$, and $\bar{I}_m \doteq \bar{h}_m/\Omega_m$ can be understood as the action of the mode $\bar{Z}_m(t)$.\footnote{The sign of the action is a matter of convention only. To shorten the notation, we define it such that it matches the sign of the wave energy. To understand the action as the number of quanta, one may choose it to be nonnegative instead, but we will blur the distinction for brevity.} As we can choose the amplitudes of $\bar{z}_\mu$ such that $|\bar{I}_m| = 1$ for all $m$, we hereby adopt that $G_{\mu\nu}$ is a signature matrix of the following form:
\begin{gather}
G_{\mu\nu} = \mbox{diag}\,(\sigma_1, \ldots \sigma_N, -\sigma_1, \ldots -\sigma_N), \quad \sigma_n = \mbox{sgn}\,\bar{h}_n.\nonumber
\end{gather}
This also implies adopting frequency units for the energy (but see \Sec{sec:rqm}).

\subsection{Dynamics in the reference-mode representation}
\label{sec:ydef}

What we will do next is find a representation of $L$ in terms of complex variables that are the classical counterparts of creation and annihilation operators for each of the $N$ positive-frequency modes. To do so, let us use $\bar{z}_\nu$ as the new basis in $Z$. Specifically, consider a transformation $z^\beta = {S^\beta}_\nu a^\nu$, where $a^\nu$ are the new variables. Notice that, since $z^\alpha$ is real, we as well can take $z^\alpha = {S^\alpha}^*_\nu a^{\nu *}$. Furthermore, by definition, $a^{m + N} = a^{m*}$, so we will replace sums over Greek indexes with pairs of sums over Latin indexes, denoting the corresponding quadrant of $G_{\mu\nu}$ as $\eta_{mn}$; \ie 
\begin{gather}\label{eq:g}
\eta_{mn} \doteq \mbox{diag}\,(\sigma_1, \ldots \sigma_N).
\end{gather}
Then one gets
\begin{gather}
L = \frac{i}{2}\,B_{\mu\nu} a^{\mu*}a^\nu + \frac{i}{2}\,G_{\mu\mu}a^{\mu*}\dot{a}^\mu - h.
\end{gather}
Here $B_{\mu\nu} \doteq-i \varpi_{\alpha\beta} {S^\alpha}^*_\mu {\dot{S}^\beta}_\nu$ is a matrix which is parametrized by two $N \times N$ blocks (and their conjugates), namely, an anti-Hermitian matrix $V_{mn}$ and a symmetric matrix $W_{mn}$ (\App{app:gamma}); then,
\begin{align}
& B_{\mu\nu} a^{\mu*}a^\nu \nonumber \\
& = \ V_{mn} a^{m*}a^n + W^*_{mn} a^{m*}a^{n*} - W_{mn} a^m a^n - V^*_{mn} a^{m} a^{n*} \nonumber \\
& = \ 2 V_{mn} a^{m*}a^n + W^*_{mn} a^{m*}a^{n*} - W_{mn} a^m a^n.
\end{align}
Similarly,
\begin{align}
G_{\mu\mu}a^{\mu*}\dot{a}^\mu
& = G_{mm}a^{m*}\dot{a}^m + G_{(m+N)(m+N)}a^m\dot{a}^{m*} \nonumber \\
& = \eta_{mm}a^{m*}\dot{a}^m - \eta_{mm}a^m\dot{a}^{m*},
\end{align}
and the energy $h$ takes the following form: 
\begin{align}
2h & = \Omega_\mu G_{\mu\mu}a^{\mu*} a^\mu \nonumber \\
& = \Omega_m G_{mm}a^{m*} a^m - \Omega_m G_{(m + N)(m + N)}a^{m} a^{m*} \nonumber \\
& = 2 \Omega_m \eta_{mm}a^{m*} a^m. 
\end{align}
The latter yields, in particular, that $h$ is the sum of the energies $h_n$ of individual real modes with amplitudes $a^n$; namely, $h_n = \Omega_n I_n$, where  $I_n \doteq \sigma_n|a_n|^2$. Thus, $I_n$ is the $n$th mode action, while $a_n^*$ and $a^n$ serve as the classical counterparts of creation and annihilation operators.

It is also convenient to simplify the above expressions by introducing the standard rules of index manipulation as if $\eta_{mn}$ were a metric (\App{app:metric}). Specifically,
\begin{gather}\label{eq:etaindex}
a_m \doteq \eta_{mm} a^m, \quad a^m = \eta^{mm} a_m,
\end{gather}
where $\eta^{mn}$ is the matrix inverse to $\eta_{mn}$. Similar transformations apply to matrices\footnote{Note that, absent negative-energy modes (which are, in a sense, exotic), the metric $\eta_{mn}$ is Euclidean. Then the difference between upper and lower indexes simply can be ignored.} (\App{app:metric}). Hence,
\begin{gather}
L = \frac{i}{2}\,(a^*_n \dot{a}^n - \dot{a}^*_n a^n) - a^*_m {Q^m}_n a^n + R,\label{eq:Llin}\\
{Q^m}_n \doteq \Omega^m_n - i{V^m}_n, \\
R \doteq \frac{i}{2}\,(W^{mn*} a^*_m a^*_n - W_{mn} a^m a^n),\label{eq:R}
\end{gather}
where, notably, the matrix ${V^m}_n$ coincides with the auxiliary matrix ${v^m}_n$ introduced in \App{app:gamma}.

The vectors $a' \doteq \text{Re}\,a$ and $a'' \doteq \text{Im}\,a$, where $a \doteq (a^1, \ldots a^N)$, are parameterized by $2N$ real variables, so they can be chosen as new phase space coordinates. Then \Eqs{eq:dSdz} and \eq{eq:dSdzf} give
\begin{gather}
\delta \mc{S}/\delta a' = 0, \quad \delta \mc{S}/\delta a'' = 0,\label{eq:dsa}\\
\delta a'(t_1) = \delta a'(t_2) = \delta a''(t_1) = \delta a''(t_2) = 0.\label{eq:da}
\end{gather}
On the other hand, if the action is understood as a function $\mc{S}(a(a', a''), a^*(a', a''))$, where 
\begin{gather}
a(a', a'') = a' + ia'', \quad a^*(a', a'') = a' - ia'',
\end{gather}
then application of the chain rule yields
\begin{gather}
\frac{\delta \mc{S}}{\delta a'} = \frac{\delta \mc{S}}{\delta a} + \frac{\delta \mc{S}}{\delta a^*},\quad
\frac{\delta \mc{S}}{\delta a''} = i\,\frac{\delta \mc{S}}{\delta a} -i\, \frac{\delta \mc{S}}{\delta a^*}.
\end{gather}
We now solve this set of equations for $\delta \mc{S}/\delta a$ and $\delta \mc{S}/\delta a^*$ and also apply \Eqs{eq:dsa}. This gives
\begin{gather}
\frac{\delta \mc{S}}{\delta a} = \frac{\delta \mc{S}}{\delta a'} - i\,\frac{\delta \mc{S}}{\delta a''} = 0,\\
\frac{\delta \mc{S}}{\delta a^*} = \frac{\delta \mc{S}}{\delta a'} + i\,\frac{\delta \mc{S}}{\delta a''} = 0.
\end{gather}
In other words, for the purpose of the LAP, $a$ and $a^*$ can be treated as independent variables, and, combining this with \Eqs{eq:da}, we can summarize the LAP as follows:
\begin{gather}
\delta \mc{S}/\delta a = 0, \quad \delta \mc{S}/\delta a^* = 0,\label{eq:lapsSa}\\
\delta a(t_1) = \delta a(t_2) = \delta a^*(t_1) = \delta a^*(t_2) = 0.
\end{gather}
In particular, $\delta \mc{S}/\delta a^*_n = 0$ leads to an Euler-Lagrange equation of the form
\begin{gather}\label{eq:dyn}
i \dot{a}^n = {Q^n}_m  a^m - i W^{nm*} a^*_m,
\end{gather}
where the symmetry of $W^{nm*}$ was used. Similarly, $\delta \mc{S}/\delta a^n = 0$ leads to equations that are complex conjugates of \Eqs{eq:dyn}. Substitution of these back to \Eq{eq:Llin}, notably, gives $L = 0$; \ie \textit{on the solution}, the numerical value of the Lagrangian of linear oscillations equals zero. Also, if the system is stationary, so $W^{nm*}$ and ${V^m}_n$ are zero, the energy $h$ is seen to be conserved. [The latter can be inferred as well from \Eqs{eq:Hlin} and \eq{eq:aux8} and the antisymmetry of $\varpi_{\alpha\beta}$.]

\subsection{Approximate conservation laws}
\label{sec:acl}

In a time-dependent system, nonzero $\dot{h}_{\alpha\beta}$ give rise to parametric effects driven by $W_{mn}$ and $W^{nm*}$, plus frequency shifts due to ${V^m}_n$. However, certain approximate integrals can still exist. Suppose that the time scale $\mc{T}$ of the reference-mode evolution is large compared to all $\Omega_n$. Then $W_{mn}$ and $W^{nm*}$ can be eliminated by averaging over the fast oscillations (one may recognize this as the quasioptical approximation), so one arrives at asymptotic ``slow-motion'' equations,
\begin{gather}\label{eq:aux302}
\dot{a}^n = -i \Omega_{n'} a^{n'} - {V^n}_m a^m.
\end{gather}
(We use primes to distinguish repeating indexes on which \textit{no} summation is performed; in other respects, $n' \equiv n$). One can hence derive equations for the actions $I_n$,
\begin{gather}\label{eq:din}
\dot{I}_n = - (V_{n'm} a^{n'*}a^m + V^*_{n'm} a^{n'}a^{m*}).
\end{gather}
If $\mc{T}$ is much larger than the beat periods, $|\Omega_n - \Omega_m|^{-1}$, then the right-hand side averages to zero, yielding $\dot{I}_n = 0$; \ie the actions of \textit{individual} modes are conserved. Otherwise, we sum \Eqs{eq:din} over $n$ to get that the \textit{total} action $I \doteq \sum_{n = 1}^N I_n$ is conserved:
\begin{gather}
\dot{I} = - (V_{nm} a^{n*}a^m + V^*_{nm} a^{n}a^{m*}) = 0,
\end{gather}
since $V_{mn}$ is anti-Hermitian, \ie $V^*_{nm} = - V_{mn}$. In particular, this prohibits growth of individual $I_n$ beyond $I$ that is determined by initial conditions --- if all the modes have positive energies. However, there is no such limit if at least one of the modes has negative energy. 

As a side remark, let us notice the following. The latter typically makes negative-energy waves unstable, but, \textit{absent coupling with each other}, negative- and positive-energy have identical properties; then one can simply replace $\sigma_n$ with $-\sigma_n$ without any effect on the dynamic equations. For example, a Lagrangian $L = -\dot{\xi}^2/2 + \Omega^2\xi^2/2$, which describes a negative-energy mode, produces an equation for $\xi$ identical to that yielded by $L = \dot{\xi}^2/2 - \Omega^2\xi^2/2$, which describes a positive-energy mode.

\section{Fundamental space}
\label{sec:fs}

We will further need a compact form of the above equations that would be invariant with respect to arbitrary linear variable transformations performed in the configuration space $(a, a^*)$. For that, let us think of a system state as an abstract vector $\ket{\psi} = \psi^n\ket{e_n}$ in a complex $N$-dimensional \textit{metric} space $\Psi$, which we call the fundamental space, where $\ket{e_n}$ is an arbitrary basis. Specifically, let us assign to $\Psi$ a Hermitian metric  $g_{mn}$ such that $g_{mn} = \eta_{mn}$ \textit{in the basis} $\ket{\Omega_n}$, where $\ket{\psi} = a^n \ket{\Omega_n}$; we will call this basis and the metric $\eta_{mn}$ fundamental. As the latter is only pseudo-Euclidean ($\eta_{nn} = \pm 1$), we can hence attribute vectors $\ket{\Omega_n}$ as ``space-like'' when they correspond to positive-energy modes ($\eta_{nn} > 0$) and ``time-like'' when they correspond to negative-energy modes ($\eta_{nn} < 0$).

Any multilinear form on $a^n$ and $a^*_n$, determined by some matrix $F$, can now be used to define a tensor of the appropriate rank; we do so by requiring that, when taken in the fundamental basis, the tensor components equal to those of $F$. (For details on the notation and rules of vector and tensor manipulation see \App{app:metric}.) In particular, a bilinear form $a_m^* {F^m}_n a^n$ yields a rank-$(1,1)$ tensor $\hat{F}(\ph\,, \ph)$, where ``$\ph$'' denote placeholders for a one-form and a vector, respectively. Such a tensor, in turn, determines a mapping ${\hat{F}:\Psi \to \Psi}$, or an \textit{operator}, via
\begin{gather}
\ket{\hat{F}\psi} \equiv \hat{F}\ket{\psi} \doteq \hat{F}(\ph\,, \ket{\psi}).
\end{gather}
Hence, for any $\ket{\alpha}$ and $\ket{\beta}$ from $\Psi$ [\Eqs{eq:ab}] we have
\begin{gather}
 \braket{\alpha|\hat{F}|\beta} = \braket{\alpha|\hat{F}\beta} = \hat{F}(\bra{\alpha}, \ket{\beta}) = \alpha_k^* {F^k}_n \beta^n.
\end{gather}
The adjoint operator is then defined via
\begin{gather}
\bra{\hat{F}^\dag \psi} \equiv \bra{\psi}\hat{F} \doteq \hat{F}(\bra{\psi}, \ph),
\end{gather}
so $\braket{\hat{F}^\dag \alpha|\beta} = \braket{\alpha|\hat{F}\beta}$, and one gets
\begin{gather}
\braket{\hat{F}^\dag \alpha|\beta} = \braket{\beta|\hat{F}^\dag \alpha}^* = \beta_k {(F^\dag)^k}^*_m \alpha^{m*}.
\end{gather}
Further notice that
\begin{multline}
 \alpha_k^* {F^k}_n \beta^n = g_{km}^* {F^k}_n \alpha^{m*} \beta^n \\
 = g_{mk} {F^k}_n \alpha^{m*} \beta^n = F_{mn} \alpha^{m*} \beta^n,
\end{multline}
\begin{multline}
 \beta_k {(F^\dag)^k}^*_m \alpha^{m*} = g_{kn} {(F^\dag)^k}^*_m \alpha^{m*} \beta^n 
 \\ = g_{nk}^* {(F^\dag)^k}^*_m \alpha^{m*} \beta^n = {(F^\dag)}^*_{nm} \alpha^{m*} \beta^n,
\end{multline}
where indexes are manipulated in a usual manner. Hence ${(F^\dag)}_{nm} = F^*_{mn}$; \ie these two matrices, \textit{with both indexes lowered}, are mutually adjoint. As always, an operator will hence be called Hermitian if $\hat{F}^\dag = \hat{F}$, or $F^*_{mn} = F_{nm}$, which property is, of course, invariant with respect to coordinate transformations (\App{app:ct}). Linear Hermitian operators will also be called ``observables''.

As an example, consider any $\ket{\alpha} = \alpha_n \ket{e_n}$ and define $\hat{A} \doteq \ket{\alpha}\bra{\alpha}$. The matrix elements of this operator are ${A^m}_n = \alpha^m \alpha_n^*$, and lowering the index yields
\begin{gather}
A_{mn} = g_{mk}{A^k}_n = g_{mk}\alpha^k \alpha_n^* = \alpha_m \alpha_n^* = A_{nm}^*.
\end{gather}
This shows that $\hat{A}$ is Hermitian, and, even more generally, so is $\hat{F}\hat{A}\hat{F}^\dag$ for any $\hat{F}$.

\section{Invariant equations}
\label{sec:invrep}

\subsection{Master Lagrangian}

The formulation developed in \Sec{sec:fs} yields $\hat{\Omega}$, $i\hat{V}$, and $\hat{Q} \doteq \hat{\Omega} - i\hat{V}$ as rank-$(1,1)$ tensors, and the homonymous operators are clearly Hermitian. Likewise, $\hat{W}$ and $\hat{W}^\dag$ are defined as symmetric tensors of rank $(0,2)$ and $(2,0)$, correspondingly, via
\begin{gather}
\hat{W} (\ket{\alpha}, \ket{\beta}) \doteq W_{mn} \alpha^m \beta^n,\\
\hat{W}^\dag (\bra{\alpha}, \bra{\beta}) \doteq W^{mn*} \alpha^*_m \beta^*_n
\end{gather}
in the fundamental basis. In particular, we then obtain
\begin{gather}
\braket{\psi|\hat{Q}|\psi} = a^*_m {Q^m}_n a^n, \\
\hat{W} (\ket{\psi}, \ket{\psi}) = W_{mn} a^m a^n,\\
\hat{W}^\dag (\bra{\psi}, \bra{\psi}) = W^{mn*} a^*_m a^*_n,
\end{gather}
which can be used as an invariant representation of the corresponding terms in the Lagrangian \eq{eq:Llin}. Similarly, $\braket{\psi|\psi} = a_n^* a^n$, where the right-hand side is recognized as the total action $I$. Hence $I$ is a true scalar on $\Psi$, given~by
\begin{gather}\label{eq:ipsi}
I = \braket{\psi|\psi}.
\end{gather}

To rewrite the rest of \Eq{eq:Llin} in an invariant form, we now proceed as follows. Consider a general coordinate transformation
\begin{gather}\label{eq:Uy}
a^n \doteq U^n{}_m \psi^m,
\end{gather}
where ${U^n}_m$ is an arbitrary (not necessarily unitary) nondegenerate matrix. Allowing ${U^n}_m$ to be time-dependent, let us introduce the covariant time derivative as the vector $\ket{\hat{\mc{D}}\psi}$ whose components $(\hat{\mc{D}}\psi)^n$ equal $\dot{a}^n$ when taken in the fundamental basis. Since $\ket{\hat{\mc{D}}\psi}$ must satisfy the transformation rule $\dot{a}^n = {U^n}_m (\hat{\mc{D}}\psi)^m$, one gets
\begin{gather}\label{eq:D}
(\hat{\mc{D}}\psi)^n = \dot{\psi}^n + (U^{-1})^n{}_k \dot{U}^k{}_m \psi^m.
\end{gather}
This leads to an invariant representation
\begin{gather}
a^*_n \dot{a}^n - \dot{a}^*_n a^n = \braket{\psi|\hat{\mc{D}}\psi} - \braket{\hat{\mc{D}}\psi|\psi},
\end{gather}
where we used that the metric on $\Psi$ transforms as $g_{mn} = \eta_{kk} {U^k}^*_m {U^k}_n$ (\App{app:metric}). The resulting Lagrangian,
\begin{gather}
L = \frac{i}{2}\,\big[\braket{\psi|\hat{\mc{D}}\psi} - \braket{\hat{\mc{D}}\psi|\psi}\big] - \braket{\psi|\hat{Q}|\psi} + R,\label{eq:Linv}\\
R = \frac{i}{2}\,\big[\hat{W}^\dag(\bra{\psi}, \bra{\psi}) - \hat{W}(\ket{\psi}, \ket{\psi})\big],
\end{gather}
hence automatically has an invariant form. Equation~\eq{eq:dyn} is then expressed as
\begin{gather}\label{eq:aLSE}
i\ket{\hat{\mc{D}}\psi} = \hat{Q} \ket{\psi} - i \ket{W},
\end{gather}
where $\ket{W}$ is the vector obtained by applying the tensor $\hat{W}^{\dag}$ to the one-form $\bra{\psi}$.  (Equation \eq{eq:aLSE} can be considered as a generalization of the result reported earlier in \Ref{arX:kauffmann10}.) Specifically, that and its duals are given by
\begin{gather}
\ket{W} \doteq \hat{W}^{\dag}(\bra{\psi}, \ph), 
\quad
\bra{W} \doteq \hat{W} (\ket{\psi}, \ph).\label{eq:Winv}
\end{gather}

External forces, which we eliminated from the beginning, can be included in \Eq{eq:Linv} by adding $\braket{Y|\psi} + \braket{\psi|Y}$ to $L$, where $\ket{Y}$ is a given, possibly time-dependent, vector. Damping can be included too, namely, by adding an anti-Hermitian operator to $\hat{Q}$. Below, we continue to neglect those corrections.

\subsection{Action as an operator}

Considering the metric-induced isomorphism between vectors and one-forms on $\Psi$, one may also want to symmetrize the dynamic equations with respect to $\ket{\psi}$ and $\bra{\psi}$. To do so, let us introduce the ``density operator'',
\begin{gather}
\hat{\rho} \doteq \ket{\psi}\bra{\psi}.
\end{gather}
[One may find this reminiscent of the \textit{quantum} action operator as it was defined in \Ref{ref:cargo05}, except that we do not introduce near-identity transformations here and thus do not rely on the existence of a small parameter.] The equation for $\hat{\rho}$ is obtained from $\hat{\mc{D}}\hat{\rho} = \ket{\hat{\mc{D}}{\psi}}\bra{\psi} + \ket{\psi}\bra{\hat{\mc{D}}{\psi}}$ and reads as follows:
\begin{gather}\label{eq:vN}
\hat{\mc{D}}{\hat{\rho}} - i[\hat{\rho}, \hat{\Omega}] = - \ket{\psi}\bra{W} - \ket{W}\bra{\psi},
\end{gather}
where $[\hat{\rho}, \hat{\Omega}] \doteq  \hat{\rho}\hat{\Omega} - \hat{\Omega} \hat{\rho}$ is the commutator. 

In particular, notice that
\begin{gather}
\mbox{tr}\,\hat{\rho} = I
\end{gather}
(where ``tr'' denotes trace), so the total action satisfies
\begin{gather}\label{eq:aux51}
\dot{I} = - 2 \,\mbox{Re}\braket{\psi|W}.
\end{gather}
Finally,
\begin{gather}
I = \frac{\braket{\psi|\hat{\rho}|\psi}}{\braket{\psi|\psi}}.
\end{gather}
Thus $I$ is understood as the expectation value of $\hat{\rho}$, and $\hat{\rho}$ can be viewed as the action operator. Hence we will use the terms ``action operator'' and ``density operator'' interchangeably. As another side note, it is natural to identify $\ket{\psi}/\sqrt{I}$ as the state vector of the wave elementary excitation (``photon wave function''), a concept that is often, and unjustly, considered controversial in other theories \cite{tex:bialynicki96}.

\subsection{Time-independent basis}
\label{sec:simpleeq}

For simplicity, we henceforth restrict variable transformations to stationary ${U^n}_m$, so $\hat{\mc{D}}$ becomes the usual time derivative. We also assume that $h$ is time-independent, so $\ket{\Omega_n}$ are fixed, and $\hat{W}$, $\hat{W}^\dag$, and $\hat{V}$ are zero. For generality, however, we will allow for an \textit{additional} linear-coupling term in the Lagrangian, say, $\mcc{H} = \mcc{H}_{\alpha\beta}(t) z^\alpha z^\beta/2$, so the total Hamiltonian is now $H = h + \mcc{H}$. (In other words, only a part of $H$ is now used to define reference modes, whereas the rest of it is treated as the interaction Hamiltonian.) Then $H$ produces a Hermitian operator $\hat{H}$ that is generally time-dependent, and
\begin{gather}\label{eq:Lreduced}
L = \frac{i}{2}\,\big[\braket{\psi|\dot{\psi}} - \braket{\dot{\psi}|\psi}\big] - \braket{\psi|\hat{H}|\psi}.
\end{gather}
As $\hat{H}$ is Hermitian, the resulting dynamic equation,
\begin{gather}\label{eq:bLSE}
i\ket{\dot{\psi}} = \hat{H} \ket{\psi},
\end{gather}
happens to be the (generalized) LSE in its invariant representation, with $\hat{H}$ serving as a Hamiltonian. The LSE has a unitary propagator, $\exp(-i \int^t \hat{H}\,dt)$, and thereby manifestly conserves the action $I$. (Dissipation could be accommodated by adding an anti-Hermitian operator to $\hat{H}$. The implications are straightforward and will not be discussed further.) Correspondingly, \Eq{eq:aux51} turns into
\begin{gather}
\dot{I} = 0,
\end{gather}
and \Eq{eq:vN} becomes the von Neumann equation,
\begin{gather}\label{eq:vNn}
i\dot{\hat{\rho}} + [\hat{\rho}, \hat{H}] = 0.
\end{gather}
Equation \eq{eq:vNn} can be understood as the ACT in its most general, operator form. Remember, however, that \Eq{eq:vNn} is yet limited to systems that are not subject to external forces or parametric effects, such as due to $\hat{W}$ and $\hat{W}^\dag$; otherwise, a source term must be added.

\section{Oscillations on a manifold}
\label{sec:ps}

\subsection{Coordinate operator}
\label{sec:coord}

Consider an arbitrary orthogonal basis on $\Psi$, comprised of some vectors $\ket{e_n}$. By normalizing those appropriately, one always can choose $\braket{e_n|e_{n'}}$ to be unity up to a sign, depending on whether $\ket{e_n}$ is space- or time-like. To simplify the notation, we will assume below that the fundamental metric is not pseudo-Euclidean but rather strictly Euclidean, except when explicitly stated otherwise.\footnote{Note that this simplification merely excludes negative-energy modes (not to be confused with negative-frequency waves), which are typically unstable and somewhat exotic in any case. Inclusion of such modes would only complicate the notation but not affect the qualitative results presented below.} Thus, from now on, $\braket{e_n|e_{n'}} = \delta_{nn'}$ for all $n$. 

The vectors $\ket{e_n}$ can now be used to construct the following family of observables. If linear ordering of $\ket{e_n}$ is adopted, one can start with $\hat{n} \doteq \sum_n n \ket{e_n}\bra{e_n}$, which serves as the index operator. More generally, we will assume $n \equiv \mvec{n}$ to be a direct sum of linearly ordered indexes, $\mvec{n} \doteq (n^1, \ldots n^D)$, so we introduce $D$ operators instead,
\begin{gather}
\hat{n}^r = \sum_\mvec{n} n^r \ket{e_\mvec{n}}\bra{e_\mvec{n}}.
\end{gather}
Using these, we now define $\hat{x}^r \doteq \mc{X}^r(\hat{n}^r)$, where the functions $\mc{X}^r$ are bijective but otherwise arbitrary. We can hence replace summation over $\mvec{n}$ with that over $\mvec{x} \doteq (x^1, \ldots x^D)$, where $x^r \doteq \mc{X}^r(n^r)$. This gives $D$ mutually commuting operators,
\begin{gather}\label{eq:xrd}
\hat{x}^r = \sum_\mvec{x} x^r \ket{\mvec{x}}\bra{\mvec{x}},
\end{gather}
where $\ket{\mvec{x}} \equiv \ket{x^1, \ldots x^D}$ is just an alternative notation for $\ket{e_\mvec{n}}$. The direct sum of $\hat{x}^r$,
\begin{gather}
\oper{x} \doteq (\hat{x}^1, \ldots \hat{x}^D),
\end{gather}
will hence be called a generalized $D$-dimensional coordinate operator on $\Psi$. The set $X$, comprised of all $\mvec{x}$, can be understood as the eigenvalue space of $\oper{x}$ and thus will be called a generalized coordinate set. The word ``generalized'' here refers to the fact that the definition of this set depends on how $\mc{X}^r$ are chosen.

\subsection{Continuous coordinates and momenta} 
\label{sec:cmom}

\subsubsection{Coordinate}
\label{sec:coord2}

From now on, let us assume that $N$ is infinite and (until \Sec{sec:vector}) that the index $\mvec{n}$ is continuous, so the set of all $\mvec{n}$ is a differentiable manifold of dimension $D$. We will also assume that $\mc{X}^r$ are smooth, so $X$ is a differentiable manifold of dimension $D$ too. We hereupon reserve the term ``spatial'' for this emergent manifold and will refer to $\oper{x}$ as a ``canonical coordinate''. Unlike $\Psi$, the coordinate space $X$ can have an arbitrary real metric $\oper{\gamma}(\mvec{x})$, so the volume element in $X$ is given by
\begin{gather}\label{eq:volel}
\mcu{D}x = \sqrt{\gamma(\mvec{x})} \,\msf{d}x,
\end{gather}
where $\gamma = |\mbox{det}\,\oper{\gamma}|$, and $\msf{d}x \equiv dx^1\ldots dx^D$. Using the notation $\delta(\mvec{x}) \equiv \delta(x^1)\ldots \delta(x^D)$, let also define the ``generalized delta function'' \cite{ref:dewitt52}
\begin{gather}
\delta(\mvec{x}, \mvec{x}') \doteq \delta(\mvec{x} - \mvec{x}')/\sqrt{\gamma(\mvec{x})}
\end{gather}
and adopt the following normalization for $\ket{\mvec{x}}$:
\begin{gather}\label{eq:xnorm}
\braket{\mvec{x}|\mvec{x}'} = \delta(\mvec{x}, \mvec{x}').
\end{gather}
Then, the sum in \Eq{eq:xrd} is replaced with an integral, yielding
\begin{gather}\label{eq:operx}
\oper{x} = \int \mvec{x} \ket{\mvec{x}}\bra{{\mvec{x}}}\,\mcu{D}x,
\end{gather}
and, similarly, the unit operator $\hat{1}$ can be represented as
\begin{gather}\label{eq:xunit}
\hat{1} = \int \ket{\mvec{x}}\bra{{\mvec{x}}}\,\mcu{D}x.
\end{gather}

For any $\ket{\psi}$, we now can define its ``$\mvec{x}$-representation'', or a scalar field on $X$,
\begin{gather}
\psi(\mvec{x}) \doteq \braket{{\mvec{x}}|\psi}.
\end{gather}
(Possible time dependence is also assumed, tacitly, throughout the paper.) In particular, the $\mvec{x}$-representation of $\ket{\mvec{x}'}$ is $\psi_{\mvec{x}'}(\mvec{x}) = \delta(\mvec{x}, \mvec{x}')$; cf. \Eq{eq:xnorm}. Likewise, any operator $\hat{F}$ acting on $\ket{\psi}$ generates a field
\begin{gather}\label{eq:fpsi}
\hat{F}\psi(\mvec{x}) \equiv (\hat{F}\ket{\psi})(\mvec{x}) \doteq \braket{{\mvec{x}}|\hat{F}|\psi},
\end{gather}
\eg $\oper{x}\psi(\mvec{x}) = \mvec{x}\psi(\mvec{x})$. Assuming the notation $F(\mvec{x}, \mvec{x}') \doteq \braket{{\mvec{x}}|\hat{F}|\mvec{x}'}$, we can also rewrite this as
\begin{gather}
\hat{F}\psi(\mvec{x}) = \int F(\mvec{x}, \mvec{x}')\,\psi(\mvec{x}')\, \mcu{D}x'.
\end{gather}

\subsubsection{Momentum} 
\label{sec:momentum}

Suppose, for now, that $\hat{F}$ is local in $X$, so $\hat{F}\ket{\mvec{x}'}$ is close to $\ket{\mvec{x}'}$. Then, due to \Eq{eq:xnorm}, $F(\mvec{x}, \mvec{x}')$ must be a narrow function of $\tilde{\mvec{x}} \doteq \mvec{x}' - \mvec{x}$ and slow function of $\mvec{x}$, which we denote as $\mc{F}(\mvec{x}, \tilde{\mvec{x}})$. Hence one can Taylor-expand $\psi(\mvec{x}')$ in $\tilde{\mvec{x}}$; \ie $\psi(\mvec{x}') = \psi(\mvec{x}) + \tilde{\mvec{x}} \cdot \nabla_{\mvec{x}} \psi(\mvec{x}) {+ \ldots}$, which leads to
\begin{gather}
\hat{F}\psi(\mvec{x}) = \big[\mc{F}_1(\mvec{x}) + \mc{F}_2^r(\mvec{x})\,\partial_{x^r} + \ldots\big]\,\psi(\mvec{x}),
\end{gather}
where $\mc{F}_1(\mvec{x}) \doteq \int \mc{F}(\mvec{x}, \tilde{\mvec{x}})\,d\tilde{X}$, $\mc{F}^r_2(\mvec{x}) \doteq \int \mc{F}(\mvec{x}, \tilde{\mvec{x}})\,\tilde{x}^r\,d\tilde{X}$, etc. In other words, we get
\begin{gather}
\hat{F} = \mc{F}_1(\oper{x}) + i \mc{F}_2^r(\oper{x})\,\hat{\kappa}_r + \ldots,
\end{gather}
where $\hat{\kappa}_r$ are defined such that their $\mvec{x}$-representation is $\hat{\kappa}_r = -i\partial_{x^r}$, or, in a vector form, $\oper{\kappa} = -i\nabla_{\mvec{x}}$. This shows that any local operator is expressed as a function of $\oper{x}$ and $\oper{\kappa}$ (also see below for a more formal argument), so, like the coordinate, $\oper{\kappa}$ is a fundamental operator in our theory. It is not Hermitian unless $\gamma$ is constant, and there is no way to fix this for general $X$.\footnote{One may recognize this as a long-standing problem in quantization theory \cite{ref:gotay96}.} Nevertheless, this inconvenience can be evaded as follows.

Let us assume, from now on, that $X$ has topological properties of $\mathbb{R}^D$. (Locally, such $X$ can still mimic any other manifold within an arbitrarily large region.) Then, a Hermitian operator can be constructed out of $\oper{\kappa}$ via $\oper{k} \doteq \oper{\kappa} + \Delta\mvec{\kappa}(\oper{x})$, where
\begin{gather}
\Delta\mvec{\kappa}(\mvec{x}) = -i\nabla_\mvec{x}\ln[\gamma^{\frac{1}{4}}(\mvec{x})].
\end{gather}
In the $\mvec{x}$-representation, this gives
\begin{gather}\label{eq:operkx}
\oper{k}\psi(\mvec{x}) = - i \gamma^{-\frac{1}{4}}(\mvec{x})\, \nabla_\mvec{x} \big[\gamma^{\frac{1}{4}}(\mvec{x}) \psi(\mvec{x})\big],
\end{gather}
or, symbolically, $\oper{k} = -i\eth_\mvec{x}$. The symbol $\eth_\mvec{x}$ will be called a regularized gradient on $X$, and $\oper{k}$ will be termed ``canonical momentum'' \cite{ref:dewitt52,ref:domingos84}. It represents the direct sum of $D$ mutually commuting operators $\hat{k}_r$ satisfying
\begin{gather}\label{eq:maincommut}
[\hat{x}^r, \hat{k}_s] = i \delta^r_s.
\end{gather}
Due to the obvious similarity with QM, we will, for brevity, attribute dynamics as ``quantum'' if this commutator is nonnegligible in a given problem. Otherwise the dynamics will be attributed as ``classical'', as in \Sec{sec:ll} and \Eqs{eq:ppart} and \eq{eq:clL}, or ``quasiclassical'', as in \Secs{sec:hydrobasic} and \ref{sec:golimit}.

As $\oper{k}$ is Hermitian, its eigenvectors, $\ket{\mvec{k}}$, form an orthogonal basis on $\Psi$, and the corresponding eigenvalues, $\mvec{k} = (k_1, \ldots k_D)$, are real and comprise some set $K$. If $K$ were discrete, then $\oper{x}$'s $\mvec{k}$-representation, $x^r_{\mvec{k}, \mvec{k}'} \doteq \braket{\mvec{k}|\hat{x}^r|\mvec{k}'}$, would have had to satisfy
\begin{align}
(k_s' - k_s)\,x^r_{\mvec{k}, \mvec{k}'} & = \braket{\mvec{k}|\hat{x}^r k_s' - k_s \hat{x}^r|\mvec{k}'} \nonumber\\
& = \braket{\mvec{k}|[\hat{x}^r, \hat{k}_s]|\mvec{k}'}\nonumber \\
& = i\delta^r_s \braket{\mvec{k}|\mvec{k}'} \propto \delta^r_s\,\delta_{\mvec{k},\mvec{k}'},
\end{align}
which is impossible at $k_s' = k_s$ for $r = s$.\footnote{In traditional QM, this makes it problematic to define a phase operator canonically conjugate a quantized action. For review, see, \eg \Ref{phd:pellonpaa02}.} We thus assume hereupon that $K$ is a differential manifold.

For any $\mvec{k}$ and $\mvec{x}$, we introduce a ``dot product'' as
\begin{gather}\label{eq:dotpr}
\mvec{k} \cdot \mvec{x} \equiv k_r x^r,
\end{gather}
which is not an inner product \textit{per~se}, because $\mvec{x}$ and $\mvec{k}$ are not necessarily vectors. We will also allow a general metric $\oper{\vartheta}(\mvec{k})$ on $K$, which can be chosen arbitrarily. Then a volume element in $K$ is given by
\begin{gather}
\mcu{D}k = \sqrt{\vartheta(\mvec{k})} \,\msf{d}k,
\end{gather}
where $\vartheta = |\mbox{det}\,\oper{\vartheta}|$, and $\msf{d}k \equiv dk_1\ldots dk_D$. Assuming the notation $\delta(\mvec{k}) \equiv \delta(k_1)\ldots \delta(k_D)$, so
\begin{gather}\label{eq:deltaf}
\delta(\mvec{k}) = (2\pi)^{-D} \int e^{i\mvec{k}\cdot \mvec{x}}\,\msf{d} x,
\end{gather}
let also define
\begin{gather}
\delta(\mvec{k}, \mvec{k}') \doteq \delta(\mvec{k} - \mvec{k}')/\sqrt{\vartheta(\mvec{k})}
\end{gather}
and adopt the following normalization for $\ket{\mvec{k}}$:
\begin{gather}\label{eq:knorm}
\braket{\mvec{k}|\mvec{k}'} = \delta(\mvec{k}, \mvec{k}').
\end{gather}
Then, $\oper{k}$ can be expressed as
\begin{gather}
\oper{k} = \int \mvec{k}\ket{\mvec{k}}\bra{\mvec{k}}\,\mcu{D}k,
\end{gather}
and we also note that
\begin{gather}\label{eq:kunit}
\hat{1} = \int \ket{\mvec{k}}\bra{\mvec{k}}\,\mcu{D}k.
\end{gather}

\subsubsection{Fundamental matrix} 

The fundamental matrix, $\braket{{\mvec{x}}|\mvec{k}} \equiv \psi_{\mvec{k}}(\mvec{x})$, is found as a solution of the following equation:
\begin{gather}
-i\eth_\mvec{x} \psi_{\mvec{k}}(\mvec{x}) = \mvec{k} \psi_{\mvec{k}}(\mvec{x}).
\end{gather}
This yields $\psi_{\mvec{k}}(\mvec{x}) = C_\mvec{k}\gamma^{-\frac{1}{4}}(\mvec{x})\, \exp(i\mvec{k} \cdot \mvec{x})$, where $C_\mvec{k}$ is the integration constant. Then, due to \Eq{eq:knorm}, we get
\begin{gather}\label{eq:fme}
\braket{\mvec{x}|\mvec{k}} = \braket{{\mvec{k}}|\mvec{x}}^* = 
\frac{\exp(i \mvec{k} \cdot \mvec{x})}{(2\pi)^{\frac{D}{2}}\,[\gamma(\mvec{x})\,\vartheta(\mvec{k})]^{\frac{1}{4}}}.
\end{gather}

In particular, this yields the $\mvec{k}$-representation of $\oper{x}$:
\begin{align}
\braket{\mvec{k}|\oper{x}|\mvec{k}'} & = \int \braket{\mvec{k}|\mvec{x}} \mvec{x} \braket{\mvec{x}|\mvec{k}'}\,\mcu{D}x \nonumber\\
& = (2\pi)^{-D}[\vartheta(\mvec{k})\,\vartheta(\mvec{k}')]^{-\frac{1}{4}}  \int \mvec{x}\,e^{i (\mvec{k}' - \mvec{k}) \cdot \mvec{x}}\,\msf{d}x \nonumber\\
& = i \nabla_{\mvec{k}} \vphantom{\int}
\delta(\mvec{k}', \mvec{k}),
\end{align}
where \Eq{eq:operx} was used. Hence,
\begin{gather}\label{eq:operxk}
\oper{x} = i\eth_{\mvec{k}},
\end{gather}
where $\eth_\mvec{k}$ is the regularized gradient on $K$, defined similarly to $\eth_\mvec{x}$; namely,
\begin{gather}\label{eq:operxk2}
\oper{x}\psi(\mvec{k}) \doteq i \vartheta^{-\frac{1}{4}}(\mvec{k})\, \nabla_\mvec{k} \big[\vartheta^{\frac{1}{4}}(\mvec{k}) \psi(\mvec{k})\big].
\end{gather}
In conjunction with \Eq{eq:operkx}, this permits, if needed, to reattribute $-\oper{k}$ as a canonical coordinate and $\oper{x}$ as a canonical momentum (cf. Ref.~\cite[Sec.~45]{book:landau1}).  

\subsection{Translation operators} 
\label{app:transl}

For any function $\mc{F}(u)$, one can formally write
\begin{gather}
\mc{F}(u + w) = \sum^\infty_{n = 0} \frac{(w\, \partial_u)^n}{n!}\, \mc{F}(u)
= e^{w\, \partial_u} \mc{F}(u),
\end{gather}
so $e^{w\, \partial_u}$ acts as a translation operator. For $D$ arguments, one similarly gets
\begin{align}
\mc{F}(\mvec{u} + \mvec{w}) & = (e^{w_1\, \partial_{u_1}}) \ldots (e^{w_D\, \partial_{u_D}})\, \mc{F}(\mvec{u}) \nonumber \\
& = e^{w_1\, \partial_{u_1} + \ldots + w_D\, \partial_{u_D}} \mc{F}(\mvec{u}) \nonumber \\
& = e^{\mvec{w} \cdot \nabla_{\mvec{u}}} \mc{F}(\mvec{u}).\label{eq:psiexp}
\end{align}
so the translation operator is $e^{\mvec{w} \cdot \nabla_{\mvec{u}}}$, and a gradient can be viewed as the generator of translations. 

Let us also introduce translations generated by the regularized gradients $\eth_{\mvec{x}}$ and $\eth_{\mvec{k}}$, \ie by $\oper{x}$ and $\oper{k}$. For $\hat{T}_{\mvec{q}} \doteq e^{ - i \oper{k} \cdot \mvec{q}}$ one gets, using \Eq{eq:kunit}, that
\begin{align}
\hat{T}_{\mvec{q}} \ket{\mvec{x}} 
& = \int  e^{ - i \oper{k} \cdot \mvec{q}}\ket{\mvec{k}}\braket{{\mvec{k}}|\mvec{x}}\,\mcu{D}k \nonumber\\
& = \int e^{- i \mvec{k} \cdot \mvec{q}}\ket{\mvec{k}}\braket{{\mvec{k}}|\mvec{x}}\,\mcu{D}k \nonumber \\
& = \Gamma_\gamma (\mvec{x}, \mvec{q})\int \ket{\mvec{k}}\braket{{\mvec{k}}|\mvec{x} + \mvec{q}}\,\mcu{D}k,
\end{align}
where $\Gamma_\gamma (\mvec{x}, \mvec{q}) \doteq [\gamma(\mvec{x} + \mvec{q})/\gamma(\mvec{x})]^{\frac{1}{4}}$. This gives
\begin{gather}\label{eq:txk1}
\hat{T}_{\mvec{q}}\ket{\mvec{x}}  = \Gamma_\gamma (\mvec{x}, \mvec{q})\ket{\mvec{x} + \mvec{q}}.
\end{gather}
One can also express the effect of $\hat{T}_{\mvec{q}}$ in the scalar form,
\begin{align}
\hat{T}_{\mvec{q}} \psi(\mvec{x}) \equiv \braket{\mvec{x}|\hat{T}_{\mvec{q}}|\psi} 
& = \braket{\hat{T}_{\mvec{q}}^\dag\mvec{x}|\psi} \nonumber \\
& = \Gamma_\gamma (\mvec{x}, -\mvec{q})\braket{\mvec{x} - \mvec{q}|\psi} \nonumber \\
& = \Gamma_\gamma (\mvec{x}, -\mvec{q})\psi(\mvec{x} - \mvec{q}),
\end{align}
where $\hat{T}^\dag_{\mvec{q}} = \hat{T}_{-\mvec{q}}$ was used. Then, from \Eq{eq:psiexp}, one gets
\begin{gather}
\hat{T}_{\mvec{q}} \psi(\mvec{x}) = \gamma^{-\frac{1}{4}}(\mvec{x})\,e^{-\mvec{q} \cdot \nabla_\mvec{x}} [\gamma^{\frac{1}{4}}(\mvec{x}) \psi(\mvec{x})].
\end{gather}
Similar formulas apply to $\hat{T}_{\mvec{p}} \doteq e^{i \mvec{p} \cdot \oper{x}}$; namely,
\begin{gather}\label{eq:txk2}
\hat{T}_{\mvec{p}}\ket{\mvec{k}}  = \Gamma_\vartheta (\mvec{k}, \mvec{p})\ket{\mvec{k} + \mvec{p}},
\end{gather}
where $\Gamma_\vartheta (\mvec{k}, \mvec{p}) \doteq [\vartheta(\mvec{k} + \mvec{p})/\vartheta(\mvec{k})]^{\frac{1}{4}}$, and
\begin{gather}
\hat{T}_{\mvec{p}} \psi(\mvec{k}) = \vartheta^{-\frac{1}{4}}(\mvec{k})\,e^{- \mvec{p} \cdot \nabla_\mvec{k}} [\vartheta^{\frac{1}{4}}(\mvec{k}) \psi(\mvec{k})].
\end{gather}

More generally, consider a family of operators
\begin{gather}\label{eq:T}
\hat{T}_{\zeta} \doteq \exp(i\zeta \wedge \hat{\msf{z}}),
\end{gather}
where $\zeta \doteq (\mvec{q}, \mvec{p})$ is the family parameter, $\hat{\msf{z}} \doteq (\oper{x}, \oper{k})$ is a direct product of the coordinate and momentum operators, and the wedge product is understood formally~as
\begin{gather}
\zeta \wedge \hat{\msf{z}} \doteq \mvec{p} \cdot \hat{\mvec{x}} - \hat{\mvec{k}} \cdot \mvec{q}
\end{gather}
(cf. \App{sec:pspgeom}). One can show that\footnote{This is seen from $e^{\hat{A} + \hat{B}} = e^{\hat{A}}\, e^{\hat{B}}\,e^{-[\hat{A},\hat{B}]/2}$, which holds for any $\hat{A}$ and $\hat{B}$ commuting with $[\hat{A},\hat{B}]$ \cite{ref:magnus54}. Take $\hat{A} = i \zeta_1 \wedge \hat{\msf{z}}$ and $\hat{B} = i \zeta_2 \wedge \hat{\msf{z}}$; then, $[\hat{A},\hat{B}] = i \zeta_1 \wedge \zeta_2$ is a scalar.}
\begin{gather}\label{eq:aux112}
\hat{T}_{\zeta_1 + \zeta_2} = \hat{T}_{\zeta_1}\, \hat{T}_{\zeta_2}\,e^{-i \zeta_1 \wedge \zeta_2/2},
\end{gather}
so translations determined by $\hat{T}_{\zeta}$ are generally not commutative. A special case of \Eq{eq:aux112} is
\begin{gather}
\hat{T}_{\zeta} 
= \hat{T}_{\mvec{q}}\, \hat{T}_{\mvec{p}}\, e^{i \mvec{p} \cdot \mvec{q}/2} 
= \hat{T}_{\mvec{p}}\,  \hat{T}_{\mvec{q}}\, e^{-i \mvec{p} \cdot \mvec{q}/2},
\end{gather}
which, together with \Eqs{eq:txk1} and \eq{eq:txk2}, also yields
\begin{gather}
T_{\zeta} (\mvec{x}_1, \mvec{x}_2)
= \frac{\delta(\mvec{x}_1 - \mvec{x}_2 - \mvec{q})}{[\gamma(\mvec{x}_1)\,\gamma(\mvec{x}_2)]^{\frac{1}{4}}}\,e^{i\mvec{p} \cdot (\mvec{x}_1 + \mvec{x}_2)/2}, \label{eq:aux61}\\
T_{\zeta} (\mvec{k}_1, \mvec{k}_2)
= \frac{\delta(\mvec{k}_1 - \mvec{k}_2 - \mvec{p})}{[\vartheta(\mvec{k}_1)\,\vartheta(\mvec{k}_2)]^{\frac{1}{4}}}\,e^{-i(\mvec{k}_1 + \mvec{k}_2) \cdot \mvec{q}/2}. \label{eq:aux62}
\end{gather}

Similarly to how the $\mvec{x}$- and $\mvec{k}$-representations were introduced in \Sec{sec:ps}, one can now define the $\zeta$-representation, $\msf{M}$, for any operator $\hat{M}$, namely, via
\begin{gather}\label{eq:fz}
\msf{M}(\zeta) \doteq \mbox{tr}\,\big(\hat{T}_{-\zeta} \hat{M}\big).
\end{gather}
Then one can express $\hat{M}$ as follows:
\begin{gather}
\hat{M} = (2\pi)^{-D}\int \msf{M}(\zeta)\, \hat{T}_{\zeta}\, \mcu{D}\zeta,
\end{gather}
where $\mcu{D}\zeta \doteq \msf{d} x\,\msf{d} k$ (note the difference from $\mcu{D}x\,\mcu{D}k$); this is proved, \eg by substituting \Eq{eq:fz} and
\begin{gather}
\hat{T}_{\zeta}  = \int \msf{T}_{\zeta} (\mvec{x}_1, \mvec{x}_2)\, \ket{\mvec{x}_1}\bra{{\mvec{x}}_2}\,\mcu{D}x_1 \mcu{D}x_2
\end{gather}
together with \Eq{eq:aux61}. It is seen then that $\hat{T}_{\zeta}$ \textit{form a complete basis for operators on} $\Psi$. On the other hand, $\hat{T}_{\zeta}$ themselves are functions of $\hat{\msf{z}}$ (and, possibly, time), so any operator on $\Psi$ can be represented as a function of $\hat{\msf{z}}$. This generalizes the argument given in \Sec{sec:momentum}.

\subsection{Schr\"odinger equation on a manifold}

Like any other operator on $\Psi$ (\Sec{sec:coord2}), the Hamiltonian $\hat{H}$ can now be cast in the form 
\begin{gather}\label{eq:mch}
\hat{H} = H(t, \oper{x}, \oper{k}),
\end{gather}
where $H$ is some function. (The latter must not be confused with the function $H$ introduced in \Sec{sec:basic}; we merely recycle the notation here.) Note that \Eq{eq:mch} is \textit{dictated} by the geometry of $X$, so it does not need to be postulated separately, contrary to how it is often done in literature.\footnote{It is the fact that $X$ is a continuum that renders a \textit{pair} of operators necessary. If $X$ is discrete instead, one can always introduce an alternative coordinate $\oper{J} = (J^1, \ldots J^D)$ along the lines of \Sec{sec:coord} such that each of its eigenvectors equals some $\ket{H_n}$. Then $\hat{H} = H(\oper{J})$, so the very notion of a canonically conjugate (to $\oper{J}$) operator becomes redundant. That said, if the space $J$ formed by all eigenvalues of $\mvec{J}$ is dense enough, it can be \textit{approximated} with a differential manifold. Then such a conjugate operator, say, $-\oper{\phi}$, can be defined via $\oper{\phi} \doteq i\eth_{\mvec{J}}$, and, correspondingly, $\oper{J} = - i\eth_\mvec{\phi}$. One may recognize these as operators of ``quasiclassical'' angle-action variables.}

We can now switch from the vector form of dynamic equations (\Sec{sec:simpleeq}) to their scalar representations. Consider Taylor-expanding $H$, so $\hat{H}$ becomes a series of polynomials of the form $\mc{P}(t, \oper{x}, \oper{k}) \rdoteq \hat{\mc{P}}$. Each polynomial of order $\msf{n} > 0$ can be expressed either as $\oper{x}\hat{\mc{P}'}$ or as $\oper{k}\hat{\mc{P}'}$, where $\hat{\mc{P}}' \doteq \mc{P}'(t, \oper{x}, \oper{k})$, and $\mc{P}'$ is a polynomial of order $\msf{n} - 1$. In particular, consider adopting the $\mvec{x}$-representation. Then it is readily seen that $\bra{\mvec{x}}\hat{\mc{P}} = \mvec{x}\bra{\mvec{x}}\hat{\mc{P}}'$ in the former case, and $\bra{\mvec{x}}\hat{\mc{P}} = (- i\eth_{\mvec{x}})\bra{\mvec{x}}\hat{\mc{P}}'$ in the latter case. By induction, one then obtains that $\bra{\mvec{x}} \mc{P}(t, \oper{x}, \oper{k}) = \mc{P}(t, \mvec{x}, -i\eth_{\mvec{x}}) \bra{\mvec{x}}$ for any $\mc{P}$. Therefore, the $\mvec{x}$-representation of $\hat{H}$ is 
\begin{gather}\label{eq:Omxr}
\hat{H} = H(t, \mvec{x}, -i\eth_{\mvec{x}}).
\end{gather}
(That said, unless the dependence on the momentum here is not polynomial, such $\hat{H}$ is not a local operator \textit{per~se}, as it includes spatial derivatives of unlimitedly high orders.) 

The Lagrangian \eq{eq:Lreduced} now becomes
\begin{gather}\label{eq:integrL}
L = \int \mcc{L}\,\mcu{D}x, 
\end{gather}
where its spatial density, $\mcc{L}$, is given by
\begin{gather}\label{eq:Lred2}
\mcc{L} = \frac{i}{2}\,\big[\psi^*(\partial_t\psi) - (\partial_t\psi^*)\psi\big] - \psi^*\,H(t, \mvec{x}, -i\eth_{\mvec{x}})\,\psi,
\end{gather}
and $\psi \equiv \psi(t, \mvec{x})$. The corresponding action integral is
\begin{gather}\label{eq:SmccL}
\mc{S} = \int^{t_2}_{t_1} \mcc{L}\,\mcu{D}x\,dt.
\end{gather}
The associated PDEs hence flow from the LAP,
\begin{gather}\label{eq:leastactpsi}
\delta_{\psi^*}\mc{S} = 0, \quad \delta_{\psi}\mc{S} = 0,
\end{gather}
which serve as equations for $\psi$ and $\psi^*$, correspondingly. Alternatively, one can simply project \Eq{eq:bLSE} on the spatial basis. In either case, one arrives at
\begin{gather}\label{eq:psix}
i\partial_t\psi = H(t, \mvec{x}, -i\eth_{\mvec{x}})\,\psi,
\end{gather}
which describes a field on a $D$-dimensional manifold. One may notice that \Eq{eq:psix} is similar to the QM LSE for a scalar particle in the spatial representation. Note also that, in contrast to QM, we derived it here without a reference to the wave-particle duality but rather from relatively weak assumptions on the geometry of $\Psi$.

\section{Projected equations for the action operator}
\label{sec:dmeq}

In contrast to the vector equation for $\ket{\psi}$, the operator equation for the action [\Eq{eq:vNn}] can be projected on two bases simultaneously and then describes a field on a $2D$-dimensional manifold such as $X \times X$, $K \times K$, or $X \times K$. Consider deriving a scalar equation for an \textit{invariant linear measure}, $\varrho$, of the action operator, $\hat{\rho}$, acted upon by some projection operator $\hat{\Pi}$. Only one such measure exists for given $\hat{\Pi}\hat{\rho}$, namely,
\begin{gather}\label{eq:psidef}
\varrho \doteq \mbox{tr}\,(\hat{\Pi}\hat{\rho}) = \mbox{tr}\,(\hat{\rho}\hat{\Pi}) = \braket{\psi|\hat{\Pi}|\psi},
\end{gather}
which we term the $\hat{\Pi}$-image (or ``symbol''; cf. \Ref{ref:mcdonald88}) of $\hat{\rho}$ and which can be interpreted as the expectation value of $\hat{\Pi}$ multiplied by~$I$. (One may notice parallels between this approach and what is known in QM as the frame formalism \cite{ref:ferrie11}.) By applying $\hat{\Pi}$ to \Eq{eq:vNn} and taking the trace, one gets the following equation for $\varrho$:
\begin{gather}\label{eq:pvNe}
i\partial_t \varrho  + \mbox{tr}\,(\hat{\Pi}[\hat{\rho}, \hat{H}]) = 0,
\end{gather}
where we assume, for simplicity, that $\hat{\Pi}$ is time-independent. For certain projectors, \Eq{eq:pvNe} can be cast in a tractable form, some examples of which will now be discussed.

\subsection{Dynamics on ${\boldsymbol{X \times X}}$}
\label{sec:preqsp}

\subsubsection{Action operator in the spatial representation} 

First, consider a projector
\begin{gather}
\hat{\Pi} = \ket{\mvec{x}''}\bra{\mvec{x}'},
\end{gather}
so the $\hat{\Pi}$-image function becomes $\varrho = \braket{\mvec{x}'|\psi}\braket{\psi|\mvec{x}''}$. The latter is simply the spatial representation of the action operator, $\rho(t, \mvec{x}', \mvec{x}'')$, so \Eq{eq:pvNe} takes the following form:
\begin{gather}\label{eq:rb}
i\partial_t \rho(t, \mvec{x}', \mvec{x}'') + \mc{R}(t, \mvec{x}', \mvec{x}'') - \mc{R}^*(t, \mvec{x}'', \mvec{x}') = 0,
\end{gather}
where we introduced
\begin{gather}
\mc{R}(t, \mvec{x}', \mvec{x}'') \doteq \int \rho(t, \mvec{x}', \mvec{x})\,H(t, \mvec{x}, \mvec{x}'')\,\mcu{D}x.
\end{gather}
Since $H(t, \mvec{x}, \mvec{x}'')$ can be expressed as
\begin{gather}
\braket{\mvec{x}|\hat{H}|\mvec{x}''} = H(t, \mvec{x}, -i\eth_{\mvec{x}})\,\delta(\mvec{x} - \mvec{x}''),
\end{gather}
and since $\hat{H}$ is Hermitian, one gets
\begin{align}
\mc{R}(t, \mvec{x}', \mvec{x}'') 
& = \int [H(t, \mvec{x}, -i\eth_{\mvec{x}})\,\rho(t, \mvec{x}, \mvec{x}')]^* \delta(\mvec{x} - \mvec{x}'')\,\mcu{D}x \nonumber \\
& = \big[H(t, \mvec{x}, -i\eth_{\mvec{x}})\,\rho(t, \mvec{x}, \mvec{x}')\big]^*_{\mvec{x} = \mvec{x}''}.
\end{align}
This yields a differential form of \Eq{eq:rb}:
\begin{align}\label{eq:rhoxpxpp}
i\partial_t \rho(t, \mvec{x}', \mvec{x}'') & + \big[H(t, \mvec{x}, -i\eth_{\mvec{x}})\,\rho(t, \mvec{x}, \mvec{x}')\big]^*_{\mvec{x} = \mvec{x}''}\nonumber\\
& - \big[H(t, \mvec{x}, -i\eth_{\mvec{x}})\,\rho(t, \mvec{x}, \mvec{x}'')\big]_{\mvec{x} = \mvec{x}'} = 0
\end{align}
(cf., \eg Ref.~\cite[Sec.~40]{book:landau10}).

\subsubsection{ACT in the spatial representation} 
\label{sec:actspat}

In particular, taking $\mvec{x}' = \mvec{x}''$ in \Eq{eq:rhoxpxpp} leads to
\begin{gather}\label{eq:rhoxx}
\partial_t A^2 - 2\mbox{Im}\,
\big[H(t, \tilde{\mvec{x}}, -i\eth_{\tilde{\mvec{x}}})\,\rho(t, \tilde{\mvec{x}}, \mvec{x})\big]_{\tilde{\mvec{x}} = \mvec{x}} = 0,
\end{gather}
where we substituted $\rho(t, \mvec{x}, \mvec{x}) = |\psi(t, \mvec{x})|^2 \rdoteq A^2$, recognized as the action spatial density. This equation, notably, permits a divergence form,\footnote{For a similar theorem in QM, see, \eg Ref.~\cite[Sec.~19]{book:landau3} or the Madelung fluid representation of QM \cite{ref:madelung27}.}
\begin{gather}\label{eq:claw}
\partial_t A^2 + \nabla \cdot \mvec{j} = 0
\end{gather}
(here the current $\mvec{j}$ is some vector field on $X$, and $\nabla \equiv \nabla_\mvec{x}$), which can be understood as the ACT in the spatial representation. Below, we discuss this in detail.

To derive \Eq{eq:claw}, let us express $\psi$ as $A e^{i\theta}$, where $A$ and $\theta$ are real, and substitute this into \Eq{eq:Lred2}. The Lagrangian density is then easily seen to have the form
\begin{gather}\label{eq:mcuL}
\mcc{L} = - A^2\,\partial_t\theta - \mcu{H}(A, \partial_{x^r}A, \partial_{x^r} \theta, \ldots; t, \mvec{x}),
\end{gather}
where $A$ and $\theta$ must now be treated as independent functions, $\mcu{H}$ is some function, and the ellipses denote higher-order derivatives. As a spin-off, we also obtain
\begin{gather}
\mcu{H} = \omega A^2, \quad A^2 = \partial_\omega \mcc{L},
\end{gather}
where we substituted $\omega \doteq - \partial_t \theta$ and $\mcc{L} = 0$ (\Sec{sec:ydef}). 

The LAP leads to the Euler-Lagrange equations
\begin{gather}\label{eq:laplp}
\delta_A \mc{S} = 0, \quad \delta_\theta \mc{S} = 0.
\end{gather}
Let us consider the latter equation in detail, using \Eq{eq:SmccL} for $\mc{S}$. The variation of $\mc{S}$ with respect to $\theta$ is
\begin{gather}\nonumber
\delta_\theta \mc{S} = \int [- A^2\,\delta(\partial_t\theta) - \mcu{J}^r\,\delta\theta_{,r} - \mcu{J}^{rs}\,\delta\theta_{,rs} - \ldots]\,\mcu{D}x\,dt,
\end{gather}
where $\mcu{J}^r \doteq \partial\mcu{H}/\partial \theta_{,r}$, the symbol $_{,r}$ denotes a partial derivative with respect to $x^r$, and similarly for higher-order derivatives. (The time integral is again taken from $t_1$ to $t_2$, but the limits are henceforth omitted.) Let us write this as a sum of independent integrals, $\delta_\theta \mc{S} = \sum \delta \mcu{S}_n$. Assuming the amplitude vanishes at infinity, and thus so are all $\mcu{J}$, they are taken by parts as follows. 

The first one, $\delta\mcu{S}_1 = - \int A^2\,\delta(\partial_t\theta)\,\mcu{D}x\,dt$, is simple:
\begin{gather}
\delta\mcu{S}_1 = \int (\partial_t A^2)\,\delta\theta\,\mcu{D}x\,dt.
\end{gather}
To take the second integral, $\delta\mcu{S}_2 = - \int \mcu{J}^r\,\delta\theta_{,r}\,\mcu{D}x\,dt$, we use \Eq{eq:volel} for $\mcu{D}x$; then,
\begin{align}
\delta\mcu{S}_2 & = \int [- (\sqrt{\gamma} \mcu{J}^r \delta\theta)_{,r} + (\sqrt{\gamma} \mcu{J}^r)_{,r}\,\delta\theta]\,\msf{d}x\,dt \nonumber\\
 & = \int (\sqrt{\gamma} \mcu{J}^r)_{,r}\,\delta\theta\,\msf{d}x\,dt \nonumber\\
 & = \int {\mcu{J}^r}_{;r}\,\delta\theta\,\mcu{D}x\,dt,
\end{align}
where ${\mcu{J}^r}_{;r}$ is the divergence of $\mcu{J}^r$ \cite[Sec.~4.7]{book:weinberg},
\begin{gather}\label{eq:div}
{\mcu{J}^r}_{;r} = \frac{1}{\sqrt{\gamma}}\,\left(\sqrt{\gamma} \mcu{J}^r\right)_{, r}.
\end{gather}
The third integral, $\delta \mcu{S}_3 = -\int \mcu{J}^{rs}\,\delta\theta_{,rs} \,\mcu{D}x\,dt$, can be cast as follows:
\begin{align}
\delta \mcu{S}_3 
& = \int [-(\sqrt{\gamma}\mcu{J}^{rs} \delta\theta_{,r})_{,s} + (\sqrt{\gamma}\mcu{J}^{rs})_{,s}\,\delta\theta_{,r}]\,\msf{d}x\,dt \nonumber \\
& =  \int (\sqrt{\gamma}\mcu{J}^{rs})_{,s}\,\delta\theta_{,r}\,\msf{d}x\,dt \nonumber \\
& = \int {\mcu{J}^{rs}}_{;s}\,\delta\theta_{,r}\,\mcu{D}x\,dt.
\end{align}
This is identical to $\delta\mcu{S}_2$ up to replacing $-\mcu{J}^r$ with ${\mcu{J}^{rs}}_{;s}$. We then readily cast $\delta \mcu{S}_3$ in a divergence form, 
\begin{gather}
\delta\mcu{S}_3 = - \int ({\mcu{J}^{rs}}{}_{;s}){}_{;r}\,\delta\theta\,\mcu{D}x\,dt,
\end{gather}
and similarly for the remaining $\delta \mcu{S}_n$. Hence one gets
\begin{gather}\label{eq:aux91}
\delta_\theta \mc{S} = \int (\partial_t A^2 + {j^r}_{;r})\,\delta\theta\,\mcu{D}x\,dt,
\end{gather}
where we introduced
\begin{gather}\label{eq:auxjr}
j^r \doteq \mcu{J}^r - \mcu{J}^{rs}{}_{;s} + \ldots
\end{gather}
[Note that, if $\mcu{H}$ depends on derivatives of $\mcu{J}$ of unlimitedly high orders, then \Eq{eq:auxjr} is an infinite series, so the current $j^r$ is nonlocal.]

Since \Eq{eq:aux91} must be zero for any $\delta\theta$, the expression in brackets must be zero, so \Eq{eq:claw} is obtained. In particular, this generalizes the result of \Ref{ref:brizard93} to arbitrary linear nondissipative oscillations on an arbitrary manifold. It is also seen that \Eq{eq:claw} is caused by the fact that $\mcc{L}$, given by \Eq{eq:mcuL}, does not depend on $\theta$ explicitly. This means that $\mvec{j}$ can be understood as the spatial part of the Noether current associated with the phase invariance of~$\mcc{L}$.

\subsection{Dynamics on ${\boldsymbol{K \times K}}$}

The above equations, as well as any other equations derived in this paper in the $\mvec{x}$-representation, are also mirrored in the $\mvec{k}$-representation up to the transformation
\begin{gather}\label{eq:corres}
(\mvec{x}, -i\eth_{\mvec{x}})\, \leftrightarrow\, (i\eth_{\mvec{k}}, \mvec{k}).
\end{gather}
The only essential difference is that $\hat{H}$ may be a polynomial in $\eth_{\mvec{x}}$ but an infinite series in $\eth_{\mvec{k}}$ (or vice versa), so what appears as local dynamics in one representation may not be local in another one.

\subsection{Dynamics on phase space, ${\boldsymbol{X \times K}}$}
\label{sec:ppdyn}

\subsubsection{Phase-space image function}
\label{sec:psi}

Now let us consider
\begin{gather}\label{eq:projpipi}
\hat{\Pi} = \hat{\Pi}_{\mvec{k}}\hat{\Pi}_\mvec{x}.
\end{gather}
Here $\hat{\Pi}_\mvec{k} \doteq \ket{\mvec{k}}\bra{\mvec{k}}$ can be understood as the operator of projection on a state with a certain momentum $\mvec{k}$, and $\hat{\Pi}_\mvec{x} \doteq \ket{\mvec{x}}\bra{\mvec{x}}$ can be understood as the operator of projection on a state with a certain coordinate $\mvec{x}$. Thus, \textit{in a sense}, such $\hat{\Pi}$ represents a projector on the ``phase space'', $\msf{Z} \doteq X \times K$, comprised of eigenvalues $\msf{z} = (\mvec{x}, \mvec{k})$ of~$\hat{\msf{z}}$. But remember that, strictly speaking, projecting on a state with a certain $\msf{z}$ is impossible, as eigenstates of $\oper{x}$ and $\oper{k}$ are different; \ie the dynamics on $\msf{Z}$ is noncommutative.

The corresponding $\hat{\Pi}$-image of $\hat{\rho}$,
\begin{gather}\label{eq:pdmk}
\varrho(t, \mvec{x}, \mvec{k}) = \mbox{tr}\,(\hat{\Pi}_{\mvec{k}}\hat{\Pi}_\mvec{x}\hat{\rho}) = \braket{\mvec{k}|\mvec{x}}\braket{\mvec{x}|\hat{\rho}|\mvec{k}},
\end{gather}
will hence be attributed as a phase-space image function, or a PSI. (One may recognize this $\varrho$ as so-called Dirac's distribution; see, \eg \Ref{ref:bamber14}, references therein, and also \Refs{ref:roos60,ref:takabayasi54}.) This PSI happens to be the Fourier transform of $\rho(t, \mvec{x}, \mvec{x}')$, up to a phase factor, and has the following properties. First of all, it is easy to see, from \Eqs{eq:xunit} and \eq{eq:kunit}, that
\begin{gather}
\int \varrho(t, \mvec{x}, \mvec{k})\,\mcu{D}k = |\psi(t, \mvec{x})|^2, \label{eq:rhop1}\\ 
\int \varrho(t, \mvec{x}, \mvec{k})\,\mcu{D}x = |\psi(t, \mvec{k})|^2.
\end{gather}
Also, for any $\hat{\mc{F}} \doteq \int \ket{\mvec{x}} \mc{F}(t, \mvec{x}) \bra{\mvec{x}}\,\mcu{D}x$ and  $\hat{\mc{G}} \doteq \int \ket{\mvec{k}} \mc{G}(t, \mvec{k}) \bra{\mvec{k}}\,\mcu{D}k$, the following equality is satisfied:
\begin{gather}\nonumber
\braket{\psi|\hat{\mc{F}} + \hat{\mc{G}}|\psi} = \int [\mc{F}(t, \mvec{x}) + \mc{G}(t, \mvec{k})]\, \varrho(t, \mvec{x}, \mvec{k})\,\mcu{D}x\,\mcu{D}k,
\end{gather} 
and, in particular,
\begin{gather}
\int \varrho(t, \mvec{x}, \mvec{k})\,\mcu{D}x\,\mcu{D}k = \braket{\psi|\psi}.\label{eq:rhonorm}
\end{gather}
That said, it is also convenient (for reasons to become apparent shortly) to introduce a rescaled PSI,
\begin{gather}\label{eq:fdef}
f(t, \mvec{x}, \mvec{k}) \doteq \varrho(t, \mvec{x},\mvec{k})\,\sqrt{\gamma(\mvec{x})\,\vartheta(\mvec{k})},
\end{gather}
which satisfies
\begin{gather}
\int f(t, \mvec{x}, \mvec{k})\,\msf{d}x\,\msf{d}k = \braket{\psi|\psi}.
\end{gather}

\subsubsection{Kinetic equation}
\label{sec:gkineq}

Equation \eq{eq:pvNe} now takes the following form:
\begin{gather}\label{eq:aux89}
i \partial_t \varrho(t, \mvec{x}, \mvec{k}) + \mcc{F}_+(t, \mvec{x}, \mvec{k}) - \mcc{F}_-(t, \mvec{x}, \mvec{k}) = 0,
\end{gather}
where we introduced
\begin{gather}\label{eq:aux97}
\mcc{F}_+(t, \mvec{x}, \mvec{k}) \doteq \mbox{tr}\,(\hat{\Pi}_{\mvec{k}}\hat{\Pi}_\mvec{x}\hat{\rho}\hat{H}) = 
\braket{\mvec{k}|\mvec{x}}\braket{\mvec{x}|\hat{\rho}\hat{H}|\mvec{k}},
\end{gather}
and $\mcc{F}_-(t, \mvec{x}, \mvec{k})$ differs only by the order of $\hat{\rho}$ and $\hat{H}$. Equation \eq{eq:aux97} can be rewritten as
\begin{multline}
\mcc{F}_+ = \int
\braket{\mvec{x}|\hat{\rho}|\mvec{k} + \mvec{p}}
\braket{\mvec{k} + \mvec{p}|\mvec{x} + \mvec{q}} \\
\braket{\mvec{x} + \mvec{q}|\hat{H}|\mvec{k}}\braket{\mvec{k}|\mvec{x}}\,\mcu{D}q\,\mcu{D}p,
\end{multline}
where $\mcu{D}q = \sqrt{\gamma(\mvec{x} + \mvec{q})}\,\msf{d}q$, $\msf{d}q \equiv dq^1 \ldots dq^D$, $\mcu{D}p = \sqrt{\gamma(\mvec{k} + \mvec{p})}\,\msf{d}p$, and $\msf{d}p \equiv dp_1 \ldots dp_D$. This leads to
\begin{gather}\nonumber
\mcc{F}_+ = \int \varrho(t, \mvec{x},\mvec{k} + \mvec{p})\,\lambda(\mvec{p}, \mvec{q})\,
\mc{W}(t, \mvec{x} + \mvec{q}, \mvec{k})\,\mcu{D}q\,\mcu{D}p,
\end{gather}
where we introduced
\begin{gather}
\lambda(\mvec{p}, \mvec{q}) \doteq 
\frac{\braket{\mvec{k} + \mvec{p}|\mvec{x} + \mvec{q}}\braket{\mvec{k}|\mvec{x}}}{\braket{\mvec{k} + \mvec{p}|\mvec{x}}\braket{\mvec{k}|\mvec{x} + \mvec{q}}} = e^{-i\mvec{p}\cdot \mvec{q}}
\end{gather}
and the function $\mc{W}(t, \mvec{x}, \mvec{k}) \doteq \braket{\mvec{k}|\mvec{x}}\braket{\mvec{x}|\hat{H}|\mvec{k}}$, which is simply the $\hat{\Pi}$-image of $\hat{H}$. It is convenient to express $\mc{W}$ as
\begin{gather}
\mc{W}(t, \mvec{x}, \mvec{k}) = \frac{\mc{H}(t, \mvec{x}, \mvec{k})}{(2\pi)^{D} \sqrt{\gamma(\mvec{x})\,\vartheta(\mvec{k})}},
\end{gather}
where $\mc{H}$ is given by
\begin{gather}\label{eq:hhh}
\mc{H}(t, \mvec{x}, \mvec{k}) \doteq \braket{\mvec{x}|\hat{H}|\mvec{k}}/\braket{\mvec{x}|\mvec{k}}
\end{gather}
and is termed the effective Hamiltonian. Note that, for any $\hat{H}$ of the form $\hat{H}(t, \oper{x}, \oper{k}) = \mc{F}(t, \oper{x}) + \mc{G}(t, \oper{k})$, one has $\mc{H}(t, \mvec{x}, \mvec{k}) = H(t, \mvec{x}, \mvec{k})$. Another useful representation is
\begin{gather}\label{eq:Hcompl}
\mc{H}(t, \mvec{x}, \mvec{k}) = e^{-i\mvec{k}\cdot\mvec{x}}\,H(t, \mvec{x}, -i\eth_{\mvec{x}})\, e^{i\mvec{k}\cdot\mvec{x}},
\end{gather}
as obtained from \Eq{eq:Omxr}. For $k$ large compared to the spatial scale of $H$, one can replace $-i\eth_{\mvec{x}}$ here with $\mvec{k}$, so $\mc{H}(t, \mvec{x}, \mvec{k}) \approx H(t, \mvec{x}, \mvec{k})$ for any $\hat{H}$. This also means that $\mc{H}(t, \mvec{x}, \mvec{k})$ is real in that regime, since $\hat{H}$ is Hermitian.

Now the expression for $\mcc{F}_+$ becomes
\begin{multline}
 \sqrt{\gamma(\mvec{x})\vartheta(\mvec{k})}\, \mcc{F}_+(t, \mvec{x}, \mvec{k}) \\
  = \int f(t, \mvec{x},\mvec{k} + \mvec{p})
\mc{H}(t, \mvec{x} + \mvec{q}, \mvec{k})\,e^{-i\mvec{p}\cdot \mvec{q}}\,\frac{\msf{d} q \,\msf{d}p}{(2\pi)^D},\nonumber
\end{multline}
and $\mcc{F}_-(t, \mvec{x}, \mvec{k})$ is similar. Consider the following symbolic representation of the integrand, with arrows pointing to functions on which the gradients act [cf. \Eq{eq:psiexp}]:
\begin{multline}
f(t, \mvec{x},\mvec{k} + \mvec{p})\,\mc{H}(t, \mvec{x} + \mvec{q}, \mvec{k}) \\
= f(t, \mvec{x},\mvec{k})\,e^{\mvec{p} \cdot \overset{\leftarrow}{\nabla}_\mvec{k}}
e^{\mvec{q} \cdot \overset{\rightarrow}{\nabla}_\mvec{x}}\mc{H}(t, \mvec{x}, \mvec{k}).
\end{multline}
Since $\overset{\leftarrow}{\nabla}_\mvec{k}$ and $\overset{\rightarrow}{\nabla}_\mvec{x}$ apply to different functions, these operators commute, so we can write
\begin{gather}\nonumber
\mcc{F}_+(t, \mvec{x}, \mvec{k}) = [\gamma(\mvec{x})\vartheta(\mvec{k})]^{-1/2}\,f(t, \mvec{x},\mvec{k})\,\star\,\mc{H}(t, \mvec{x},\mvec{k}),
\end{gather}
where a binary operator $\star$ is defined as follows:
\begin{gather}
\star\, \doteq \int e^{
- i\mvec{p}\cdot\mvec{q}
+ \mvec{p} \cdot \overset{\leftarrow}{\nabla}_\mvec{k} 
+ \mvec{q} \cdot \overset{\rightarrow}{\nabla}_\mvec{x}
}\,\frac{\msf{d} q \,\msf{d}p}{(2\pi)^D}.
\end{gather}
When taking \textit{this} integral, one may treat $i\nabla_{\mvec{k}}$ and $-i\nabla_{\mvec{x}}$ as constants, so it is easy to see that
\begin{gather}\label{eq:stardef}
\star\, = e^{-i \overset{\leftarrow}{\nabla}_\mvec{k} \cdot \overset{\rightarrow}{\nabla}_\mvec{x}}.
\end{gather}
[Note that, because of the definition of the dot product, \Eq{eq:dotpr}, the star product is invariant with respect to how the metrics $\oper{\gamma}$ and $\oper{\vartheta}$ are chosen.] Let us also introduce a bracket associated with~$\star$,
\begin{gather}\label{eq:ourbracket}
i\moyal{f}{\mc{H}} \doteq f\,\star\,\mc{H} - \mc{H} \star f,
\end{gather}
which we will term the star bracket. Then, \Eq{eq:aux89} takes the following compact form:
\begin{gather}\label{eq:moyaleq1}
\partial_t f + \moyal{f}{\mc{H}} = 0.
\end{gather}

As an equation describing the action operator projection on phase space, \Eq{eq:moyaleq1} is called a (full) kinetic equation, or FKE. That said, \Eq{eq:moyaleq1} can also be viewed as merely the Fourier transform of \Eq{eq:rhoxpxpp}; \ie the two carry the same information. Note also that adopting $\hat{\Pi} = \hat{\Pi}_{\mvec{x}}\hat{\Pi}_\mvec{k}$ instead of $\hat{\Pi} = \hat{\Pi}_{\mvec{k}}\hat{\Pi}_\mvec{x}$ leads to the equation that is complex-conjugate to \Eq{eq:moyaleq1}.

\subsection{Other projectors. Wigner function}
\label{sec:wigner}

While using the projectors $\hat{\Pi}_{\mvec{k}}\hat{\Pi}_\mvec{x}$ and $\hat{\Pi}_{\mvec{x}}\hat{\Pi}_\mvec{k}$ offers advantages of simplicity and tractability, one may instead adopt other $\hat{\Pi}$ to generate $\varrho$ (and $f$) and yet arrive at similar equations. For instance, at $\gamma = \vartheta = 1$, \Eqs{eq:rhop1}-\eq{eq:rhonorm} hold as well for the Wigner function \cite{ref:wigner32,ref:case08,book:zachos},
\begin{gather}\nonumber
\varrho_W(t, \mvec{x}, \mvec{k}) \doteq \int \braket{\mvec{x} + \mvec{q}/2|\,\hat{\rho}\,|\mvec{x} - \mvec{q}/2}\,e^{-i\mvec{k}\cdot\mvec{q}}\,
\frac{\msf{d}x}{(2\pi)^{D}},
\end{gather} 
which is a real (but not sign-definite) function, unlike a general PSI, and is generated by $\hat{\Pi} = \hat{\Pi}_W$, where
\begin{gather}
\hat{\Pi}_W \doteq \int 
\hat{T}_{-\mvec{q}/2}\,\hat{\Pi}_\mvec{x}\,\hat{T}_{\mvec{q}/2}\,\hat{\Pi}_\mvec{k}\,\hat{T}_{\mvec{q}/2}\,
\hat{\Pi}_\mvec{x}\,\hat{T}_{-\mvec{q}/2}
\,\msf{d} x.
\end{gather} 
(For an alternative representation of $\hat{\Pi}_W$ see \Ref{ref:tracy93}.) The corresponding dynamic equation has the same form as \Eq{eq:moyaleq1}. Specifically,
\begin{gather}\label{eq:wigmoy}
\partial_t f_W + \moyal{f_W}{\mc{H}_W} = 0,
\end{gather}
where $f_W \doteq \varrho_W$, $\mc{H}_W \doteq \mbox{tr}\,(\hat{H}\hat{\Pi}_W)$, and the new star product, also known as the Moyal product \cite{ref:moyal49}, is
\begin{gather}
\star_W = \exp\left(\frac{i}{2}\big[
\overset{\leftarrow}{\nabla}_\mvec{x} \cdot \overset{\rightarrow}{\nabla}_\mvec{k}
-\overset{\leftarrow}{\nabla}_\mvec{k} \cdot \overset{\rightarrow}{\nabla}_\mvec{x}
\big]\right).
\end{gather} 

Other $\hat{\Pi}$-images and star products are also possible and, with enough effort, can always be attributed with \textit{some} physical meaning; see, \eg \Ref{ref:ozorio98}. (For an overview of some of such functions, see also \Ref{ref:lee95}.) In particular, based on the similarity with the classical probability distribution, the Wigner function and similar PSI are often identified as ``quasiprobabilities'', or ``quantum probability distributions'' (see also \Sec{sec:ll}). However, such dramatic interpretations are superfluous, as there is no definition of quasiprobability other than $\varrho$; \ie, if we want to define the term ``quasiprobability'', we can allow it to have \textit{any} properties, and being real does not have to be one of them. Each PSI is just $I$ times the expectation value of a certain projector, $\hat{\Pi}$ [\Eq{eq:psidef}], which can be used to calculate moments of the action operator (\Sec{sec:psi}). Changing the projector merely changes an action-operator representation, \ie results in variable transformation in the FKE. In this sense, the only fundamental FKE is the operator equation for $\hat{\rho}$ itself [\Eq{eq:vNn}], whereas any of its projections is inherently artificial and, as such, does not necessarily have a natural physical meaning.

\section{Basic waves}
\label{sec:waves}

Now that we have described general oscillations on a manifold, let us consider a special class of such oscillations, namely, wave processes.

\subsection{Definitions}
\label{sec:wavesdef}

There may exist a choice of $\oper{x}$ and $\oper{k}$ such that the phase space is foliated, yielding $\hat{H} = H(\oper{k})$. Let us consider oscillations in such a system (which can be called integrable in the classical-mechanics sense), namely,
\begin{gather}\label{eq:weq}
i\ket{\dot{\psi}} = H(\oper{k}) \ket{\psi},
\end{gather}
assuming now that $H$ is time-independent. We will call \Eq{eq:weq} homogeneous stationary wave equation, and its solutions of the form $\ket{\psi} = \ket{\mvec{k}} e^{-iH(\mvec{k}) t}$ will be called homogeneous stationary waves.

More generally, the solution of \Eq{eq:weq} can be sought as follows. Suppose that $\ket{\psi}$ is locally close to a monochromatic wave with some wave vector $\bar{\mvec{k}}$ (\ie the spectrum width $\Delta k$ satisfies $\Delta k \ll \bar{k}$) and frequency $\bar{\omega} \doteq H(\bar{\mvec{k}})$ yet has a slowly varying envelope $\ket{\phi}$; namely, 
\begin{gather}
\ket{\psi} = \exp(- i\bar{\omega} t + i \bar{\mvec{k}} \cdot \oper{x}) \ket{\phi}.
\end{gather}
Hence one gets $\ket{\dot{\phi}} = \hat{\mc{U}} \ket{\phi}$, where
\begin{gather}\label{eq:aux111}
\hat{\mc{U}} \doteq e^{-i \bar{\mvec{k}} \cdot \oper{x}}H(\oper{k})\,e^{i \bar{\mvec{k}} \cdot \oper{x}} - \bar{\omega}.
\end{gather}
Using that $e^{i \bar{\mvec{k}} \cdot \oper{x}} = \hat{T}_{\bar{\mvec{k}}}$ is an operator of translation in the momentum space (\Sec{app:transl}), the effect of the first term in \Eq{eq:aux111} on any $\ket{\mvec{k}}$ can be expressed as follows:
\begin{align}
\hat{T}_{-\bar{\mvec{k}}} H(\oper{k}) \hat{T}_{\bar{\mvec{k}}} \ket{\mvec{k}} 
& = \Gamma_\vartheta (\mvec{k}, \bar{\mvec{k}}) \hat{T}_{-\bar{\mvec{k}}} H(\oper{k}) \ket{\mvec{k} + \bar{\mvec{k}}} \nonumber \\
& = \Gamma_\vartheta (\mvec{k}, \bar{\mvec{k}}) H(\mvec{k} + \bar{\mvec{k}}) \hat{T}_{-\bar{\mvec{k}}} \ket{\mvec{k} + \bar{\mvec{k}}} \nonumber \\
& = H(\mvec{k} + \bar{\mvec{k}}) \ket{\mvec{k}}.
\end{align}
On the other hand, $H(\mvec{k} + \bar{\mvec{k}}) \ket{\mvec{k}} = H(\oper{k} + \bar{\mvec{k}}) \ket{\mvec{k}}$, so
\begin{gather}
\hat{\mc{U}} = H(\oper{k} + \bar{\mvec{k}}) - \bar{\omega}.
\end{gather}
In the $\mvec{x}$-representation, $\oper{k}$ becomes a gradient, which, for an envelope, is small. Hence, we can take
\begin{gather}
H(\oper{k} + \bar{\mvec{k}}) \approx \bar{\omega} + \frac{\partial H(\bar{\mvec{k}})}{\partial \bar{k}_r}\,\hat{k}_r
+ \frac{1}{2}\,\frac{\partial H(\bar{\mvec{k}})}{\partial \bar{k}_r\, \partial \bar{k}_s}\,\hat{k}_r \hat{k}_s.
\end{gather}
We now substitute the dispersion relation and introduce
\begin{gather}
\mvec{v}_{\rm g}(\mvec{k}) \doteq \nabla_\mvec{k}H(\mvec{k}),
\end{gather}
called the group velocity, and $\bar{\mvec{v}}_{\rm g} \doteq \mvec{v}_{\rm g}(\bar{\mvec{k}})$. (That being said, $\mvec{v}_{\rm g}$ does not necessarily have units of velocity, as $\mvec{k}$ is a \textit{generalized} momentum.) We also introduce a matrix
\begin{gather}\label{eq:Ups}
\Upsilon^{rs} \doteq \frac{1}{2}\,\frac{\partial \bar{v}_{\rm g}^r}{\partial \bar{k}_s}
= \frac{1}{2}\,\frac{\partial H(\bar{\mvec{k}})}{\partial \bar{k}_r\, \partial \bar{k}_s},
\end{gather}
which determines a real symmetric constant spatial tensor $\oper{\Upsilon} \equiv \bar{\mvec{v}}'_{\rm g}/2$. The envelope equation then becomes
\begin{gather}\label{eq:aux71}
i\ket{\dot{\phi}} = (\bar{\mvec{v}}_{\rm g} \cdot \oper{k} + {\textstyle \frac{1}{2}}\,\oper{k} \cdot \bar{\mvec{v}}'_{\rm g} \cdot \oper{k}) \ket{\phi}.
\end{gather}
The $\mvec{x}$-representation of \Eq{eq:aux71}, say, at $\gamma = \mbox{const}$, is
\begin{gather}\label{eq:lsevv}
i(\partial_t  \phi + \bar{\mvec{v}}_{\rm g} \cdot \nabla \phi) + {\textstyle \frac{1}{2}}\,\nabla \cdot \bar{\mvec{v}}'_{\rm g} \cdot \nabla \phi = 0,
\end{gather}
where $\nabla \cdot \bar{\mvec{v}}'_{\rm g} \cdot \nabla/2$ can be understood as the Laplace-Beltrami operator in space with metric~$\oper{\Upsilon}$. The former two terms here describe ballistic propagation of an envelope with velocity $\bar{\mvec{v}}_{\rm g}$, and the latter term describes dispersive spreading. Equation \eq{eq:lsevv} is known as the linear quasioptical equation, or the LSE in the narrow sense [as opposed to \Eq{eq:bLSE}, which is the generalized LSE]. However, in contrast to the traditional derivation employing the Fourier transform \cite[Sec.~27]{book:karpman}, now \Eq{eq:lsevv} flows from \Eq{eq:aux71} that originates from purely geometric, representation-independent arguments.

\subsection{Liouville limit. Wave kinetic equation}
\label{sec:ll}

Let us now allow the Hamiltonian to vary with $\mvec{x}$ on some scale $\ell$ large compared to $\bar{k}^{-1}$, so $\mc{H}(t, \mvec{x}, \mvec{k})$ is real (\Sec{sec:gkineq}). We will yet require 
\begin{gather}\label{eq:gocond}
\ell^{-1} \ll \Delta k \ll \bar{k},
\end{gather}
so \Eq{eq:moyaleq1} can be simplified. Specifically, notice that 
\begin{gather}
f\, (-i \overset{\leftarrow}{\nabla}_\mvec{k} \cdot \overset{\rightarrow}{\nabla}_\mvec{x}) \mc{H} \sim \mc{H}f/(\Delta k \ell),\\
\mc{H}\, (-i \overset{\leftarrow}{\nabla}_\mvec{k} \cdot \overset{\rightarrow}{\nabla}_\mvec{x}) f \sim\mc{H}f\, (\Delta k/\bar{k}),
\end{gather}
so the star products in \Eq{eq:ourbracket} can be treated as asymptotic series in $(\Delta k \ell)^{-1}$ and $(\Delta k/\bar{k})$. (It is assumed here that $|\nabla_\mvec{k}\mc{H}| \sim \mc{H}/k_0$, but keep in mind that this scaling is not universal and is adopted only for specificity.) We will retain only the first two terms in each case; \ie
\begin{gather}
f\, \star \mc{H} \approx 1 - i (\nabla_\mvec{k} f) \cdot (\nabla_\mvec{x} \mc{H}), \label{eq:961}\\
\mc{H}\, \star f \approx 1 - i (\nabla_\mvec{k} \mc{H}) \cdot (\nabla_\mvec{x} f), \label{eq:962}
\end{gather}
so the bracket \eq{eq:ourbracket} becomes the Poisson bracket, 
\begin{gather}
\{f, \mc{H}\} \doteq ({\nabla}_\mvec{x} f) \cdot ({\nabla}_\mvec{k}\mc{H}) - ({\nabla}_\mvec{k} f) \cdot ({\nabla}_\mvec{x} \mc{H}).
\end{gather}
This leads to the Liouville equation
\begin{gather}\label{eq:pois2}
\partial_t f + \{f, \mc{H}\} = 0,
\end{gather}
which is more explicitly written as
\begin{gather}\label{eq:liouv}
\partial_t f + (\nabla_\mvec{k} \mc{H}) \cdot (\nabla_\mvec{x} f)  - (\nabla_\mvec{x} \mc{H}) \cdot (\nabla_\mvec{k} f) = 0.
\end{gather}
[See \App{sec:pspgeom} for additional details. Also note that \Eq{eq:wigmoy} for $f_W$ leads to similar results.] Characteristics, or ``rays'', of these equations are given by
\begin{gather}\label{eq:char}
\dot{x}^r = \partial_{k_r} \mc{H}, \quad \dot{k}_r = - \partial_{x^r} \mc{H}.
\end{gather}
We will call this regime classical, or the LL, and attribute \Eq{eq:pois2} as the WKE, as opposed to the FKE \eq{eq:moyaleq1}. Note that the LL must not be confused with the GO limit (\Sec{sec:go}), as commonly done in literature. Also notably, \Eqs{eq:pois2}-\eq{eq:char} are invariant with respect to variable transformation $(\mc{H}, \mvec{k}) \to (\msf{c}\mc{H}, \msf{c}\mvec{k})$, where $\msf{c}$ is an arbitrary constant. (It is this property that renders classical mechanics independent of the specific value of $\hbar$.)

The traditional interpretation of \Eqs{eq:char} is that they govern trajectories of individual quanta constituting a wave, so $f$ is interpreted as the phase space density of those quanta. However, in general, this picture is misleading, and the fact that $f$ can be complex serves as a hint. (This issue is blurred when the real $f_W$ is used for $f$.) Unless a wave consists of small isolated particle-like envelopes\footnote{Those would travel along the characteristics \eq{eq:char}, and each of such ``particles'', occupying a phase volume $\Delta X_i\,\Delta K_i$, would conserve its own action $I_i \doteq \rho(\mvec{x}_i, \mvec{k}_i)\,\Delta X_i\,\Delta K_i$, which is positive-definite except for negative-energy waves. Then the density of quanta is unambiguously defined as $\sum_i I_i$ per unit phase-space volume, \ie as a coarse-grained $f$.}, wave quanta cannot be attributed with definite coordinates $\msf{z}$, so their local phase space density cannot be defined in principle, not to mention connected with $f$. Equations \eq{eq:char} should then be understood as a ``statistical'' effect, rather than ``single-particle'' effect, as also will become apparent in \Sec{sec:go}. Accordingly, $f$ has no fundamental meaning in general (and neither does $f_W$); it is merely a tool for calculating moments of the action operator. See also \Sec{sec:wigner}.

\subsection{Hydrodynamics of waves}
\label{sec:go}

Let us now consider a more general case, allowing
\begin{gather}
\ell^{-1} \sim \Delta k \ll \bar{k}.
\end{gather}
Although \Eq{eq:962} still holds, \Eq{eq:961} now may be invalid. (This fact goes unnoticed when one formally expands the star bracket in a single small parameter; \eg similar QM calculations often incorrectly adopt $\hbar$ as such.) Then \Eq{eq:moyaleq1} becomes intractable as is, but it yields tractable equations for \textit{moments} of $f$, which serve as ``fluid'' equations. Those are derived as follows.

\subsubsection{Basic equations}
\label{sec:hydrobasic}

First of all, consider \Eq{eq:moyaleq1} in the form
\begin{gather}\label{eq:vrR}
\partial_t f + \partial_{x^r} (f\,\partial_{k_r} \mc{H}) - \partial_{k_r} (f\,\partial_{x^r} \mc{H}) + \mcc{G}  = 0,
\end{gather}
where we introduced 
\begin{gather}
\mcc{G} \doteq - if \, (e^{-i \overset{\leftarrow}{\nabla}_\mvec{k} \cdot \overset{\rightarrow}{\nabla}_\mvec{x}} - 1) \mc{H} +
(\nabla_\mvec{x} \mc{H}) \cdot (\nabla_\mvec{k} f),
\end{gather}
or, more explicitly,
\begin{gather}
\mcc{G} = (i/2)(\partial^2_{k_r}f)\,\partial^2_{x^r}\mc{H}
+ (1/6)(\partial^3_{k_r}f)(\partial^3_{x^r}\mc{H})
+ \ldots
\end{gather}
Equation \eq{eq:vrR} is not of a hyperbolic type, so, unlike in the LL, phase space rays cannot be defined now. But consider integrating \Eq{eq:vrR} over $\msf{d} k$ (not to be confused with $\mcu{D}k$), using that $\nu \doteq (\bar{k}\ell)^{-1} \ll 1$. The integral of $\mcc{G}$, taken by parts, is estimated as follows:
\begin{gather}\nonumber
\int \mcc{G}\,\msf{d}k = \int [(i/2)\,(\partial^2_{x^r}\partial^2_{k_r}\mc{H}) + \ldots]\,f\,\msf{d}k = \mc{O}(\nu^{-2}),
\end{gather}
so it is negligible, even though $\mcc{G}$ itself is not small; somewhat following \Ref{ref:berry77}, we call such $\mcc{G}$ a ghost term (see also \Sec{sec:ghost}). Hence one obtains
\begin{gather}
\partial_t\int f\,\msf{d} k + \partial_{x^r} \int (\partial_{k_r} \mc{H})\, f\,\msf{d} k  = 0.
\end{gather}
Substituting \Eq{eq:fdef} then yields
\begin{gather}\label{eq:actij}
\partial_t \mc{I} + \nabla \cdot \mvec{\mc{J}} = 0,
\end{gather}
where we used \Eq{eq:div} for the divergence and introduced
\begin{gather}
\mc{I} \doteq \int \varrho\,\mcu{D}k, \\
\mvec{\mc{J}} \doteq \int (\nabla_{\mvec{k}}\mc{H})\,\varrho\,\mcu{D}k.
\end{gather}
According to \Eq{eq:rhop1}, one has $\mc{I} = |\psi(t, \mvec{x})|^2$, so $\mc{I}$ is simply the action spatial density, and thus $\mvec{\mc{J}}$ is the action flux density. Equation \eq{eq:actij} therefore represents the ACT in its spatial representation, analogous to \Eq{eq:claw}. (Here, we use the notation $\mc{I}$ instead of $A^2$ and $\mvec{\mc{J}}$ instead of $\mvec{j}$ only to emphasize that we no longer consider the general case but rather limit our consideration to small $\nu$.) 

One can similarly introduce higher moments of \Eq{eq:vrR}, just like it is done for gases and plasmas in arbitrary metric \cite{my:metric}. For instance, multiplying \Eq{eq:vrR} by $k_s$ and integrating it over $\msf{d}k$ yields
\begin{gather}\label{eq:aux99}
\partial_t \int k_s f\,\msf{d} k 
+ \partial_{x^r} \int (\partial_{k_r} \mc{H})\, k_s f\,\msf{d} k 
+ \int (\partial_{x^r}\mc{H})\, f\,\msf{d} k = 0,
\end{gather}
as the integral over $\mcc{G}$ is, again, negligible. The physical meaning of this equation will now be explained.

\subsubsection{Geometrical-optics limit. Whitham's equations}
\label{sec:golimit}

Let us consider the so-called GO regime, when $\Delta k$ is negligibly small, so one can use an approximation
\begin{gather}
f(t, \mvec{x}, \mvec{k}) \approx \sqrt{\gamma(\mvec{x})}\,\mc{I}(t, \mvec{x})\,\delta(\mvec{k} - \bar{\mvec{k}}(t, \mvec{x})),
\end{gather}
termed the cold limit. Then, \Eqs{eq:actij} and \eq{eq:aux99} become
\begin{gather}
\partial_t \mc{I} + \nabla \cdot (\mc{I}\,\bar{\nabla}_{\mvec{k}} \mc{H}) = 0,\label{eq:aux101} \\
\partial_t (\bar{k}_s \mc{I}) + \nabla \cdot (\bar{k}_s \mc{I}\,\bar{\nabla}_{\mvec{k}} \mc{H}) + \mc{I}\, \bar{\partial}_{x^s} \mc{H} = 0,\label{eq:aux102}
\end{gather}
where bars above partial derivatives denote that those derivatives are evaluated at $\mvec{k} = \bar{\mvec{k}}(t, \mvec{x})$. (One may notice similarity of these equations with the standard continuity and momentum equations for ideal cold fluids \cite{book:landau6}.) By combining the two, one can also reduce \Eq{eq:aux102} down to
\begin{gather}
\partial_t \bar{k}_s + (\bar{\nabla}_{\mvec{k}} \mc{H} \cdot \nabla) \bar{k}_s + \bar{\partial}_{x^s} \mc{H} = 0.\label{eq:aux103}
\end{gather}
The characteristics of these equations are then seen to be
\begin{gather}\label{eq:charf}
d_t \bar{x}^r = \bar{\partial}_{k_r} \mc{H}, \quad d_t \bar{k}_r = - \bar{\partial}_{x^r} \mc{H},
\end{gather}
so they are akin to \Eqs{eq:char}. But remember that, unlike the latter, \Eqs{eq:charf} allow only one wavevector at a given spatial location. Thus, notwithstanding the similarity of the ray equations, the GO limit and the LL are not the same. Contrary to a common presumption (\eg Ref.~\cite[Sec.~4.8]{book:stix}), the Liouville equation generally does \textit{not} hold for GO waves, because those can have $\Delta k$ as small as $\ell^{-1}$, which invalidates \Eq{eq:961}. 

Notice now that, in the GO regime, $\psi \equiv Ae^{i\theta}$ is such that the eikonal, $\theta$, is rapid compared to $A$.~Then, 
\begin{gather}\label{eq:ktheta}
\bar{\mvec{k}} \approx \nabla_{\mvec{x}} \theta,
\end{gather}
so $\bar{\mvec{k}}(t, \mvec{x})$ is curl-free; \ie $\partial_{x^r} \bar{k}_s = \partial_{x^s} \bar{k}_r$. This permits combining the last two terms in \Eq{eq:aux103}~as
\begin{gather}\nonumber
(\bar{\nabla}_{\mvec{k}} \mc{H} \cdot \nabla) \bar{k}_s + \bar{\partial}_{x^s} \mc{H} 
= (\bar{\partial}_{k_r} \mc{H})\, \partial_{x^s} \bar{k}_r + \bar{\partial}_{x^s} \mc{H}
= \partial_{x^s} \bar{\omega},
\end{gather}
where $\bar{\omega}(t, \mvec{x}) \doteq \mc{H}(t, \mvec{x}, \bar{\mvec{k}}(t, \mvec{x}))$. Hence, \Eq{eq:aux103} can be written simply as
\begin{gather}\label{eq:consist2}
\partial_t \bar{\mvec{k}} + \nabla \bar{\omega} = 0.
\end{gather}
Finally, one also has, similarly to \Eq{eq:ktheta}, that
\begin{gather}\label{eq:wtheta}
\bar{\omega} \approx - \partial_t \theta.
\end{gather}
Equation \eq{eq:wtheta} can be used in combination with \Eq{eq:ktheta} for an alternative derivation of \Eq{eq:consist2}.

Let us now introduce $\bar{\mvec{v}}_{\rm g} \doteq \bar{\nabla}_{\mvec{k}} \mc{H}$ by analogy with \Sec{sec:wavesdef} but also drop the bars for clarity. Then, the GO equations derived above are summarized as follows:
\begin{gather}
\partial_t \theta + \omega = 0,\label{eq:aux1005}\\
\partial_t \mc{I} + \nabla \cdot (\mc{I}\mvec{v}_{\rm g}) = 0, \label{eq:aux105}\\
\partial_t \mvec{k} + \nabla \omega = 0, \quad \nabla \times \mvec{k} = 0. \label{eq:aux106}
\end{gather}
Note that \Eqs{eq:aux1005}-\eq{eq:aux106} match the well-known Whitham's equations \cite[Sec.~11.7]{book:whitham}, which are commonly derived directly from the GO variational principle [except \Eqs{eq:aux106}; those stem merely from the definitions of $\omega$ and $\mvec{k}$]. Indeed, in the GO limit, \Eq{eq:mcuL} turns into
\begin{gather}\label{eq:aux201}
\mcc{L} = [(\underbrace{-\partial_t \theta}_{\omega}) - \mc{H}(t, \mvec{x}, \underbrace{\nabla \theta}_{\mvec{k}})]\mc{I}.
\end{gather}
Then the requirement $\delta_\theta \mc{S} = 0$ readily yields \Eq{eq:aux105}, and $\delta_\mc{I} \mc{S} = 0$ immediately leads to \Eq{eq:aux1005}, which also can be interpreted as the local dispersion relation, 
\begin{gather}\label{eq:aux501}
 \omega = \mc{H}(t, \mvec{x}, \mvec{k}).
\end{gather}
For detailed derivations and a comprehensive discussion regarding the general properties of GO waves that flow from the approximation $\mcc{L} = \mcc{L}(\mcu{A}, \omega, \mvec{k}; t, \mvec{x})$, where $\mcu{A}$ is an arbitrarily defined wave amplitude, see \Ref{my:amc}. The mentioned paper also discusses $\mcc{L}$ specific to linear GO electromagnetic waves, as well as the corresponding definitions of the wave action, energy-momentum tensor, and angular momentum. For $\mcc{L}$ describing \textit{nonlinear} electromagnetic waves in collisionless plasma, see \Refs{my:itervar,my:bgk,my:acti,my:actii,my:actiii}.

\subsubsection{Particle-like dynamics of waves}
\label{sec:wp}

As a side remark, let us note that \Eqs{eq:aux1005} and \eq{eq:aux501} can be cast as well in the form
\begin{gather}\label{eq:HJ}
\partial_t \theta + \mc{H}(t, \mvec{x}, \nabla \theta) = 0.
\end{gather}
Even though we allow GO waves to have arbitrarily large spatial extent, \Eq{eq:HJ} is \textit{exactly} the same as the commonly known Hamilton-Jacobi equation for point particles. This implies that methods of particle manipulations readily extend to GO waves ``as is''. For example, one can develop a ``photon-oscillation-center'' formalism to describe wave propagation in media whose parameters vary quasiperiodically in space and (or) time, such as in photonic crystals \cite{book:joannopoulos}; see a companion paper, \Ref{tex:mylens}. One can also apply \Eq{eq:HJ} to describe, \eg photon acceleration \cite{book:mendonca}, photon Landau damping \cite{ref:bingham97}, the all-optical bump-on-tail instability \cite{ref:dylov08}, and wave trapping in autoresonance \cite{ref:shalibo12,tex:murch10}. Details of this approach will be discussed in separate publications.

Furthermore, one can apply the GO formalism presented here to ``actual'' particles [the cause being that particles \textit{are} waves on the quantum level, and \Eq{eq:aux201} is equivalent to the classical limit of the quantum-particle Lagrangian density]. Like waves, particles need not be localized for \Eq{eq:HJ} to hold. If, however, the action density \textit{is} localized, \ie
\begin{gather}\label{eq:ppart}
\mc{I} = \sum_{j = 1}^{\mcc{N}} \delta(\mvec{x}, \bar{\mvec{x}}_j(t)),
\end{gather}
then the \textit{full} Lagrangian, \Eq{eq:integrL}, becomes
\begin{gather}\label{eq:clL}
L = \sum_{j = 1}^{\mcc{N}} \big[\bar{\mvec{k}}_j \cdot \dot{\bar{\mvec{x}}}_j - \mc{H}(t, \bar{\mvec{x}}_j, \bar{\mvec{k}}_j)],
\end{gather}
where $\bar{\mvec{k}}_j \doteq \mvec{k}(t, \bar{\mvec{x}}_j)$. One may recognize this as the standard Lagrangian that describes the (generally nonlinear) dynamics of $\mcc{N}$ identical classical particles. Note also that, for \Eq{eq:clL} to hold, we assumed each particle to have exactly unit action [$\int \delta(\mvec{x}, \bar{\mvec{x}}_j(t)) \,\mcu{D}x = 1$].

It is seen then that classical mechanics can be viewed as a special case of linear GO. This agrees with \Sec{sec:ll}, where we already derived the classical Liouville equation for waves in the appropriate limit. On the other hand, classical mechanics is not identical to linear GO, for the latter can as well describe extended waves and, as such, should rather be identified with \textit{quasi}classical mechanics.

\subsection{Statistical kinetic equation}

For completeness, let us also introduce a statistical description of waves. Assuming smooth $\hat{H}$, it is often convenient to perform ensemble averaging to derive the evolution of $\msf{W}(t, \mvec{x}, \mvec{k}) \doteq \favr{f(t, \mvec{x}, \mvec{k})}$, or
\begin{multline}\nonumber
\msf{W}(t, \mvec{x}, \mvec{k}) 
= \braket{\mvec{k}|\mvec{x}}\braket{\mvec{x}|\favr{\hat{\rho}}|\mvec{k}} \\
= (2\pi)^{-D} \int e^{i\mvec{x} \cdot(\mvec{k}' - \mvec{k})} \favr{\psi_{\mvec{k}'}(t, \mvec{x})  \psi^*_{\mvec{k}}(t, \mvec{x})}\, \msf{d}k',
\end{multline}
where we adopted $\gamma = 1$ and $\vartheta = 1$ for simplicity. This gives $\partial_t \msf{W} + \moyal{\msf{W}}{\mc{H}} = 0$, where $\mc{H}$ is connected to $\hat{H}$ via \Eq{eq:hhh}. Assuming that $\msf{W}$ is smooth in both $\mvec{x}$ and $\mvec{k}$ (but $f$ itself now does not need to be such; otherwise see \Sec{sec:ll}), the corresponding LL is obtained. Then \Eq{eq:moyaleq1} leads to a WKE-like equation for $\msf{W}$,
\begin{gather}\label{eq:msfWst}
\partial_t \msf{W} + \{\msf{W}, \mc{H}\} = 0,
\end{gather}
where $\mc{H}(t, \mvec{x}, \mvec{k}) = H(t, \mvec{x}, \mvec{k})$. However, note that \Eq{eq:msfWst} is fundamentally different from the original WKE [\Eq{eq:pois2}] as it describes only statistical properties of a field rather than its precise instantaneous image. We thus term \Eq{eq:msfWst} SKE.

The utility of the SKE relies on ergodicity, so it is limited compared to that of the FKE. This is seen already from the fact that \Eq{eq:msfWst} effectively treats waves as classical particles, \ie neglects their interference and therefore misses certain qualitative effects that can be important. For instance, the SKE can predict ray chaos (\eg see \Ref{ref:bonoli82,ref:kupfer93}), whereas the FKE \eq{eq:moyaleq1} is quantumlike, so it does not allow chaos \textit{per~se}. (See also \Ref{ref:mcdonald88b} for waves in classically-chaotic stadiums.) This becomes particularly an issue at wavelengths that are not-too-small compared to the system size and at $t \to \infty$, as has been long recognized for truly quantum systems \cite{ref:chirikov88}.

\subsection{Sample applications}
\label{sec:applic}

In practice, $\mcc{L}$ may not be known in advance, and one has to start out with a PDE instead. For waves of the basic type that we have been discussing so far, such equations have a generic form
\begin{gather}\label{eq:aux202}
\partial_t \varphi + \hat{M}\varphi = 0,
\end{gather}
where $\varphi$ is some field on a given manifold, which we attribute as $X$, and $\hat{M} = {M}(t, \oper{x}, \oper{k})$ is some linear operator. It is generally nontrivial to identify the corresponding FKE and WKE in an \textit{ad~hoc} manner, which sometimes leads to controversies \cite{ref:krommes00}. However, these problems do not emerge if one starts out with finding a variable transformation $\varphi = \hat{U}\psi$ that turns \Eq{eq:aux202} into the LSE with Hermitian $\hat{H} = H(t, \oper{x}, \oper{k})$; specifically,
\begin{gather}\label{eq:mcM}
i \partial_t \psi = \hat{H}\psi, \quad \hat{H} \doteq -i\hat{U}^{-1}(\partial_t \hat{U} + \hat{M}\hat{U}).
\end{gather}
Then, the corresponding $\mcc{L}$ is given by \Eq{eq:Lred2}, and all the above machinery applies immediately, so one does not need to rederive the SKE for each new $\hat{M}$ separately. To illustrate this, we will now discuss applications of our theory to sample equations, specifically, linear Hasegawa-Mima equations that are often used to model the drift-wave turbulence in inhomogeneous magnetized plasmas \cite{ref:smolyakov99,ref:krommes00}. 

\subsubsection{Pure Hasegawa-Mima model} The ``pure'' linear Hasegawa-Mima model is
\begin{gather}\label{eq:aux203}
\partial_t (1 - \nabla^2) \varphi + (\mvec{V}_* \cdot \nabla) \varphi + (\mvec{V}_E \cdot \nabla) (- \nabla^2)\varphi = 0,
\end{gather}
where $\mvec{V}_*$ is a constant vector, and $\mvec{V}_E = \mvec{V}_E(t, \mvec{x})$ is a prescribed field such that ${\nabla \cdot \mvec{V}_E = 0}$. One can cast \Eq{eq:aux203} into the form \eq{eq:aux202} by taking $\hat{M} = (1 - \nabla^2)^{-1}[(\mvec{V}_* \cdot \nabla) + (\mvec{V}_E \cdot  \nabla) (- \nabla^2)]$, or $-i\hat{M} = (1 + \oper{k}^2)^{-1}[\mvec{V}_* \cdot \oper{k} + (\mvec{V}_E \cdot \oper{k}) \oper{k}^2]$. Then let us adopt $\hat{U}^{-1} = \oper{k}(1 +\oper{k}^2)^{1/2}$, so \Eq{eq:mcM} is yielded with
\begin{gather}
\hat{H} = \underbrace{\frac{\mvec{V}_* \cdot \oper{k}}{1 + \oper{k}^2}}_{\hat{H}_*} + 
 \underbrace{\frac{\oper{k}}{\sqrt{1 + \oper{k}^2}}}_{\hat{\varkappa}}\,
 \underbrace{\big[\mvec{V}_E(t, \oper{x}) \cdot \oper{k}\big] \vphantom{\frac{\oper{k}}{\sqrt{1 + \oper{k}^2}}} }_{\hat{\Delta}}\,
 \underbrace{\frac{\oper{k}}{\sqrt{1 + \oper{k}^2}}}_{\hat{\varkappa}}.
\end{gather}
The first term here, $\hat{H}_*$, is manifestly Hermitian, as $\oper{k}$ is Hermitian, and $\mvec{V}_*$ is constant. The second term, $\hat{H}_E = \hat{\varkappa}\hat{\Delta}\hat{\varkappa}$, satisfies $\hat{H}_E^\dag = \hat{\varkappa}^\dag \hat{\Delta}^\dag \hat{\varkappa}^\dag$. But $\hat{\varkappa}$ is Hermitian as a superposition of commuting Hermitian operators, and so is $\hat{\Delta}$, because $\mvec{V}_E$ is divergence-free; thus, $\hat{H}_E^\dag = \hat{H}_E$. In other words, $\hat{H}$ has the sought Hermitian form, and hence one knows the FKE and WKE explicitly.

In particular, smooth $\hat{H}$ corresponds~to 
\begin{gather}
\mc{H}(t, \mvec{x}, \mvec{k}) = \frac{\mvec{k} \cdot \mvec{V}_*}{1 + k^2} + \frac{k^2[\mvec{k} \cdot \mvec{V}_E(t, \mvec{x})]}{1 + k^2}.
\end{gather}
This expression for $\mc{H}$ can certainly could have been inferred directly from \Eq{eq:aux203}, but now we also know the appropriate $\hat{U}$ that explicitly defines $\msf{W}$ entering \Eq{eq:msfWst}. In particular, if $\varphi_{\mvec{k}'}(t, \mvec{x})$ is correlated with $\varphi^*_{\mvec{k}}(t, \mvec{x})$ only at $\mvec{k}'$ close to $\mvec{k}$, one has
\begin{gather}\label{eq:aux206}
\msf{W} (t, \mvec{x}, \mvec{k}) \approx I_\mvec{k}(t, \mvec{x})\,U^{-2}(\mvec{k}),
\end{gather}
where we employed $\hat{U} = U(\oper{k})$, and
\begin{gather}\nonumber
I_\mvec{k} (t, \mvec{x}) 
\doteq (2\pi)^{-D} \int e^{i\mvec{x} \cdot(\mvec{k}' - \mvec{k})}\favr{\varphi_{\mvec{k}'}(t, \mvec{x}) \,\varphi^*_{\mvec{k}}(t, \mvec{x})}\, \msf{d}k'.
\end{gather}
(Our definition of $I_\mvec{k}$ is similar to that \Ref{ref:smolyakov99}, but, contrary to \Ref{ref:smolyakov99}, $I_\mvec{k}$ is \textit{not} the Wigner function.) In precise agreement with \Refs{ref:smolyakov99,ref:krommes00}, that gives
\begin{gather}
\msf{W}(t, \mvec{x}, \mvec{k}) = I_\mvec{k}(t, \mvec{x})\,k^2(1 + k^2).
\end{gather}

\subsubsection{Generalized Hasegawa-Mima model} Consider also the ``generalized'' Hasegawa-Mima model,
\begin{gather}\nonumber
\partial_t (1 - \nabla^2) \varphi + (\mvec{V}_* \cdot \nabla) \varphi + (\mvec{V}_E \cdot \nabla) (1 - \nabla^2)\varphi = 0. 
\end{gather}
In this case we take $\hat{U}^{-1} = 1 + \oper{k}^2$, so the Hamiltonian~is
\begin{gather}
\hat{H} = \frac{\mvec{V}_* \cdot \oper{k}}{1 + \oper{k}^2} + \mvec{V}_E(t, \oper{x}) \cdot \oper{k}.
\end{gather}
Like above, for the LL one hence obtains $\mc{H} = \mvec{k} \cdot \mvec{V}_*/(1 + k^2) + \mvec{k} \cdot \mvec{V}_E$ and, using \Eq{eq:aux206}, arrives at
\begin{gather}
\msf{W}(t, \mvec{x}, \mvec{k}) = I_\mvec{k}(t, \mvec{x})\, (1 + k^2)^2.
\end{gather}
This again agrees with \Refs{ref:smolyakov99,ref:krommes00}; see also \Ref{ref:mendonca11b}. 

It is hoped then, considering how concise the above calculation is, that the utility of approaching wave equations through their LSE form is hereby made evident. The above examples also illustrate the fact (often overlooked in literature; cf., \eg Ref.~\cite[Chap.~13]{book:stix}) that \textit{knowing a local dispersion relation is insufficient for deriving a field equation in inhomogeneous and (or) nonstationary medium}. Indeed, the local dispersions for $\varphi$ and $\psi$ are the same, but $\partial_t \varphi$ and $\partial_t \psi$ differ by $\mc{O}([\hat{M}, \hat{U}])$. The commutator may be small, but its effect accumulates, so the scalings $\varphi(t, \mvec{x})$ and $\psi(t, \mvec{x})$ can be very different. This is seen already from the fact that $\braket{\psi|\psi}$ is conserved, whereas $\braket{\varphi|\varphi}$ generally is not. In particular, see, \eg \Ref{ref:kravtsov74} for the terms that must be added in the complex-amplitude equation for an electromagnetic GO wave, as compared to a strictly sinusoidal wave, to avoid the nonphysical dissipation that is effectively introduced otherwise. For stationary waves, the corrections to the commonly used electric-field equation are determined by the spatial dispersion and thus are often negligible when modeling cold electromagnetic oscillations; yet they can become noticeable at mode conversion to electrostatic oscillations, for which the spatial dispersion is significant \cite{tex:jaeger11}. For nonstationary waves, the corrections are determined also by the temporal dispersion and thus are \textit{always} significant. See, for instance, the relevant discussions in \Refs{my:mquanta,my:dense,my:iorec} on waves in plasmas undergoing compression, ionization, and recombination.

\section{General linear waves}
\label{sec:glw}

\subsection{Kinetic equations for the extended space}
\label{sec:extspace}

\subsubsection{Basic equations} 

The above results can be applied also to systems whose Lagrangian densities depend on $\hat{\omega} \doteq i\partial_t$ differently than in \Eq{eq:Lred2}. (That said, the eigenfrequencies must remain real; this permits, \eg certain quadratic and biquadratic functions of $\hat{\omega}$, but not arbitrary polynomials of $\hat{\omega}$ of power higher than two.) Suppose a Lagrangian density of the most general form consistent with linear dynamics on a manifold subjected to the same restrictions as described in \Sec{sec:momentum}. Specifically, we will assume
\begin{gather}\label{eq:KGL0}
\mcc{L} = - \psi^*\hat{\extd{H}}\psi,
\end{gather}
where $\hat{\extd{H}}$ is a Hermitian operator of the form
\begin{gather}
\hat{\extd{H}} = \extd{H}(t, \mvec{x}, -i\partial_t, - i\eth_\mvec{x}).
\end{gather}
Let us treat $t$ as yet another canonical coordinate, with $-\hat{\omega}$ serving as the corresponding canonical momentum. The new coordinates, $\extd{\mvec{x}} \doteq (t, \mvec{x})$, form the extended space, $\extd{X}$. The new momentum is defined as $-i\eth_{\extd{\mvec{x}}}$, so its eigenvalues, $\extd{\mvec{k}}$, form the extended momentum space, $\extd{K}$, and the extended phase space is defined as $\extd{\msf{Z}} \doteq \extd{X} \times \extd{K}$ with coordinates
\begin{gather}
\extd{\msf{z}} = (t, x^1, \ldots x^D, - \omega, k_1, \ldots k_D).
\end{gather}

Equation \eq{eq:KGL0} can be written as
\begin{gather}\label{eq:KGL3}
\mcc{L} = - \psi^*\extd{H}(\extd{\mvec{x}}, -i\eth_{\extd{\mvec{x}}})\psi,
\end{gather}
so $\mcc{L}$ contains no ``time'' variable whatsoever. To put it in the form \eq{eq:Lred2}, let us introduce an auxiliary system governed by the Lagrangian density
\begin{gather}
\extd{\mcc{L}} = \frac{i}{2}\,[\psi^*(\partial_\mcc{s}\psi) - (\partial_\mcc{s}\psi^*)\psi] - \psi^*\extd{H}(\extd{\mvec{x}}, -i\eth_{\extd{\mvec{x}}})\psi,
\end{gather}
where $\mcc{s}$ is an additional time variable. The corresponding LSE can be written as
\begin{gather}\label{eq:aux88}
i\partial_\mcc{s} \psi =  \extd{H}(\extd{\mvec{x}}, -i\eth_{\extd{\mvec{x}}})\psi,
\end{gather}
and the FKE for the extended system is 
\begin{gather}
\partial_\mcc{s} \extd{f} + \moyal{\extd{f}}{\extd{\mc{H}}} = 0.
\end{gather}
The new PSI, $\extd{f}$, and the new effective Hamiltonian, $\extd{\mc{H}}$, are functions of $\extd{\msf{z}}$, and the bracket is adjusted accordingly. 

Recall now that $\extd{H}(\extd{\mvec{x}}, -i\eth_{\extd{\mvec{x}}})\psi$ equals the functional derivative of $\mc{S}$ with respect to $\psi^*$, and, according to the LAP, $\delta_{\psi^*} \mc{S} = 0$ [\Eq{eq:leastactpsi}]. Thus, according to \Eq{eq:aux88}, those can be found as a subset of the trajectories of the new system corresponding to $\partial_\mcc{s} = 0$. This means that the evolution of the original system satisfies
\begin{gather}\label{eq:extmoy}
\moyal{\extd{f}}{\extd{\mc{H}}} = 0,
\end{gather}
which can be adopted as a new FKE. 

\subsubsection{Liouville limit} 

In the LL, \Eq{eq:extmoy} turns into
\begin{multline}
-(\partial_\omega \extd{\mc{H}})\,(\partial_t \extd{f}) + (\partial_t \extd{\mc{H}})\,(\partial_\omega \extd{f})\\
+ \nabla_\mvec{k} \extd{\mc{H}} \cdot \nabla_\mvec{x} \extd{f}  - \nabla_\mvec{x} \extd{\mc{H}} \cdot \nabla_\mvec{k} \extd{f} = 0.
\end{multline}
The corresponding ray equations then are
\begin{gather}
\dot{\omega} = \frac{\partial_t \extd{\mc{H}}}{-\partial_\omega \extd{\mc{H}}}, \quad
\dot{x}^r = \frac{\partial_{k_r} \extd{\mc{H}}}{-\partial_\omega \extd{\mc{H}}}, \quad
\dot{k}_r = - \frac{\partial_{x^r} \extd{\mc{H}}}{-\partial_\omega \extd{\mc{H}}}.
\end{gather}
One can also introduce an auxiliary time, $\tau$, along the ray (not to be confused with $\mcc{s}$) and cast these as follows:
\begin{align}
d_\tau t = - \partial_\omega \extd{\mc{H}}, \quad
& d_\tau\omega = \partial_t \extd{\mc{H}}, \\
d_\tau x^r = \partial_{k_r} \extd{\mc{H}}, \quad
& d_\tau k_r = -\partial_{x^r} \extd{\mc{H}},
\end{align}
where $d_\tau \equiv d/d\tau$ (cf., \eg Ref.~\cite[Sec.~4.7]{book:stix}). Finally, the equations derived earlier in \Sec{sec:ll} are recovered from here as a special case corresponding to $\extd{\mc{H}} = \mc{H} - \omega$.

\subsubsection{Example: Klein-Gordon equation}

The equations derived above generalize, for instance, the existing kinetic formulation \cite{ref:mendonca11} of the KGE for a \textit{complex} field $\psi$,\footnote{For a \textit{real} field governed by the KGE, say, $\chi = \mbox{Re}\,\psi$, the Lagrangian density would be given by \Eq{eq:Lred2} with $\hat{H} = (\hat{\mcc{P}}^2 + \mcc{M}^2)^{1/2}$, in agreement with \Ref{arX:kauffmann10}. However, this would be a different physical system than that described by \Eq{eq:KGL0} with $\hat{H}$ given by \Eq{eq:kgeOm}; \eg the former does not allow $\psi$ (not $\psi^*$) to oscillate at negative frequency.}
\begin{gather}\label{eq:KGEfull}
\big[-(i\partial_t + \mc{A}_0)^2 + (i\nabla_\mvec{x} + \mvec{\mc{A}})^2 + \mcc{M}^2\big]\psi = 0,
\end{gather}
which, in particular, describes a scalar relativistic particle with mass $\mcc{M}$ interacting with a four-vector potential $\extd{\mvec{\mc{A}}} = (\mc{A}_0, \mvec{\mc{A}})$ (in appropriate units). Specifically, the KGE corresponds to $\mcc{L}$ of the form \Eq{eq:KGL3} with
\begin{gather}\label{eq:kgeOm}
\extd{\hat{H}} = (i\partial_t + \mc{A}_0)^2 - (i\nabla_\mvec{x} + \mvec{\mc{A}})^2 - \mcc{M}^2,
\end{gather}
or, alternatively, $\extd{\hat{H}} = - \hat{\mcc{P}}^2 - \mcc{M}^2$, where we assume an extended spatial metric\footnote{Note that we specify the extended spatial metric here only to cast the Hamiltonian in a compact form. The star bracket itself is insensitive to the spatial metric by definition.} of the form $\mbox{diag}\,(-1, \oper{\gamma})$, and $\mvec{\mcc{P}} \doteq - i\extd{\nabla} - \extd{\mvec{\mc{A}}}$. [The operator $\mcc{P}^2$ is Hermitian as a products of two identical, and thus commuting, Hermitian operators, $\mvec{\mcc{P}}$.] Then \Eq{eq:extmoy} applies immediately. 

Note also that the KGE corresponding to $\extd{\mvec{\mc{A}}} = 0$,
\begin{gather}
(\partial^2_t - \nabla^2_\mvec{x} + \mcc{M}^2)\psi = 0,
\end{gather}
can be viewed as a special case of the LSE as given by \Eq{eq:lsevv}. Indeed, if in \Eq{eq:lsevv} one takes $\bar{\mvec{v}}_{\rm g} = 0$ and $\phi \propto \exp(i \mcc{M}^2 t)$ with $\mcc{M} = \mbox{const}$, then one gets
\begin{gather}\label{eq:lsevv2}
(-\nabla \cdot \oper{\Upsilon} \cdot \nabla + \mcc{M}^2) \phi = 0.
\end{gather}
As $\oper{\Upsilon}$ is real and symmetric, we can always choose a basis in $X$ where this matrix has a diagonal form, say, $\Upsilon^{rs} = \mbox{diag}\,(\Upsilon^1, \ldots \Upsilon^D)$. Then, \Eq{eq:lsevv2} becomes
\begin{gather}\label{eq:lsevv3}
( - \Upsilon^r \partial^2_r + \mcc{M}^2) \phi = 0.
\end{gather}
If one of $\Upsilon^r$ is negative (\ie the signature of $\oper{\Upsilon}$ as a metric is not positive-definite), then $-\Upsilon^r \partial^2_r$ is effectively a d'Alembertian, and \Eq{eq:lsevv3} turns into the KGE.

\subsection{Vector waves}
\label{sec:vector}

The above results can be generalized to the case when $X$, as opposed to being simply connected, is rather a \textit{disjoint union} of $B$ simply connected manifolds, $X_b$. Like in \Sec{sec:momentum}, let us assume that each of the latter has topological properties of $\mathbb{R}^D$. Let us also view all $X^b$ as copies of some $X_\diamond$. Then coordinates can be understood as ordered pairs, $\mvec{x} = (\mvec{x}_\diamond, b)$, where $\mvec{x}_\diamond$ belongs to $X_\diamond$, and $b$ is an integer spanning from 1 to $B$. 

For any $\ket{\psi}$, its $\mvec{x}$-representation is defined as $\braket{\mvec{x}|\psi} \equiv \braket{\mvec{x}_\diamond, b|\psi} \rdoteq \psi^b(\mvec{x}_\diamond)$. The momentum operator is defined such that, in the $\mvec{x}$-representation, $\oper{k} \psi^b(\mvec{x}_\diamond) \doteq -i\eth_{\mvec{x}_\diamond} \psi^b(\mvec{x}_\diamond)$. This equation yields eigenvectors parameterized by ordered pairs $\mvec{k} \doteq (\mvec{k}_\diamond, b)$, where the eigenvalues $\mvec{k}_\diamond$ comprise some manifold $K_\diamond$, and $b$ spans from 1 to $B$. Hence, the momentum space $K$, comprised of all $\mvec{k}$, is understood as a disjoint union of $B$ copies of $K_\diamond$.

Using the identity 
\begin{gather}\label{eq:aux301}
\hat{1} =  \sum_{b = 1}^B \int \ket{\mvec{x}_\diamond, b} \bra{\mvec{x}_\diamond, b}\,\mcu{D}x_\diamond,
\end{gather}
we can now represent the Lagrangian \eq{eq:Lreduced} in the form $L = \int \mcc{L}\,\mcu{D}x_\diamond$, where
\begin{align}
\mcc{L} = \frac{i}{2}\,&\big\{\psi_b^*(t, \mvec{x}_\diamond)\,[\partial_t\psi^b(t, \mvec{x}_\diamond)] - [\partial_t\psi^*_b(t, \mvec{x}_\diamond)]\,\psi^b(t, \mvec{x}_\diamond) \big\} \nonumber \\ 
& - \psi^*_b(t, \mvec{x}_\diamond) \int \braket{\mvec{x}_\diamond, b| \hat{H} | \mvec{x}'_\diamond, b'}\,\psi^{b'}(t, \mvec{x}'_\diamond)\,\,\mcu{D}x'_\diamond, \nonumber
\end{align}
and $\psi_b^* \doteq \braket{\psi|\mvec{x}_\diamond, b}$. For each given $b$ and $b'$, the bracket $\braket{\mvec{x}_\diamond, b| \hat{H} | \mvec{x}'_\diamond, b'}$ is a kernel of an operator acting on functions of $\mvec{x}'_\diamond$. As usual then, one can write
\begin{multline}
\int \braket{\mvec{x}_\diamond, b| \hat{H} | \mvec{x}'_\diamond, b'}\,\psi^{b'}(t, \mvec{x}'_\diamond)\,\,\mcu{D}x'_\diamond
\\ = {\Theta^b}_{b'}(t, \mvec{x}_\diamond, -i\eth_{\mvec{x}_\diamond})\,\psi^{b'}(t, \mvec{x}_\diamond).
\end{multline}
To simplify the notation, let us interpret the family of $\psi^b$ as a $B$-component vector field $\mvec{\psi}(t, \mvec{x}_\diamond) \doteq (\psi^1(t, \mvec{x}_\diamond), \ldots \psi^B(t, \mvec{x}_\diamond))$ and ${\Theta^b}_{b'}$ as elements of the matrix $\oper{\Theta}$. Then $\mcc{L}$ finally becomes
\begin{multline}\label{eq:Lvecwaves}
\mcc{L} = \frac{i}{2}\,\big[\mvec{\psi}^* \cdot (\partial_t \mvec{\psi})- (\partial_t\mvec{\psi}^*) \cdot \mvec{\psi}\big] 
\\ - \mvec{\psi}^* \cdot \oper{\Theta}(t, \mvec{x}_\diamond, -i\eth_{\mvec{x}_\diamond}) \cdot \mvec{\psi}.
\end{multline}

Waves described by \Eq{eq:Lvecwaves} can be attributed as vector waves. (Typical examples are electromagnetic waves that allow for more than one polarization.) Due to \Eq{eq:aux301}, the state vector for such waves can be decomposed as $\ket{\psi} = \sum_b \ket{\psi_b}$, and $\hat{\rho} = \sum_{bb'} \hat{\rho}_{b'}{}^{b}$, $\hat{\rho}_{b'}{}^{b} \doteq \ket{\psi_{b'}}\bra{\psi^{b}}$, where each $\ket{\psi_b}$ is comprised of $\ket{\mvec{x}_\diamond, b}$ with specific $b$. Coupled kinetic equations for $\hat{\rho}_{b'}{}^{b}$ can then be derived by projecting the operator equations for each $\hat{\rho}_{b'}{}^{b}$ on $X_\diamond$ and $K_\diamond$. This yields $B^2$ equations, or, alternatively, one can think of the FKE as of a matrix equation. As usual, there is no unique way to write this FKE, but otherwise the procedure is similar to the one used in \Sec{sec:dmeq}, so we will not repeat it here. 

Let us yet mention that \Eq{eq:Lvecwaves} also has a simple GO limit, obtained when $\mvec{\psi} = \mvec{\phi} e^{i\theta}$ such that $\mvec{\phi}$ is a slowly-varying envelope, and $\theta$ is a rapid phase common for all the wave components. In this limit, one has a formula akin to \Eq{eq:aux201}, namely,
\begin{align}
\mcc{L} & = -\mvec{\phi}^* \cdot \big[\partial_t \theta + \oper{\Theta}(t, \mvec{x}_\diamond, \nabla_\diamond \theta)\big] \cdot \mvec{\phi} \nonumber
\\ & = - \mbox{tr}\,\Big\{\big[\oper{1}\,\partial_t \theta + \oper{\Theta}(t, \mvec{x}_\diamond, \nabla_\diamond \theta)\big] \hat{\mvec{\rho}}\Big\},\label{eq:701}
\end{align}
where $\oper{1}$ is a unit operator, and $\hat{\mvec{\rho}} \doteq \mvec{\phi}^*\mvec{\phi}$ is understood as a ``reduced'' action operator. It is seen then that, as long as wave components are uncoupled (or coupled adiabatically, as in \Ref{tex:mylens}), they cannot be assigned individual actions even in the GO limit.

Similar equations were studied earlier in connection with the problem of linear mode conversion; see \eg \Refs{ref:tracy93,ref:friedland87,ref:cook92} and references cited therein. The difference is, however, that we deal with \textit{generalized} coordinates and momenta here (our $X_\diamond$ is not necessarily ``the physical space''), and we also keep them in the operator form. This eliminates the need for \textit{ad~hoc} application of metaplectic transformations used, \eg in \Ref{ref:tracy93}, as in our case such transformation are already embedded, effectively, in the definition of a wave. Note also that our equations can be generalized, along the lines of \Sec{sec:extspace}, to allow arbitrary dependence on the time derivative in $\mcc{L}$. Then, or if one allows for a pseudo-Euclidean metric instead, our framework also accommodates more general waves such as those as Dirac's electron \cite[Sec.~26]{book:landau4}. (The particle spin is hence understood simply a geometric feature of the classical coordinate space $X$, not to be confused with $X_\diamond$; \ie contrary to the traditional point of view, there is nothing ``inherently quantum'' in spin.) For a related discussion see \Ref{arX:cabrera12} and references therein.

\section{Nonlinear waves}
\label{sec:nlin}

\subsection{Approximate conservation laws}
\label{sec:aclnl}

Let us also discuss, briefly, how the theory is modified by the presence of \textit{nonlinear} coupling of reference modes via some Hamiltonian $\mcc{H}$. Under the assumptions used to obtain \Eq{eq:aux302}, we now arrive at
\begin{gather}
\dot{a}^n = -i \Omega_{n'} a^{n'} - {V^n}_m a^m -i\,\frac{\partial \mcc{H}}{\partial a_n^*},
\end{gather}
and \Eq{eq:din} turns into
\begin{multline}
\dot{I}_n = - (V_{n'm} a^{n'*}a^m + V^*_{n'm} a^{n'}a^{m*})
\\ + i\left(a^{n'}\frac{\partial \mcc{H}}{\partial a^{n'}} - a^*_{n'}\frac{\partial \mcc{H}}{\partial a_{n'}^*} \right).\label{eq:aux310}
\end{multline}
(Remember that we use primes to distinguish repeating indexes on which \textit{no} summation is performed; in other respects, $n' \equiv n$.) If $\mcc{H}$ can be expressed as a function of actions only, so $\partial \mcc{H}/\partial a_n^* = (\partial \mcc{H}/\partial I_{n'}) a^{n'}$ and $\partial \mcc{H}/\partial a^n = (\partial \mcc{H}/\partial I_{n'}) a^*_{n'}$, one arrives at \Eq{eq:din} and the same conservation laws as in \Sec{sec:acl}. Otherwise, the approximate invariants are derived as follows.

\subsubsection{Manley-Rowe relations} If $\mcc{H}$ is a (real) combination of arbitrary multilinear functions of $a_n$ and $a^*_n$, then conserved are so-called Manley-Rowe integrals \cite{my:manley,my:beatf,ref:brizard95}. For example, consider a general three-wave resonant coupling,
\begin{gather}
\mcc{H} = \chi a_1 a_2^* a_3^* + \chi^* a_1^* a_2 a_3,
\end{gather}
where $\chi$ is some parameter, possibly a slow function of time. Then \Eq{eq:aux11} yields
\begin{gather}
\dot{I}_1 = i\left(\chi a_1 a_2^* a_3^* - \chi^* a_1^* a_2 a_3 \right),\\
\dot{I}_2 = i\left(\chi^* a_1^* a_2 a_3 - \chi a_1 a_2^* a_3^* \right),\\
\dot{I}_3 = i\left(\chi^* a_1^* a_2 a_3 - \chi a_1 a_2^* a_3^* \right),
\end{gather}
so it is seen that
\begin{gather}\nonumber
I_1 + I_2 = \mbox{const}, \quad I_1 + I_3 = \mbox{const}, \quad I_2 - I_3 = \mbox{const}.
\end{gather}
Similar conservation laws will hold also at resonant coupling of other types. For the general, \textit{geometric} formulation of Manley-Rowe conservation laws see \Refs{my:manley,ref:tennyson82}.

\subsubsection{Total action} The total action is governed by 
\begin{gather}\label{eq:aux11}
\dot{I} = i\left(a^{n}\frac{\partial \mcc{H}}{\partial a^{n}} - a^*_{n}\frac{\partial \mcc{H}}{\partial a_{n}^*} \right)
\end{gather}
[where one can substitute $a^{n}(\partial \mcc{H}/\partial a^{n}) = a_{n}(\partial \mcc{H}/\partial a_{n})$], so it is not conserved at nonlinear interactions in general. An exception is the case when $\mcc{H}$ can be written as $\mcc{H} = 2\,\mbox{Re}\,\mcc{H}_c$, where $\mcc{H}_c$ is a multilinear form that is ``symmetric'' in the following sense: it must be $p_n$-linear in $a_n$, $q_n$-linear in $a_n^*$, and $\sum^N_{n = 1} p_n = \sum^N_{n = 1} q_n \rdoteq \wp$. Then, 
\begin{gather}\nonumber
a_{n}\frac{\partial \mcc{H}}{\partial a_{n}} 
= a_{n}\frac{\partial \mcc{H}_c}{\partial a_{n}} + a_{n}\frac{\partial \mcc{H}_c^*}{\partial a_{n}} 
= \sum^N_{n = 1} \big(p_n \mcc{H}_c + q_n \mcc{H}_c^*\big) = \wp \mcc{H},
\end{gather}
and, similarly, $a^*_{n}(\partial \mcc{H}/\partial a_{n}^*) = \wp \mcc{H}$ as well. Hence \Eq{eq:aux11} yields $\dot{I} = 0$, and, of course, the same result also applies when $\mcc{H}$ is a sum of such symmetric forms.

\subsection{Nonlinear Schr\"odinger equation}
\label{sec:nlse}

In geometric terms, one can interpret each $\mcc{H}$ of the aforementioned symmetric type as a rank-$(\wp,\wp)$ tensor evaluated on $\ket{\psi}$ and $\bra{\psi}$. In other words,
\begin{gather}
\mcc{H} = \hat{\mcc{H}}^{(\wp)}\big(\underbrace{\bra{\psi} \ldots \bra{\psi}}_{\wp};\, \underbrace{\ket{\psi} \ldots \ket{\psi}}_{\wp}\big),
\end{gather}
or, using also that $\mcc{H}$ is real,
\begin{gather}\label{eq:aux342}
\mcc{H} = \mcc{H}^{(\wp)}_{(m;\,n)} \prod_{i = 1}^\wp (\psi^{m_i*} \psi^{n_i}), \quad \mcc{H}^{(\wp)}_{(m;\,n)} = \mcc{H}^{(\wp)*}_{(n;\,m)},
\end{gather}
where we introduced the shorthand notation $(m;\,n) \doteq (m_1 \ldots m_\wp;\, n_1 \ldots n_\wp)$. Instead of \Eq{eq:aLSE}, we then get
\begin{gather}\label{eq:aux351}
i\ket{\hat{\mc{D}}\psi} = (\hat{Q} + \hat{\mc{N}}) \ket{\psi} - i \ket{W}.
\end{gather}
Here the nonlinear operator $\ket{\mc{N}} \doteq \delta\mcc{H}/\delta \bra{\psi}$ is given by
\begin{gather}\nonumber
\hat{\mc{N}} \doteq \sum_{j = 1}^\wp \hat{\mcc{H}}^{(\wp)}\big(\bra{\psi} \ldots, \ph_{\,j}, \ldots \bra{\psi};\,\ket{\psi} \ldots, \ph_{\,j}, \ldots\ket{\psi}\big),
\end{gather}
where $j$th and $(\wp+j)$th arguments of $\hat{\mcc{H}}$ are not evaluated. (Remember that the symbol ``$\ph$'' denotes a placeholder, and the index $j$ is added to show which specific arguments of $\hat{\mcc{H}}$ it replaces.) With all indexes lowered, this yields
\begin{gather}
\mc{N}_{pq} = \sum_{j = 1}^\wp \mcc{H}^{(\wp)}_{(m|p;\,n|q)_j} \prod_{i \ne j}^\wp (\psi^{m_i*} \psi^{n_i}),
\end{gather}
where $(m|p;\,n|q)_j$ is the same as $(m;\,n)$ yet with $m_j$ replaced with $p$, and $n_j$ replaced with $q$. Notice then that
\begin{align}
\mc{N}^*_{qp} 
& = \sum_{j = 1}^\wp \mcc{H}^{(\wp)*}_{(m|q;\,n|p)_j} \prod_{i \ne j}^\wp (\psi^{m_i} \psi^{n_i*})\nonumber\\
& = \sum_{j = 1}^\wp \mcc{H}^{(\wp)*}_{(n|q;\,m|p)_j} \prod_{i \ne j}^\wp (\psi^{n_i} \psi^{m_i*}) = \mc{N}_{pq},\label{eq:aux341}
\end{align}
where \Eq{eq:aux342} was used. This means that $\hat{\mc{N}}$ is Hermitian, which explains why symmetric $\mcc{H}$ conserve $I$.

Equation \eq{eq:aux351} can be attributed as the generalized nonlinear Schr\"odinger equation (NLSE). In a stationary metric, it simply becomes
\begin{gather}\label{eq:cLSE300}
i\ket{\dot{\psi}} = \hat{H}_{\rm NL} \ket{\psi},
\end{gather}
where $\hat{H}_{\rm NL} \doteq \hat{\Omega} + \hat{\mc{N}}$ is the new, nonlinear Hamiltonian. The NLSE as it is known most commonly \cite[Sec.~17.7]{book:whitham}, 
\begin{gather}\label{eq:aux371}
i(\partial_t \psi + \bar{\mvec{v}}_{\rm g} \cdot \nabla \psi) 
+ {\textstyle \frac{1}{2}}\,\nabla \cdot \bar{\mvec{v}}'_{\rm g} \cdot \nabla \psi 
+ \mu |\psi|^2\psi = 0
\end{gather}
[here $\mu$ is a constant, and $\psi \equiv \psi(t, \mvec{x})$], is a special case of \Eq{eq:cLSE300} that corresponds to $\hat{\Omega} = \bar{\mvec{v}}_{\rm g} \cdot \oper{k} + \oper{k} \cdot \bar{\mvec{v}}'_{\rm g} \cdot \oper{k}/2$ (cf. \Sec{sec:wavesdef}), and $\hat{\mcc{H}}^{(2)} = - (\mu/2) \ket{\mvec{x}}\ket{\mvec{x}} \bra{\mvec{x}}\bra{\mvec{x}}$.

\subsection{Kinetic equation and nonlinear ray tracing}
\label{sec:ghost}

The FKE for nonlinear waves can be constructed much like in \Sec{sec:dmeq}. The only difference is that now a ``collision operator'' may emerge from $\mcc{H}$ \cite{book:zakharov-b}, but here we will consider only the simplest paradigmatic case, when all nonlinear interactions are included in $\hat{H}_{\rm NL}$. Specifically, let us consider the one-dimensional version of \Eq{eq:aux371} in the frame traveling with velocity~$\bar{v}_{\rm g}$,
\begin{gather}\label{eq:aux371b}
i\partial_t \psi + \partial^2_x \psi + \mu |\psi|^2\psi = 0.
\end{gather}
(We eliminated the coefficient in front of the second term by rescaling $x$. The coefficient $\mu$ can be eliminated too, by rescaling $\psi$, but only up to a sign.) This corresponds to the following nonlinear Hamiltonian:
\begin{gather}
\hat{H}_{\rm NL}(t, \hat{x}, \hat{k}) = \hat{k}^2/2  - \mu [A(t, \hat{x})]^2,
\end{gather}
with the image $\mc{H}_{\rm NL}(t, x, k) = k^2/2 - \mu A^2(t, x)$. Hence, the FKE is obtained, as usual, in the form
\begin{gather}\label{eq:nlfke}
\partial_t f + \moyal{f}{\mc{H}_{\rm NL}} = 0.
\end{gather}

The LL and ray tracing flow from \Eq{eq:nlfke} under the same restrictions as for linear waves (\Sec{sec:ll}), specifically, if $f$ is wide in both coordinate and momentum space (see, \eg \Refs{ref:hasegawa75,ref:mima78}). Contrary to a popular opinion (see, \eg \Refs{book:mendonca,ref:gao10,ref:censor82}), quasimonochromatic nonlinear waves (QNW) do not fit in this picture simply as a special case, \ie cannot be described by a WKE. This is because such waves have the spatial scale of $A(t, x)$ of the same order as that of $f(t, x, k)$; then
\begin{gather}
(\partial^n_x A^2)\,(\partial^n_k f) \sim A^2 f
\end{gather}
for any $f$ at any $n$, \ie ghost terms are \textit{never} negligible for QNW. Hence the FKE cannot be approximated with the WKE, and nonlinear phase space rays cannot be defined for QNW in principle. (This is specific to QNW; Hamiltonians of other waves have spatial scales independent of~$f$.)

One should not be confused by the fact that \textit{some} properties of QNW may nevertheless be reproduced accurately within the WKE model, as reported \eg in \Ref{ref:hall02,ref:reitsma08}. That happens when the relevant \textit{integrals} of the ghost term vanish (even though $\mcc{G}$ itself is nonnegligible), so the WKE accidentally leads to the same hydrodynamic equations as the FKE. For example, this explains the correct rate, $\gamma_{\rm MI}$, of the modulational instability that the WKE happens to yield in the cold limit \cite{ref:hall02}.\footnote{See more about such instabilities, \eg in \Refs{ref:hasegawa75,ref:mima78,ref:helczynski02,ref:onorato03,ref:marklund06,ref:hansson13,ref:santos07,ref:shukla10}. See more about other methods of describing partially coherent waves, \eg in \Ref{ref:semenov08} and references therein.} The result is caused by the following: (i)~all mixed derivatives of $\mc{H}_{\rm NL}$ are zero, which makes the FKE and WKE yield identical GO equations (\Sec{sec:go}) in the cold limit, and (ii)~those GO equations are sufficient to obtain $\gamma_{\rm MI}$ \cite{book:whitham,my:itervar}. 

Since the GO model does not capture effects caused by the wave nonzero spectral width, the nonlinear FKE is not easily applicable beyond the cold limit to QNW, contrary to the existing literature, and same applies to the nonlinear ray tracing in general. Having said that, in the cold limit nonlinear rays are well defined and can be used for practical applications, including reduced simulations of wave dynamics \cite{my:actiii}. As a matter of fact, there are generally \textit{two} sets of such rays at each location (and that is another way of seeing that WKE cannot describe QNW in principle). For details, one is referred to \Refs{my:itervar,book:whitham}.

\section{Parallels with quantum mechanics}
\label{sec:rqm}

One may notice that a particular case of the above theory is QM, yielded from exactly one axiom that the underlying physical system is a classical nondissipative linear oscillator (\ie the simplest nontrivial closed stable nondissipative system). The nonrelativistic quantum theory corresponds to the adiabatic limit, when $I$ is conserved, so one can take $\braket{\psi|\psi} = 1$.\footnote{The standard convention would be to also assume no negative-energy waves and allow only unitary coordinate transformations \eq{eq:Uy}; then $g_{mn}$ is Euclidean. But QM models with more general metrics exist too; see, \eg Lee-Wick model \cite{ref:lee69}.} The equivalence between the classical Boltzmann-Vlasov equation and the Liouville equation \eq{eq:liouv} that we happened to derive for the formally introduced $\mvec{x}$ and $\mvec{k}$ also permits assigning the traditional physical meaning to these quantities. [Alternatively, one can appeal to the Hamilton-Jacobi equation, \Eq{eq:HJ}, that flows from our formal theory.] Specifically, if the coordinate basis vectors $\ket{\mvec{x}}$ are chosen such that $\hbar\mc{H}(\mvec{x}, \hbar\mvec{k})$ happens to be the classical energy expressed as a function of the physical coordinate and physical momentum, then $\mvec{x}$ and $\hbar\mvec{k}$ must be understood as such. Note that $\hbar$ is not a fundamental constant within this approach; it merely characterizes units in which the energy-momentum is measured. Also note that there is no way to infer $\hat{H}$ from $\mc{H}$ in general; \ie the true quantum Hamiltonian can only be guessed but not derived \textit{per se} from its classical counterpart. In simple cases, such as in vacuum, one can appeal to symmetry considerations (such as in \Ref{book:landau4}) to justify a guess, but otherwise one may have to derive $\hat{H}$ independently from its classical limit.

Nonadiabatic dynamics corresponds to QFT, where $I$, measuring the number of particles, can vary in time. Since $I$ is introduced as a classical action, it is not quantized in our case as is. Its quantization is nevertheless very natural within the new formalism and can be performed exactly as the first quantization in the traditional QM. Specifically, one needs only to replace the independent variables $a_n^*$ and $a^n$ in the energy $h = \braket{\psi|\hat{H}|\psi}$ with (noncommuting) operators $\hat{a}^\dag_n$ and $\hat{a}^n$ on some new vector space $\Psi'$. Hence $h$ also becomes an operator, $\hat{h} = H_n \hat{a}^\dag_n\hat{a}^n$, and one may recognize $\hat{a}^\dag_n$ and $\hat{a}^n$ as creation and annihilation operators. The new scalar energy is then yielded in the form $h' = \braket{\psi'|\hat{h}|\psi'}$, where $\ket{\psi'}$ are vectors from $\Psi'$. This determines the new Lagrangian
\begin{gather}
L' = \frac{i}{2}\,\big[\braket{\psi'|\dot{\psi}'} - \braket{\dot{\psi}'|\psi'}\big] - \braket{\psi'|\hat{h}|\psi'},
\end{gather}
so the dynamic equation becomes $i\ket{\dot{\psi}'} = \hat{h}\ket{\psi'}$. But then, if needed, it is easy to quantize the theory once again and so on, and that also leads to the idea of ``$\msf{N}$th quantization'' (cf., \eg \Ref{arX:finkelstein10} and references therein). 

Finally, consider some other obvious reasons for why parallels between our theory and QM are important:
\begin{enumerate}
\item[(i)]
Since classical-wave physics happens to be not just analogous but, in fact, identical to (the mathematical framework of) QM, classical waves can be studied using \textit{exactly} the same methods as quantum particles. For example, one can imagine classical applications of the QM perturbation theory to studying wave propagation in inhomogeneous and (or) nonstationary media, as will be reported elsewhere. Some other insights and clarifications brought in by the quantumlike approach to classical waves were also discussed above.
\item[(ii)]
Our representation of linear wave physics may also, in principle, be useful as an axiomatic introduction into the QM/QFT formalism. The only difference is that the theory presented here does not address the problem of quantum collapse, for which discretizing $I$ would be essential. (On the other hand, the standard textbook interpretation of QM is not unlike in this sense.) This vividly shows that, for a given fundamental Hilbert space $\Psi$, the only difference between classical and quantum oscillations is the measurement process but not the dynamics \textit{per~se}. Another advantage of our approach is that neither the spaces $\Psi$ and $X$ (and, even more generally, spacetime; see \Sec{sec:glw}), nor commutation relations like \Eq{eq:maincommut} need to be postulated, as most often done in literature \cite{book:dirac}. Instead, they \textit{emerge} as convenient tools for describing solutions of \Eq{eq:hameq}, which has no geometry associated with it in the first place.
\end{enumerate}

\section{Summary}
\label{sec:conc}

The paper reports an axiomatization of the general theory of classical nondissipative waves based on understanding of these waves as multidimensional oscillators. To our knowledge, this is the first attempt of such axiomatization, even though formal studies of linearized Hamiltonian systems are certainly plenty. Specifically, our selected results are summarized, section by section, as follows:
\begin{enumerate}
\item In \Sec{sec:bg}, the wave concept is formalized. The definition may seem trivial, but it is constructive; all the linear wave physics is eventually derived from this definition alone and thus applies to waves of any nature.
\item In \Sec{sec:basic}, the natural, complex-amplitude representation of the wave Lagrangian is derived in a general time-dependent basis, rendering the ACT transparent.
\item In \Sec{sec:fs}, the wave dynamics is cast in a coordinate-invariant form, where the sign of the wave energy is absorbed by the fundamental metric.
\item In \Sec{sec:invrep}, the classical-wave action is defined as an operator (``density matrix''), whose dynamics is governed by a von Neumann equation.
\item In \Sec{sec:ps}, the generic Lagrangian is derived for a scalar wave propagating on a manifold. It is \textit{proved} that the Hamiltonian of such a wave has a form $\hat{H} = H(t, \oper{x}, \oper{k})$.
\item In \Sec{sec:dmeq}, a unified invariant method is proposed for obtaining various kinetic equations in (almost) arbitrarily curved coordinates. The ACT for noneikonal waves is derived in the spatial representation, for the first time extending the result of \Ref{ref:brizard93} to general waves. It is also made clear that similar theorems hold in any other (\eg momentum) representations too.
\item In \Sec{sec:waves}, the applicability conditions for the LL of the kinetic equation and GO equations are revised. It is emphasized that, contrary to an assumption often adopted in literature, the LL and GO are not the same. It is also explained how the new theory allows a statistical description of waves. As an example, the SKEs for two Hasegawa-Mima models are shown to flow from the general theory \textit{automatically} and unambiguously, unlike in other formulations.
\item In \Sec{sec:glw}, it is shown how the axiomatic wave theory applies to oscillations in the extended space. In particular, kinetic equations derived in \Ref{ref:mendonca11} are generalized and shown to flow naturally from the general theory, so they need not be rederived \textit{ad~hoc}. It is also shown how modifying the assumptions about the fundamental-space geometry leads to the concept of vector waves and mode-coupling equations, specifically, in a form that makes the \textit{ad~hoc} application of conventional metaplectic transformations redundant.
\item In \Sec{sec:nlin}, it is explained how nonlinear waves naturally fit in Dirac's bra-ket formalism. It is also argued that, contrary to some literature, nonlinear wave kinetics can be more subtle than linear wave kinetics.
\item In \Sec{sec:rqm}, it is argued that the classical-wave theory exhibits one-to-one correspondence with QM, so the QM machinery is applicable to classical waves ``as is''. Also discussed is a curious spin-off: the proposed formalism naturally leads to the idea of $\msf{N}$th quantization.
\end{enumerate}
It is hoped that these results facilitate understanding of classical waves in a self-contained manner and, for the first time, from invariant first principles (\ie without appealing to \textit{ad~hoc} methods such as the Wigner-Weyl-Moyal formalism). Some specific problems that the new formulation helps to solve, aside from those discussed above, will be addressed in a series of papers that will follow shortly.

\section*{Acknowledgments}
 
The author thanks J. W. Burby, E. A. Startsev, and N. J. Fisch for valuable discussions. The work was supported by the NNSA SSAA Program through DOE Research Grant No. DE274-FG52-08NA28553, by the U.S. DOE through Contract No. DE-AC02-09CH11466, and by the U.S. DTRA through Research Grant No. HDTRA1-11-1-0037.

\appendix

\section{Properties of $\boldsymbol{B_{\mu\nu}}$}
\label{app:gamma}

The matrix $B_{\mu\nu}$ can be written as $B_{\mu\nu} = G_{\mu\alpha}{A^\alpha}_\nu$, where ${A^\alpha}_\beta \doteq {(S^{-1})^\alpha}_\mu{\dot{S}^\mu}_\beta$. By differentiating $G_{\mu\nu} = \mbox{const}$, one finds then that $B_{\mu\nu}$ is anti-Hermitian, 
\begin{gather}\label{eq:gq}
B_{\mu\nu}^* = - B_{\nu\mu}.
\end{gather}
 
Other properties of this matrix are understood as follows. First, notice that the columns of ${S^\mu}_\nu$ are the polarization vectors $\bar{z}_\nu$ (\Sec{eq:ls}), which we can split into the coordinate and momentum parts, $\bar{z}_\nu = (\bar{\xi}_\nu, \bar{\pi}^\nu)$. Hence we can write  ${S^\mu}_\nu$ and its inverse as block matrices of the following form:
\begin{gather}
\hat{S}  \doteq 
\left(
\begin{array}{c @{\quad} c}
\hat{a} & \hat{b} \\[3pt]
\hat{c} & \hat{d}
\end{array}
\right),
\\
\hat{S}^{-1}  \doteq 
\left(
\begin{array}{c @{\quad} c}
\hat{q}^{-1} & -\hat{q}^{-1} \hat{b} \hat{d}^{-1} \\[3pt]
-\hat{p}^{-1} \hat{c} \hat{a}^{-1} & \hat{p}^{-1}
\end{array}
\right),
\end{gather}
where we introduced $\hat{q} \doteq \hat{a} - \hat{b}\hat{d}^{-1}\hat{c}$, $\hat{p} \doteq \hat{d} - \hat{c}\hat{a}^{-1}\hat{b}$, and
\begin{gather}
\hat{a} = \hat{b}^* \doteq \hat \Xi, \quad \hat{c} = \hat{d}^* \doteq \hat \Pi,
\end{gather}
with $\Xi^m{}_n \doteq \bar\xi_n{}^m$ and $\Pi^m{}_n \doteq \bar\pi{}^n{}_m$. Further using that $\hat{q} = \hat\Xi^* \hat\Lambda$ and $\hat{p} = - \hat\Pi \hat\Lambda^*$, and introducing 
\begin{gather}
\hat\Lambda \doteq \hat\Xi^{*-1} \hat\Xi - \hat\Pi^{*-1}\hat\Pi,\\
\hat{v} \doteq \hat{\Lambda}^{-1}(\hat\Xi^{*-1} \dot{\hat{\Xi}} - \hat\Pi^{*-1}\dot{\hat\Pi}), \\
\hat{w} \doteq \hat{\Lambda}^{*-1}(\hat\Xi^{-1} \dot{\hat{\Xi}} - \hat\Pi^{-1}\dot{\hat\Pi}),
\end{gather}
one gets that $\hat{S}^{-1}\dot{\hat{S}} \equiv \hat{A}$ equals
\begin{gather}
\hat{A} =
\left(
\begin{array}{c @{\quad} c}
\hat{v} & \hat{w}^* \\[3pt]
\hat{w} & \hat{v}^*
\end{array}
\right).
\end{gather}
Finally, we define $V_{mn} \doteq \eta_{mk}{v^k}_n$ and $W_{mn} \doteq \eta_{mk}{w^k}_n$,~so
\begin{align}
B_{\mu\nu} & =
\left(
\begin{array}{c @{\quad} c}
V_{mn} & W^*_{mn} \\[3pt]
-W_{mn} & -V_{mn}^*
\end{array}
\right) \\
& = 
\left(
\begin{array}{c @{\quad} c}
- V^*_{nm} & W^{nm*} \\[3pt]
- W_{nm} & V^{nm}
\end{array}
\right),
\end{align}
where $m = \mu\,(\mbox{mod}\, N)$ and $n = \nu\,(\mbox{mod}\, N)$. The latter equality flows from \Eq{eq:gq} and reveals that $V_{mn}$ is anti-Hermitian, and $W_{mn}$ is symmetric.

\section{General vector space with Hermitian metric}
\label{app:metric}

In this appendix we present a brief tutorial on tensor algebra for vector spaces with Hermitian metric.

\subsection{Metric, vectors, and one-forms}

In general, a Hermitian metric $\hat{g}$ in a complex vector space $\mc{V}$ is defined as a nondegenerate map ${\hat{g}:\,\mc{V} \times \mc{V} \to \mathbb{C}}$ which satisfies\footnote{For the purpose of this definition, the $N$-dimensional complex vector space $\Psi$ can be understood as a $2N$-dimensional real vector space; see, \eg \Ref{ref:rund66}.}
\begin{gather}\nonumber
\hat{g}(\lambda\ket{\alpha} + \kappa\ket{\beta}, \ket{\gamma}) = \lambda^*\hat{g}(\ket{\alpha}, \ket{\gamma}) + \kappa^*\hat{g}(\ket{\beta}, \ket{\gamma}),\\
\hat{g}(\ket{\alpha}, \ket{\beta}) = \hat{g}(\ket{\beta}, \ket{\alpha})^*, \label{eq:hermg}
\end{gather}
where $\ket{\alpha}$, $\ket{\beta}$, and $\ket{\gamma}$ are arbitrary vectors from $\mc{V}$, and $\lambda$ and $\kappa$ are arbitrary complex numbers. Hence $\hat{g}$ will determine the inner product $\braket{\alpha|\beta} \doteq \hat{g}(\ket{\alpha}, \ket{\beta})$, which also makes $\mc{V}$ a Hilbert space. Assuming some arbitrary basis $\ket{e_n}$, we can substitute here
\begin{gather}\label{eq:ab}
\ket{\alpha} = \alpha^n \ket{e_n}, \quad \ket{\beta} = \beta^n \ket{e_n},
\end{gather}
where $\alpha^n$ and $\beta^n$ are the vector components in this basis. Then, by definition of~$\hat{g}$,
\begin{gather}\label{eq:scpr}
\braket{\alpha|\beta} = g_{mn} \alpha^{m*} \beta^n, \quad g_{mn} \doteq \hat{g}(\ket{e_m}, \ket{e_n}).
\end{gather}
Due to \Eq{eq:hermg}, the matrix $g_{mn}$ is Hermitian; however, it is not necessarily a unit or diagonal matrix. It is thus convenient to introduce also the ``dual'' vectors $\ket{e^n}$ orthogonal to $\ket{e_n}$. Specifically, we define those via
\begin{gather}\label{eq:duale}
\braket{e^m|e_n} \equiv \hat{g}(\ket{e^m}, \ket{e_n}) = \delta^m_n
\end{gather}
and notice that the basis formed of $\ket{e_n}$ can be used for vector decomposition as well; \ie we can write $\ket{\alpha} = \alpha_n \ket{e^n}$. The ``contravariant'' components $\alpha^n$ and the ``covariant'' coefficients $\alpha_n$ are then found to be
\begin{gather}\label{eq:aux13}
\alpha^n = \braket{e^n|\alpha}, \quad \alpha_n = \braket{e_n|\alpha}
\end{gather}
and are connected through
\begin{gather}\label{eq:iso}
\alpha_k = \braket{e_k|\alpha} = \hat{g}(\ket{e_k}, \ket{\alpha}) = g_{mn} \delta^m_k \alpha^n = g_{kn} \alpha^n.
\end{gather}
In particular, using $g_{kn} = g_{nk}^*$, we now can rewrite the expression for the inner product as follows:
\begin{gather}\label{eq:scalar2}
\braket{\alpha|\beta} = g_{mn} \alpha^{m*} \beta^n = g^*_{nm} \alpha^{m*} \beta^n = \alpha^*_n \beta^n.
\end{gather}

In view of \Eq{eq:scalar2}, it is also convenient to think of $\bra{\alpha}$ as linear functionals, called covectors or one-forms, on~$\mc{V}$:
\begin{gather}
\bra{\alpha} \doteq \hat{g}(\ket{\alpha}, \ph\,),
\end{gather}
where ``$\ph$'' denotes a vector placeholder. One-forms comprise the dual space $\mc{V}^\dag$, where we also define the bases $\bra{e^n}$ and $\bra{e_n}$ like in $\mc{V}$. Then $\bra{\alpha} = \bra{e^n} \tilde{\alpha}_n^* = \bra{e_n} \tilde{\alpha}^{n*}$, where $\tilde{\alpha}_n^*$ and $\tilde{\alpha}^{n*}$ are some coefficients. This yields $\braket{\alpha|e_n} = \tilde{\alpha}_n^*$ and $\braket{\alpha|e^n} = \tilde{\alpha}^{n*}$. On the other hand, these must be equal to the complex-conjugate \Eqs{eq:aux13}, so $\tilde{\alpha}_n = \alpha_n$ and ${\tilde{\alpha}^{n} = \alpha^n}$. Hence, in summary,
\begin{gather}\label{eq:aa}
\ket{\alpha} = \alpha^n \ket{e_n}, \quad \bra{\alpha} = \bra{e^n} \alpha_n^*,
\end{gather}
so, via \Eq{eq:iso}, any vector $\ket{\alpha}$ is unambiguously mapped to its own one-form $\bra{\alpha}$, and vice versa.

\subsection{Tensors}

A tensor $\hat{F}$ of rank $(p,q)$, where $p$ and $q$ are nonnegative integers, is defined as a multilinear form ${\hat{F}:\,\mc{V}^{\dag p} \times \mc{V}^q \to \mathbb{C}}$ of $p$ one-forms and $q$ vectors. The result of application of $\hat{F}$ to any $\bar{p}$ one-forms and $\bar{q}$ vectors is a multilinear form too, $\mc{V}^{\dag (p-\bar{p})} \times {\mc{V}^{(q-\bar{q})} \to \mathbb{C}}$, so it is also a tensor, namely, of rank $(p-\bar{p}, q-\bar{q})$. In particular, a tensor of rank $(1,0)$ is a vector, a tensor of rank $(0,1)$ is a one-form, and a tensor of rank $(0,0)$ is a scalar. 

Of primary interest for us here are bilinear forms ($p + q = 2$). Below, we will discuss them in further detail, separating tensors into the following four classes.

\subsubsection{Class I} Any of the following four tensors
\begin{align}
^{\rm I}_1\hat{F}(\ket{\alpha}, \ket{\beta}) & = F_{mn} \alpha^{m*} \beta^n, \\
^{\rm I}_2\hat{F}(\ket{\alpha}, \bra{\beta}) & = {F_m}^n \alpha^{m*} \beta_n, \\
^{\rm I}_3\hat{F}(\bra{\alpha}, \ket{\beta}) & = {F^m}_n \alpha^*_m \beta^n,\\
^{\rm I}_4\hat{F}(\bra{\alpha}, \bra{\beta}) & = F^{mn} \alpha^*_m \beta_n
\end{align}
[where the right-hand sides are obtained much like in \Eq{eq:scpr}, using \Eqs{eq:ab}] determines the other three~via
\begin{gather}\nonumber
F_{mn} \alpha^{m*} \beta^n = {F_m}^n \alpha^{m*} \beta_n = {F^m}_n \alpha^*_m \beta^n = F^{mn} \alpha^*_m \beta_n.
\end{gather}
Indeed, provided the one-to-one mapping between vectors and one-forms, and also the fact that $g_{mn}$ is Hermitian, these equations yield the following one-to-one mapping between matrix elements of all the four types:
\begin{gather}\label{eq:aux23}
F_{mn} = {F_m}^k g_{kn} = g_{mk} {F^k}_n = g_{mk} F^{kl} g_{ln}.
\end{gather}

Of course, \Eq{eq:aux23} is applicable to $\hat{g}$ as well, for the metric was defined as a tensor of the $^{\rm I}_1\hat{F}$ type. Applying \Eq{eq:aux23} to $\hat{g}$, we then get
\begin{gather}
g_{mn} = {g_m}^k g_{kn} = g_{mk} {g^k}_n = g_{mk} g^{kl} g_{ln},
\end{gather}
which yields
\begin{gather}
{g_n}^m = {g^m}_n = g_{nk} g^{km} = \delta_m^n.
\end{gather}
In particular, this means that $g^{mn}$ is a matrix inverse to $g_{mn}$. One can then show that $g^{mn} = \braket{e^m|e^n}$ and also invert \Eq{eq:iso}, so that it gives
\begin{gather}\label{eq:au}
\alpha^m = g^{mn}\alpha_n.
\end{gather}

\subsubsection{Class II} Likewise, any of the following four tensors
\begin{align}
^{\rm II}_1\hat{F}(\ket{\alpha}, \ket{\beta}) & = F_{mn} \alpha^m \beta^n, \\
^{\rm II}_2\hat{F}(\ket{\alpha}, \bra{\beta}) & = {F_m}^n \alpha^m \beta_n, \\
^{\rm II}_3\hat{F}(\bra{\alpha}, \ket{\beta}) & = {F^m}_n \alpha_m \beta^n,\\
^{\rm II}_4\hat{F}(\bra{\alpha}, \bra{\beta}) & = F^{mn} \alpha_m \beta_n
\end{align}
determines the other three via
\begin{gather}\nonumber
F_{mn} \alpha^m \beta^n = {F_m}^n \alpha^m \beta_n = {F^m}_n \alpha_m \beta^n = F^{mn} \alpha_m \beta_n,
\end{gather}
which yields
\begin{gather}\label{eq:aux25}
F_{mn} = {F_m}^k g_{kn} = g^*_{mk} {F^k}_n = g^*_{mk} F^{kl} g_{ln}.
\end{gather}

\subsubsection{Classes III and IV} There are also other two classes of tensors, $^{\rm III}_j\hat{F}^*$ and $^{\rm IV}_j\hat{F}^*$, that are bilinear forms complex-conjugate to those of classes I and II. The index manipulation rules for them are complex-conjugate of \Eqs{eq:aux23} and \eq{eq:aux25}, correspondingly.

\subsection{Coordinate transformations}
\label{app:ct}

Now let us calculate how the components of tensors change at coordinate transformations. Suppose two bases, $\ket{e_n}$ and $\ket{e_n'}$, so, for any vector $\ket{\alpha}$, we have $\ket{\alpha} = \alpha^n \ket{e_n} = \alpha'^n \ket{e'_n}$; hence,
\begin{gather}\label{eq:t1}
\alpha^n = {U^m}_n \alpha'^n, \quad {U^m}_n \doteq \braket{e^m|e'_n}.
\end{gather}
The coordinate transformation for one-forms, $\bra{\alpha} = \bra{e^n}\alpha_n^* = \bra{e'^n}\alpha_n'^*$, is derived similarly,
\begin{gather}\label{eq:t2}
\alpha'_n = \braket{e'_n|\alpha} =  \braket{e'_n|e^m} \alpha_m = {U^{m*}}_n \alpha_m.
\end{gather}
The transformation for tensors that are bilinear forms of $\ket{\alpha}$ and (or) $\bra{\alpha}$ can then also be found, namely, by requiring that these forms remain invariant under the transformations \eq{eq:t1} and \eq{eq:t2}. Specifically, for class-I tensors (including $\hat{g}$) one needs to require $F_{mn} \alpha^{m*} \beta^n =  F_{mn}' \alpha'^{m*} \beta'^n$, which then yields $F'_{mn} = {U^k}^*_m F_{kl} {U^l}_n$. In particular, this shows that, if $F_{mn}$ is Hermitian, then $F'_{mn}$ is also Hermitian. For class-II tensors one needs to require $F_{mn} \alpha^{m} \beta^n =  F_{mn}' \alpha'^{m} \beta'^n$, which then yields $F'_{mn} = {U^k}_m F_{kl} {U^l}_n$. In particular, this shows that, if $F_{mn}$ is symmetric, then $F'_{mn}$ is also symmetric. The classes III and IV are treated similarly.

\section{Phase space geometry}
\label{sec:pspgeom}

\subsection{Metric structure}

It is convenient to equip the phase space $\msf{Z}$ with an effective metric given by the canonical symplectic form, $\hat{\varpi}$ [\Eq{eq:hameq}]. (We use the term ``effective'' here because, most commonly, metric is understood as a symmetric tensor, whereas $\varpi_{\alpha\beta}$ is antisymmetric.) Consider then the $2D$-dimensional space $\msf{Y}_\msf{z}$ tangent to $\msf{Z}$ at given $\msf{z}$; it is comprised of $\msf{y} = (\mvec{v}, \mvec{w})$, where $\mvec{v}$ is a vector tangent to $X$ at given $\mvec{x}$, and $\mvec{w}$ is tangent to $K$ at given $\mvec{k}$. For any two elements of $\msf{Y}_\msf{z}$, the bilinear map $\hat{\varpi}: \msf{Y}_\msf{z} \times \msf{Y}_\msf{z} \to \mathbb{R}$ then serves as an effective inner product; \ie
\begin{gather}
\msf{y} \wedge \msf{y}' \doteq \hat{\varpi}(\msf{y}, \msf{y}') = - \hat{\varpi}(\msf{y}', \msf{y}) = - \msf{y}' \wedge \msf{y},
\end{gather}
or [cf. \Eq{eq:omega}; Greek indexes now span from $1$ to $2D$]
\begin{gather}\label{eq:wedge}
\msf{y} \wedge \msf{y}' = \varpi_{\alpha\beta}\msf{y}^\alpha\msf{y}'^\beta = \mvec{w} \cdot \mvec{v}' - \mvec{w}' \cdot \mvec{v}.
\end{gather}
As usual, one hence can define the space $\msf{Y}^\dag_\msf{z}$ dual to $\msf{Y}$ and the rules of index manipulation,
\begin{gather}\label{eq:yindex}
\msf{y}_\alpha \doteq \varpi_{\alpha\beta}\msf{y}^\beta, \quad \msf{y}^\alpha = \varpi^{\alpha\beta} \msf{y}_\beta,
\end{gather}
where $\varpi^{\alpha\beta}$ is the matrix inverse to $\varpi_{\alpha\beta}$. [For $\msf{y}^\alpha = (v^1, \ldots v^D, w_1,\ldots w_D)$, one has $\msf{y}_\alpha = ({-w_1},\ldots {-w_D}, v^1, \ldots v^D)$.] Then the wedge product is expressed~as
\begin{gather}
\msf{y} \wedge \msf{y}' = \msf{y}^\alpha \msf{y}'_\alpha = - \msf{y}_\alpha \msf{y}'^\alpha
= -\varpi^{\alpha\beta}\msf{y}_\alpha\msf{y}'_\beta,
\end{gather}
which also defines the wedge product as a bilinear map on $\msf{Y}_\msf{z} \times \msf{Y}_\msf{z}^\dag$, $\msf{Y}_\msf{z}^\dag \times \msf{Y}_\msf{z}$, and $\msf{Y}_\msf{z}^\dag \times \msf{Y}_\msf{z}^\dag$.

Consider, for instance, the phase space gradient, $\nabla_\msf{z} \doteq (\nabla_\mvec{x}, \nabla_\mvec{k})$. According to \Eq{eq:yindex}, dual to it is the vector $\nabla^\msf{z} = (\nabla_\mvec{k}, -\nabla_\mvec{x})$. The wedge product of the two is then
\begin{gather}\label{eq:wedpoi}
(\nabla^\msf{z} \mc{G}) \wedge (\nabla_\msf{z} \mc{F}) = \{\mc{F}, \mc{G}\}
\end{gather}
for any $\mc{F}$ and $\mc{G}$. [Raising and lowering one or both indexes $\msf{z}$ in \Eq{eq:wedpoi} does not affect the right-hand side.] Correspondingly, the divergence of a vector field $\msf{y}$ on $\msf{Y}_\msf{z}$ is introduced as $\nabla_{\msf{z}} \wedge \msf{y}$, and the Laplacian of any scalar field $\mc{G}$ on $\msf{Z}$ is
\begin{gather}
\nabla_{\msf{z}} \wedge (\nabla^\msf{z} \mc{G}) \equiv 0.
\end{gather}
This also allows rewriting \Eq{eq:wedpoi} in a divergence form,
\begin{gather}\label{eq:divf}
\{\mc{F}, \mc{G}\} = - \nabla_\msf{z} \wedge (\mc{F}\,\nabla^\msf{z} \mc{G}).
\end{gather}

\subsection{Liouville equation}

The Liouville equation, \Eq{eq:pois2}, can now be cast in a vector form, namely, as follows. From \Eqs{eq:divf}, we~get
\begin{gather}\label{eq:liuvdiv}
\partial_t f = \nabla_\msf{z} \wedge (f\,\nabla^\msf{z} \mc{H}).
\end{gather}
Alternatively, \Eq{eq:wedpoi} yields
\begin{gather}\label{eq:liouvwedge}
\partial_t f + (\nabla^\msf{z} \mc{H}) \wedge (\nabla_\msf{z} f) = 0,
\end{gather}
whose rays are immediately seen to satisfy [cf. \Eqs{eq:char}]
\begin{gather}\label{eq:rays3}
\dot{\msf{z}}^\alpha = \nabla^\alpha \mc{H}.
\end{gather}
The advantage of this representation is that \Eqs{eq:liuvdiv}-\eq{eq:rays3} are invariant with respect to coordinate transformations in $\msf{Z}$ if $\hat{\varpi}$, which defines the wedge product [\Eq{eq:wedge}], is understood as a tensor, \ie if the matrix $\varpi_{\alpha\beta}$ is transformed accordingly (\Sec{app:ct}). Equation~\eq{eq:hameq} is then understood as a special case of $\varpi_{\alpha\beta}$ that corresponds to canonical coordinates as defined in \Sec{sec:cmom}.

Note also that \Eq{eq:liouvwedge} can be rewritten as\footnote{One may recognize this as the Koopman-von Neumann formulation of classical mechanics. For a review, see \Ref{phd:mauro03}.}
\begin{gather}\label{eq:liouH}
i\partial_t f = \hat{\msf{H}} f,
\end{gather}
where we introduced the Liouvillian
\begin{gather}\label{eq:Hdef}
\hat{\msf{H}} \doteq -i\{\ph, \mc{H}\} = -i(\nabla^\msf{z} \mc{H}) \wedge \nabla_\msf{z}
\end{gather}
as an operator on the space $\msf{F}$ of differentiable and square-integrable functions on $\msf{Z}$. The operator $\hat{\msf{p}} \doteq - i\nabla_\msf{z}$ is naturally understood as a momentum operator on $\msf{Z}$ (remember that the metric on $\msf{Z}$ has a determinant with unit absolute value), so one can rewrite \Eq{eq:Hdef} as
\begin{gather}\label{eq:Hdef2}
\hat{\msf{H}} = (\nabla^\msf{z} \mc{H}) \wedge \hat{\msf{p}}.
\end{gather}
It is easily seen that $\hat{\msf{H}}$ is Hermitian on $\msf{F}$, so the Liouville equation in the form \eq{eq:liouvwedge} can be viewed as an LSE, with $\hat{\msf{H}}$ serving as the Hamiltonian. The corresponding Lagrangian is given by
\begin{gather}
\msf{L} = \frac{i}{2}\,f^*\big[\partial_t f + \{f, \mc{H}\}\big] - \frac{i}{2}\,f\big[\partial_t f^* + \{f^*, \mc{H}\}\big].
\end{gather}
The functions $f$ and $f^*$ are formally viewed as independent here. For real initial values, one gets $f = f^*$ as a \textit{solution} of \Eq{eq:liouvwedge} at all times, as $\mc{H}$ in the LL is real.

Notably, \Eq{eq:Hdef2} can be viewed as the Hamiltonian of a free massless scalar particle traveling in a $2D$-dimensional symplectic vacuum with a varying ``speed of light'' $\nabla^\msf{z} \mc{H}$. Since $\hat{\msf{H}}$ is linear in $\hat{\msf{p}}$ [\ie the corresponding matrix $\hat{\Upsilon}$ is zero; cf. \Eq{eq:Ups}], this effective particle exhibits no dispersive spreading \textit{per~se}. Also notably, one can, in principle, address ``oscillations'' of $f$ in the same way as we approached the oscillations of $\ket{\psi}$; \ie one can construct a new fundamental space out of $\msf{F}$ (instead of $\Psi$), introduce a new density operator (quadratic in $f$ and thus quartic in $\psi$), define a new PSI, derive a FKE for it (just like we did for $f$), and so on. For a related discussion on how the Wigner function can itself be understood as an effective wave function, see \Ref{arX:bondar13}.\\


\noindent
\textsl{{Notice:} This manuscript has been authored by Princeton University under Contract Number 
DE-AC02-09CH11466 with the U.S. Department of Energy. The publisher, by accepting the article for publication acknowledges, that the United States Government retains a non-exclusive, paid-up, irrevocable, world-wide license to publish or reproduce the published form of this manuscript, or allow others to do so, for United States Government purposes.}

\end{document}